\documentclass[superscriptaddress,pra,aps,twocolumn,10pt,longbibliography]{revtex4-1}
\usepackage{bm}
\usepackage{amsmath}
\usepackage{amsfonts}
\usepackage{graphicx}
\usepackage{natbib}
\usepackage{epstopdf}
\usepackage[pdftex]{color}
\usepackage[dvipsnames]{xcolor}
\usepackage{hyperref}
\usepackage{verbatim}
\usepackage{float}
\hypersetup{
    colorlinks=true,       
    linkcolor=cyan,          
    citecolor=magenta,        
    filecolor=magenta,      
    urlcolor=cyan,           
    runcolor=cyan
}

\newcommand{\beq}{\begin{eqnarray}}
\newcommand{\eeq}{\end{eqnarray}}
\newcommand{\bes} {\begin{subequations}}
\newcommand{\ees} {\end{subequations}}

\newcommand{\ignore}[1]{}

\graphicspath{{figs/}}

\begin{document}

\title{Power of Pausing: Advancing Understanding of Thermalization in Experimental Quantum Annealers}

\author{Jeffrey Marshall}
\affiliation{USRA Research Institute for Advanced Computer Science, Mountain View, California 94035, USA}
\affiliation{USRA NAMS Quantum Academy R\&D Student Program, NASA Ames Research Center, Mountain View, California 94035, USA}

\affiliation{Department of Physics and Astronomy, and Center for Quantum Information Science \& Technology, University of Southern California, Los Angeles, California 90089, USA}

\author{Davide Venturelli}
\affiliation{USRA Research Institute for Advanced Computer Science, Mountain View, California 94035, USA}
\affiliation{QuAIL, NASA Ames Research Center, Moffett Field, California 94035, USA}

\author{Itay Hen}
\affiliation{Department of Physics and Astronomy, and Center for Quantum Information Science \& Technology, University of Southern California, Los Angeles, California 90089, USA}
\affiliation{Information Sciences Institute, University of Southern California, Marina del Rey, California 90292, USA}

\author{Eleanor G. Rieffel}
\affiliation{QuAIL, NASA Ames Research Center, Moffett Field, California 94035, USA}

\begin{abstract}
We investigate alternative annealing schedules on the current generation of quantum annealing hardware (the D-Wave 2000Q), which includes the use of forward and reverse annealing with an intermediate pause. 
This work provides new insights into the inner workings of these devices (and quantum devices in general), particular into how thermal effects govern the system dynamics. 
We show that a pause mid-way through the anneal can cause a dramatic change in the output distribution, and we provide evidence suggesting thermalization is indeed occurring during such a pause. 
We demonstrate that upon pausing the system in a narrow region shortly after the minimum gap, the probability of successfully finding the ground state of the problem Hamiltonian can be increased over an order of magnitude.
We relate this effect to relaxation (i.e. thermalization) after diabatic and thermal excitations that occur in the region near to the minimum gap. For a set of large-scale problems of up to 500 qubits, we demonstrate that the distribution returned from the annealer very closely matches a (classical) Boltzmann distribution of the problem Hamiltonian, albeit one with a temperature at least 1.5 times higher than the (effective) temperature of the annealer.
Moreover, we show that larger problems are more likely to thermalize to a classical Boltzmann distribution.
\end{abstract}
\maketitle

\section{Introduction}

Inspired by thermal annealing and by the adiabatic theorem of quantum mechanics, quantum annealers are designed to make use of diminishing quantum fluctuations in order to efficiently explore the solution space of particular discrete optimization problems.
In the last few years, chip sizes have grown in accordance with Moore's law, and current devices contain on the order of 2000 superconducting qubits, potentially allowing for relatively large scale problems to be solved.
Though progress has been made \cite{Albash-scaling,googleTunneling}, whether quantum annealing provides a speedup \cite{ronnow:14,mandra2016strengths} over conventional approaches to optimization remains open.
Alternatively, researchers have suggested that quantum annealers may be useful for thermal sampling tasks \cite{Amin:2015,Amin:boltzmann,perdomo,fairInUnfair}, such as the NP-hard problem of sampling from a Boltzmann distribution, which has application in machine learning and artificial intelligence. Early results and theory exist \cite{perdomo,adachi,Amin:boltzmann}, but the extent to which thermalization occurs in quantum annealing remains a hotly contested issue.

We take advantage of two advances to further understand the behavior of quantum annealers and the distributions they output. The first is the introduction of new features that support a wider variety of annealing schedules on the D-Wave 2000Q quantum annealers; the only schedule parameter earlier machines provided was setting the total anneal time, and thereby the speed with which the default annealing schedule was traversed. Specifically, we make use of the new pause feature, which allows one to pause the anneal, keeping the strengths of the driver Hamiltonian and the problem Hamiltonian constant for extended periods of time (up to $\sim 1$ms), before completing the default annealing schedule. We also make use of the reverse annealing feature, which allows one to start in a classical state with the strength of the problem Hamiltonian and the driver Hamiltonians what they would be at the end of the default anneal, and to then increase the strength of the driver Hamiltonian and reduce that of the problem Hamiltonian, following the default schedule in reverse, up until a point at which one then anneals forward according to the default schedule, possibly pausing between the reverse and forward anneals. The second advance we rely on is new entropic sampling techniques based on population annealing that enable accurate estimates (i.e., with quantifiable error) of the eigenspectum degeneracies for large-scale (e.g. $500$ qubits) planted-solution problems \cite{Barash-2018-entropic_PA,fairInUnfair}.

We surveyed the performance of the quantum annealer on two problem classes, studying the output of the device under an anneal with an intermediate pause inserted at different locations during the anneal, and for differing pause lengths. The first problem class contains 12-qubit problems for which we can compute exactly the eigenspectrum throughout the anneal. The second class contains planted-solution problems, for which entropic sampling provides accurate estimates of the spectrum degeneracies for the problem Hamiltonian, with hundreds of qubits  (see Appendix \ref{sect:cbm} for more information). We found that a pause can increase the performance by orders of magnitude when the pause occurs within a well defined, relatively narrow region of the anneal, but has little effect if placed outside that region. This effect, and the location of the best region for pausing, is remarkably robust across instances within a class, pause lengths, and total annealing times. 
We interpret the results with a phenomenological model that takes into account
the relevant time scales involved in the annealing process, and also used the reverse anneal feature to further investigate these phenomena. This picture suggests that a pause is effective after the minimum gap, and after thermal excitations start to diminish, but while the driver is still strong enough and the eigenvalues of the Hamiltonian are not too far apart so that significant dynamics can still take place. This picture differs greatly from that for closed system adiabatic quantum computing for which a pause would be most effective at the minimum gap. For the 12-qubit problems, we were able to compute the location of the minimum gap and confirm our open system picture. 

We then turned to the question of how Boltzmann the final distributions are. For the 12-qubit problems, we compared the final distribution with the projected (onto the computational basis) quantum Boltzmann distributions for the full Hamiltonian at all points along the anneal. The final distribution fits reasonably well to the projected quantum Boltzmann distribution at the optimal pause location and at the device's operating temperature (KL divergence $<0.1$), but fits even better to points later in the anneal at temperatures higher than the operating temperature. Outside this parameter range the fit was poor. For planted-solution problems, equipped with the energy spectrum degeneracies of the problem, we compared the final distribution with the classical Boltzmann distribution of the problem Hamiltonian. Here, we found a strong correlation ($R^2 > 0.9$) for all instances and pause locations, with the correlations becoming even stronger as the problem size increased and as a pause was added, with the best fit being at the optimal pause location, but for a higher temperature than might be expected. 

We conclude with a discussion of these results, and a call for a deeper theory that can make quantitative predictions about the optimal pause location, the location of the best fit projected Boltzmann distribution, the best fit location and temperature for fit with a classical Boltzmann distribution for the problem Hamiltonian, as well as experimental investigation of these phenomena for other problem classes. The observations we make are relevant not only for quantum annealers, but for any quantum device which is non-negligibly coupled to a thermal environment, thus shedding light on fundamental physical processes involved across a broad range of devices. We begin by reviewing some background material before presenting our results and discussing their interpretation and implications.

\subsection{Background}
Transverse field quantum annealing evolves the system over rescaled time $s=t/t_a \in [0,1]$, where $t$ is the time, and $t_a \in [1,2000]\mu$s the total run-time (chosen by the user). We will occasionally refer to the rate of the anneal $\mathrm{d}s/\mathrm{d}t$ which on the D-Wave 2000Q can be set to zero during the pause or take values in interval $\mathrm{d}s/\mathrm{d}t\in [0.0005,1]\mu$s$^{-1}$ otherwise. The time-dependent Hamiltonian is of the form
\beq
H(s)=A(s)H_d+B(s)H_p \,,
\label{annealing}
\eeq
where 
$H_p=\sum_{\langle i,j\rangle} J_{ij} \sigma_i^z \sigma_j^z + \sum_i h_i \sigma_i^z$ 
is the programmable Ising spin-glass problem (the final Hamiltonian) to be sampled defined
by the parameters $\{ J_{ij},h_i\}$, and $H_d=-\sum_i^N \sigma^x_i$ is a
transverse-field (or `driver') Hamiltonian which provides the quantum fluctuations (the initial Hamiltonian). Here $N$ is the total number of qubits in the problem, and $\langle i,j \rangle$ indicates the sum is only over a coupled qubits, defined by the hardware `chimera' graph (see Fig.~\ref{fig:chimera} of Appendix \ref{sect:figs}).
The device we use, the D-Wave 2000Q contains 16$\times$16 unit cells each containing 8 qubits, thus having a maximum of 2048 qubits. However, because of some faulty qubits/couplers, the actual number of operating qubits is 2031. These `dead' qubits appear randomly dispersed throughout the hardware graph.

\begin{figure}
\begin{center}
\includegraphics[width=0.9\columnwidth]{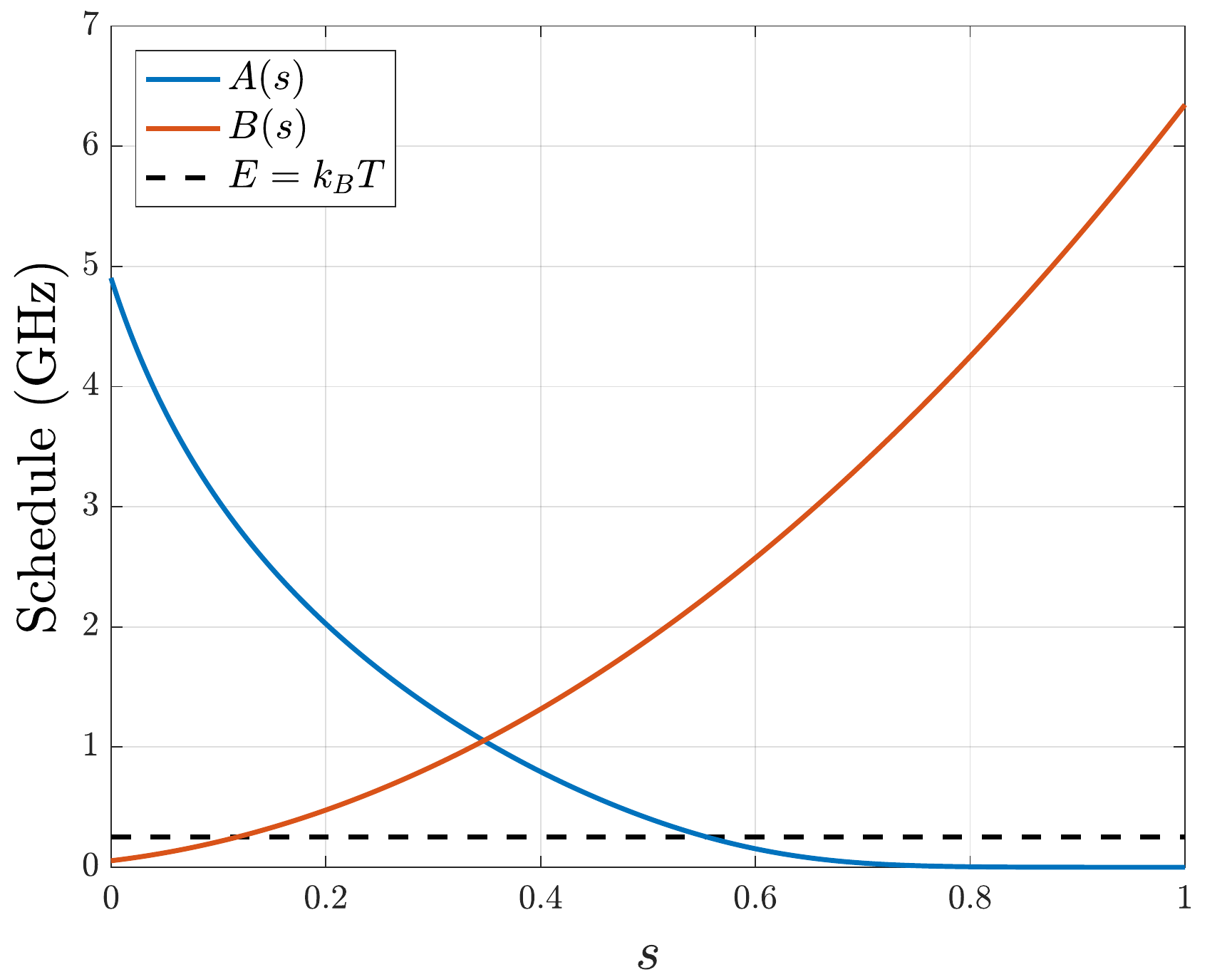}
\caption{Annealing schedule in GHz (units of $h=1$). The operating temperature ($T=12.1$ mK, or equivalently $0.25$ GHz) of the chip is also shown (black-dashed line).
}
\label{fig:schedule}
\end{center}
\end{figure}

The initial state is fixed as the ground state of $H_d$, $|\Psi(0)\rangle = |+\rangle^{\otimes N}$ where $|+\rangle = \frac{1}{\sqrt{2}}(|0\rangle + |1\rangle)$ (defined in the computational basis via $\sigma^z = |1\rangle \langle 1| - |0\rangle \langle 0|$).
The manner in which the Hamiltonian is evolved in time is determined by the annealing schedule (i.e. the time dependence of $A,B$). The default schedule for the D-Wave 2000Q is shown in Fig.~\ref{fig:schedule}.

After an annealing run, the system is measured in the computational basis. Performing many such runs allows statistics about the annealer to be collected; useful measures include the probability of successfully finding the ground-state of $H_p$ (which is the solution to a classical optimization problem) which we denote as $P_0$, or the average energy returned $\langle E\rangle$.

One way to provide more robust statistics, is by changing the `gauge' of the problem. 
This is a trivial re-mapping of the problem so as to avoid certain biases which may be present for certain couplers/qubits (e.g., some couplers may have fewer analog control errors associated with programming in $J=+1$ as compared to $J=-1$, or certain qubits may be more likely to align with $+z$ as compared to $-z$ even in the absence of any fields). The mapping involves changing the couplings/fields as $J_{ij}\rightarrow J_{ij}r_ir_j,h_i\rightarrow h_ir_i$, where $\overrightarrow{r}=(r_1,\dots ,r_N)$ is a vector of random entries $r_i \in \{-1,1\}$.
Notice any configuration $\overrightarrow{s}=(s_1,\dots,s_N)$ has a corresponding configuration of the mapped problem $\overrightarrow{s'}=(r_1s_1,\dots ,r_Ns_N)$ with the same cost, thus the problem itself is exactly the same.

\begin{figure}
\begin{center}
\includegraphics[width=0.9\columnwidth]{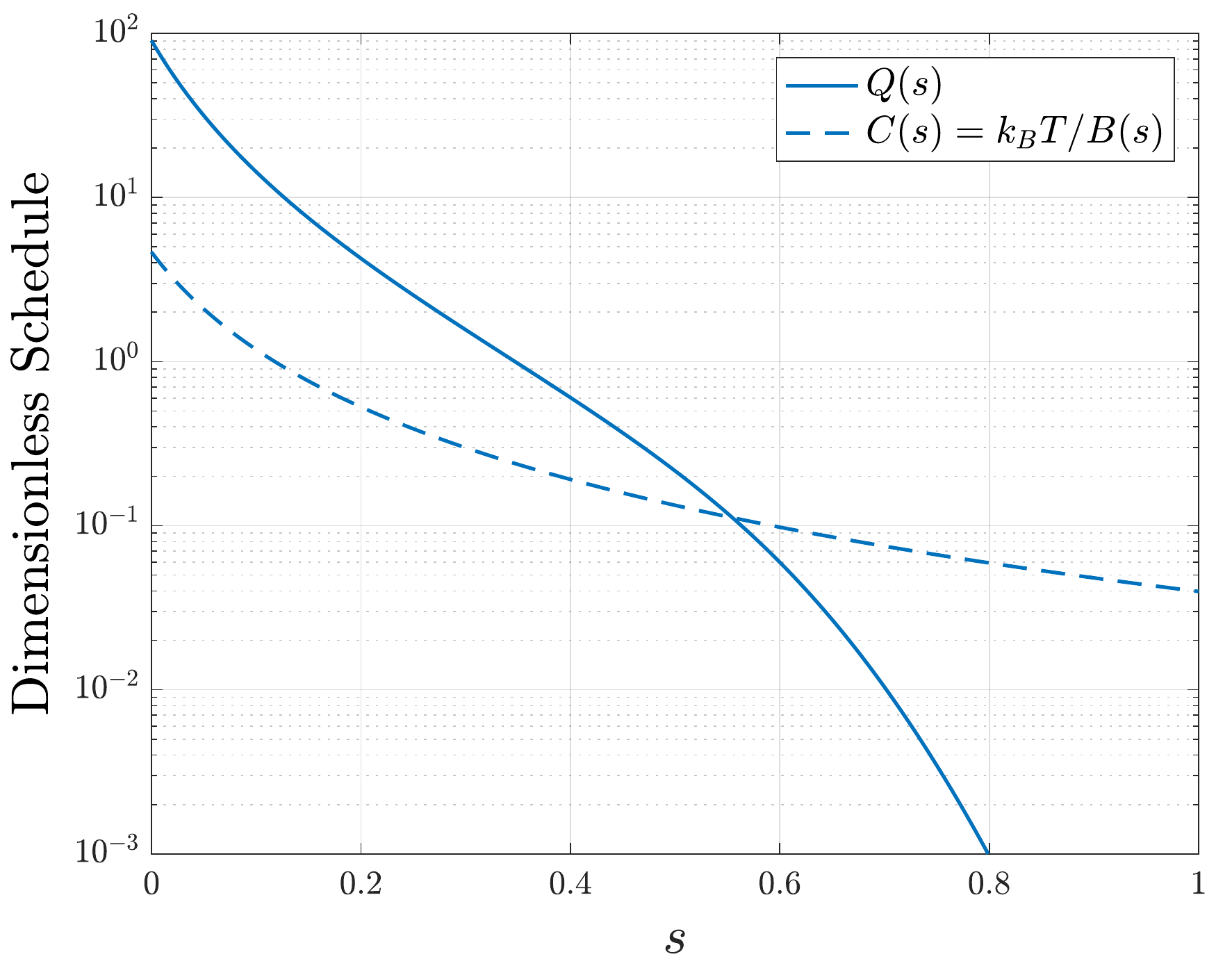}
\caption{Dimensionless annealing schedule. We plot the ratio $Q(s):=A(s)/B(s)$, and the ratio of the operating temperature ($T=12.1$mK) to the strength of the problem Hamiltonian, $C(s):=k_B T/B(s)$.}
\label{fig:dim-schedule}
\end{center}
\end{figure}

The hardware graph allows one to study a variety of problems useful for both scientific research as well as application and industry. Although embedding means in principle any graph can be encoded, this introduces additional complications and also limits the size considerably (with the overhead scaling quadratically in the worst case). For these reasons we will focus our attention on problems whose native graph is that of the D-Wave machine, allowing us to study sizes up to around 2000 qubits. We consider two natural problem classes, the first being with all random real valued couplings, uniformly distributed, and the second, problems constructed around a planted solution, allowing for knowledge of the ground state \cite{hen:15}, circumventing the need for heuristic solvers to quantify the success of the device. For more discussion on problems which are relevant for current generation quantum annealing devices, see e.g.~Ref.~\cite{test-driving}.

An important quantity identified in Ref.~\cite{marshall-rieffel-hen-2017} is the ratio between the strength of the driving Hamiltonian and the problem Hamiltonian,
$Q(s):=A(s)/B(s)$, shown in Fig.~\ref{fig:dim-schedule} for the D-Wave 2000Q schedule. Also important are the classical thermal fluctuations which are governed by the quantity $C(s):=k_B T/B(s)$, where $T$ is the temperature of the annealer.
Observing the relative scales of the characteristic energies associated to the driving terms (i.e. transverse field, environmental bath) with the energy of the problem Hamiltonian allows us to infer the existence of different regimes where a given process becomes energetically dominant. 
In particular, i) at early times when $Q\gg C > 1$, and the system mostly remains in the ground state of $H$, ii) when $Q\sim C \sim O(1)$ and non-trivial dynamics occur with $H_d$ driving various transitions between computational basis states, and iii) $Q \ll C \ll 1$ when the Hamiltonian is mostly diagonal (in the computational basis) and little population transfer occurs between the eigenstates (the `frozen' region) through diabatic transitions. Thermal transitions could occur but those depend also on the strength of the coupling to the thermal bath (see below).

\subsection{Adapting the standard annealing schedule}
\begin{figure}
\begin{center}
\includegraphics[width=0.9\columnwidth]{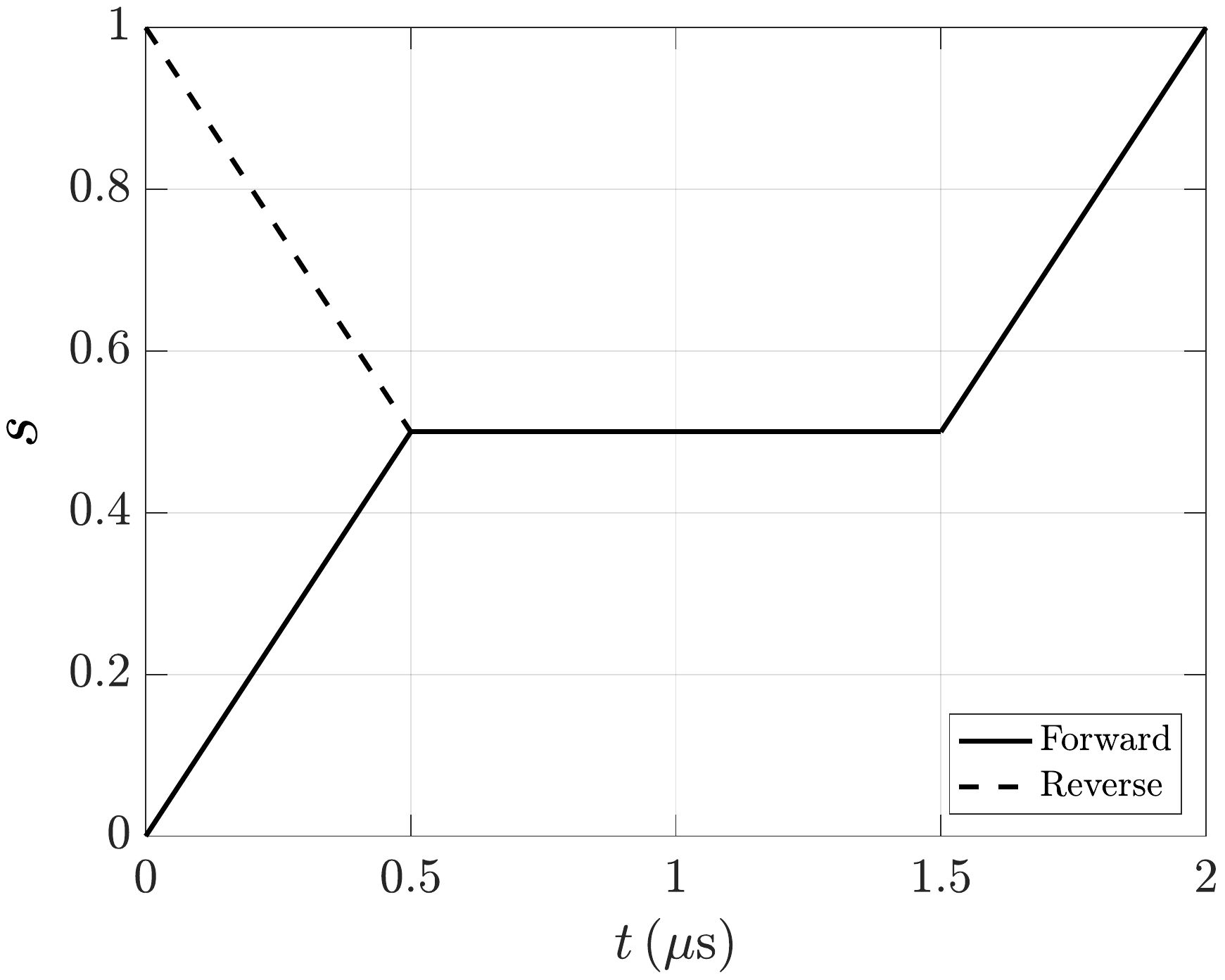}
\caption{Example of annealing parameter $s$ as a function of time $t$ for an anneal with a pause, for both forward and reverse annealing. Here a 1$\mu$s pause ($t_p=1\mu$s) is inserted into the annealing schedule at $s_p=0.5$ (i.e. at $t=0.5\mu$s), which otherwise has a total anneal time of $t_a=1\mu$s.
}
\label{fig:pause-example}
\end{center}
\end{figure}

The current generation of hardware, the D-Wave 2000Q, allows users to tweak the default schedule in various ways.
In particular this gives one the ability to:
\begin{itemize}
\item[1)] Pause the evolution at some intermediate point $s_p<1$ in the anneal, for time $t_p$.
\item[2)] Perform reverse annealing, where the system is initialized in a classical configuration at time $s=1$, evolved backwards to an earlier time $s_p<1$
before evolving the system back to time $s=1$ where a read-out occurs. Additionally, a pause can be inserted between the two evolutions.
\item[3)] Speed-up or slow-down the schedule during intermediate intervals of the anneal.
\item[4)] Provide per-qubit annealing offsets.
\end{itemize}
Fig.~\ref{fig:pause-example} shows an example of an annealing schedule with a pause, and also an example of a reverse annealing schedule.

Based on these features, new methods of sampling from an annealer have been developed and proposed, such as exploiting reverse annealing to explore the energy landscape in a novel manner \cite{king-topological,reverseAnnealing,ottaviani2018low, venturelli2018reverse,dwaveReverse}. Moreover, performance advantages have been observed  by offsetting the fields of some of the qubits, allowing one to evade spurious transitions which occur during the minimum gap \cite{offset,inhomogeneous-driving}.

This paper focuses mostly on the first capability listed above, where we embed a pause in the default annealing schedule, i.e., the Hamiltonian is fixed at $H(s_p)$ for lengths of time chosen by the user.
This approach enables us to study key mechanisms determining the output of the annealer, such as thermalization, and to identify key regimes of the anneal.
To further confirm our explanation for the behavior of the system, 
we use reverse annealing in a similar way, inserting pauses of varying lengths at differing locations.

\subsection{Theory \label{sect:theory}}

Naively, one might expect the final distribution at the end of the anneal to be a classical Boltzmann distribution for the problem Hamiltonian $H_p$ at the operating temperature of the device, specifically, $\rho \sim \exp(-\beta B(1)H_p)$, where $\beta=1/k_B T$, with $T$ the operating temperature of the annealer (on the order of 10-20mK in the various generations of D-Wave annealers). But it has long been known that is not the case. The freeze-out hypothesis due to Amin \cite{Amin:2015} suggests that while early in the anneal the system thermalizes, later in the anneal, at an instance dependent, but temperature independent, freeze-out point, the transverse field strength has diminished and the gap between eigenvalues increased to the point that the transition matrix elements are so small that essentially no dynamics happens after this ``freeze-out'' point. This hypothesis predicts that, for instances with well-defined freeze-out points, the final distribution would indeed be a classical Boltzmann distribution for $H_p$ but at a higher ``effective temperature'' corresponding to the freeze-out point. More specifically, the theory hypothesizes a freeze-out point $s^*$ that occurs once $Q(s^*)$ and $C(s^*)$ are so `small' that the time-scale upon which the transverse field drives transitions between eigenstates of $H_p$ is much longer than the system evolution time, hence little population transfer occurs.
The expected distribution would then be close to $\rho \sim \exp(-\beta B(s^*)H_p)$ \cite{Amin:2015,marshall-rieffel-hen-2017}.
The paper proposing this hypothesis \cite{Amin:2015} recognized that a well-defined freeze-out point will only exist under certain circumstances, with more recent debate as to how typically those circumstances hold.

Taking inspiration from work that addressed an open-system treatment in the weak coupling
regime for general problems
\cite{adiabaticME}, and in the non-perturbative regime for specific problems \cite{smelyanskiy2017quantum,boixo2016computational},
we look at
transitions between instantaneous energy levels $\omega_{ij}(s):= E_j(s) - E_i(s)>0$ that are governed by Fermi's Golden Rule rate $\Gamma_{i\rightarrow j} :=\Gamma_{ij}$:
\begin{equation}
 \Gamma_{ij}(s)\propto \frac{J(\omega_{ij}(s))}{\exp(\beta \omega_{ij}(s))-1}\sum_{k,\alpha} g_\alpha^2 |\langle E_i(s) |\sigma_k^\alpha |E_j(s)\rangle|^2,
\label{eq:rate}
\end{equation} 
where $g_\alpha$ is the environment coupling strength  to the $\alpha=x,y,z$ component of the system spins, and $\sigma_k^\alpha$ is the $\alpha$ Pauli operator acting on the $k$-th spin. $J(\omega)$ is the spectral density function of the bath, which is often modelled as Ohmic with high-frequency cut-off, see Ref.~\cite{adiabaticME}. For detailed balance, the rate $\Gamma_{ji}$ is the same up to a factor of $\exp(\beta \omega_{ij})$.

The explanation of freeze-out in this picture is that as $s\rightarrow 1$, energy gaps $\omega_{ij}$ open up, as well as the matrix elements $\langle E_i(s)|\sigma^z|E_j(s)\rangle \rightarrow 0$ (which is typically the dominant environment-spin coupling $g_z \gg g_{x,y}$ \cite{leggett-2state-system,Hanson-RWA,boixo2016computational}) as the Hamiltonian becomes more diagonal in the $z$-basis, thus the transition rates dramatically slow down late in the anneal.
Therefore, the two strongest (possibly competing) effects determining the relaxation rate Eq.~(\ref{eq:rate}) are the instantaneous energy gap, and $Q(s)$ (via the matrix element).

Marshall et al.~\cite{marshall-rieffel-hen-2017} used annealers operating at two different temperatures to corroborate the freeze-out hypothesis, finding consistency in the freeze-out point location for instances with well defined freeze-out points late in the anneal. On the negative side, however, this study showed that the most instances did not fall into this category, leaving open whether one would even expect the final distribution to be a classical Boltzmann distribution for $H_p$, in the typical case.

\begin{figure}
\begin{center}
\includegraphics[width=0.9\columnwidth]{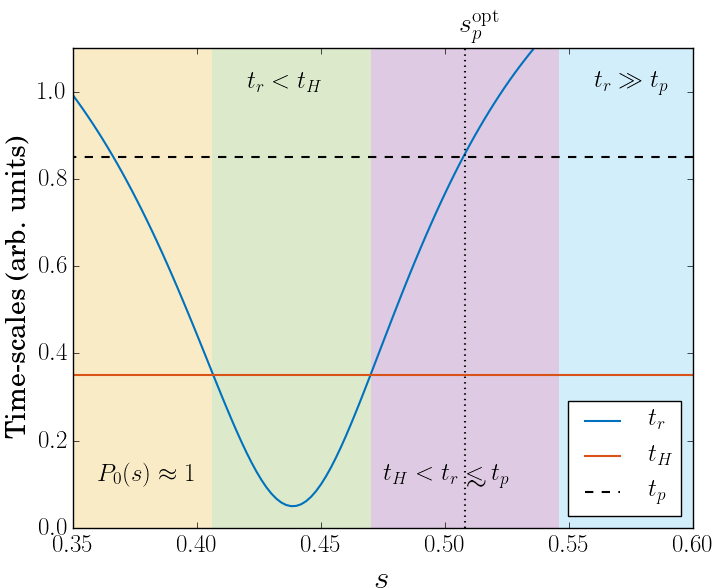}
\caption{A cartoon example illustrating time-scales relevant to analyzing a quantum annealer with coupling to a thermal bath. 
We indicate the thermal relaxation time-scale $t_r$ 
by the blue solid line, the Hamiltonian evolution time-scale 
$t_H$, by the red (horizontal) solid line, and the pause time scale, $t_p$, by the black dash line \cite{note-cartoon}.
The diagram focuses on the part of the anneal where the relaxation time scale is shortest, around the location of the minimum gap of the problem (e.g. around $s=0.44$ in this example). 
We show four characteristic regions during the anneal. 1) [yellow] at early times, when the ground state is separated by a large gap from any excited state, and the population in the ground state is approximately 1. 
2) [green] As the energy gap closes, and the thermalization time-scale decreases rapidly, thermal excitations may occur. 
3) [purple] As the energy gaps open up, the relaxation time-scale increases (exponentially). Once $t_r > t_H$ instantaneous thermalization can no longer occur. 
If however $t_r \lesssim t_p$, a pause may allow effective thermalization, which could lead to a significantly larger population in the ground state.
We indicate the point where $t_r \approx t_p$ by $s_p^{\mathrm{opt}}$. 
Finally, in region 
4) [blue], where $t_r\gg t_p$ is the longest time scale, very little population transfer will occur (even if one pauses the system for time $t_p$). The dynamics are effectively frozen.}
\label{fig:cartoon}
\end{center}
\end{figure}

With the new entropic sampling techniques, and the new pause and reverse annealing capabilities on the D-Wave 2000Q quantum annealer, it is now possible to develop a better understanding of the final distributions returned and the extent to which thermalization happens along the way. Before describing the results, we discuss a qualitative theory in terms of key time scales involved in open system quantum annealing. The key time scales we consider are 
\begin{itemize}
    \item the {\it pause time scale} $t_p$, the length of the pause inserted into the annealing schedule,
    \item the {\it relaxation (thermalization) time scale} $t_r$, related to the inverse of Eq.~(\ref{eq:rate}), and 
    \item the {\it Hamiltonian evolution time scale}, $t_H(s)\sim t_a \left\|\frac{\mathrm{d}\tilde{H}(s)}{\mathrm{d}s}\right\|^{-1}$, which is the characteristic time upon which the system Hamiltonian changes. This quantity depends on both $t_a$ and the parameters $A(s)$ and $B(s)$. For reasons of dimensionality, $\tilde{H}$ is a dimensionless version of the Hamiltonian (e.g. $\tilde{H}=H/A(0)$).
\end{itemize}
Fig.~\ref{fig:cartoon} provides a schematic illustration of four regions within an anneal which one would expect have qualitatively different dynamics. In the figure, we plot $t_H$ as a straight line only for convenience. Early in the anneal, the eigenvalues are spread, with $E_i - E_0 \gg 0$, and with the system starting the ground state state, one expects little in the way of population dynamics and the probability that the system is in the instantaneous ground state to be high, $P_0 \approx 1$. Once the eigengaps decrease sufficiently and $t_r$ is within the Hamiltonian evolution time scale, we expect to see thermal excitations, with sufficient dynamics that the system instantaneously thermalizes. As we leave this region, the energy gaps open, and the strength of the driver Hamiltonian reduces so that transition rates (up and down) begin to decrease, and instantaneous thermalization can no longer take place.
If however $t_r \lesssim t_p$, we expect a pause to aid thermalization and thereby substantially improving solution probability.
This region is may be relatively narrow since $t_r$ increases exponentially as the gaps open up. 
Near the end of the anneal, with the gap distances becoming very large, and with negligible transverse field applied, one would expect that the dynamics are effectively frozen.

\section{Results}

Throughout this work we consider problems of two different types:
\begin{enumerate}
    \item To study large problems, we work with problems of the planted-solution type \cite{hen:15}, such as the $\mathcal{I}_{783}$ instance of Fig.~\ref{fig:zoom-pause}. We chose these problems, because recent techniques enable us 
    to know in advance the general analytic form of the spectrum of $H_p$, including the ground state, as well as certain information about the degeneracy of the energy levels (see Appendix \ref{sect:cbm}). 
    \item To analyze problems with respect to spectral properties of the full Hamiltonian, $H(s)$, we used $12$-qubit problems, where $J_{ij}\in[-1,1]$ (uniformly random). 
\end{enumerate}
In both cases, we used zero local fields, $h_i=0$.

While problems with local fields or large ferromagnetic structure (e.g. embedded problems) could benefit from specific analysis, we expect that the general results and arguments presented will be generalizable to a large range of problem sets.

\subsection{Forward annealing with a pause \label{sect:FA}}

We consider the following simple adaptation to the standard annealing schedule. Allow the system to run as normal to some (re-scaled) time $s_p\in[0,1]$, upon which we pause the system for time $t_p\in[0,2000]\mu$s, after which we continue the evolution as per normal. 

\begin{figure}
\begin{center}
\includegraphics[width=0.9\columnwidth]{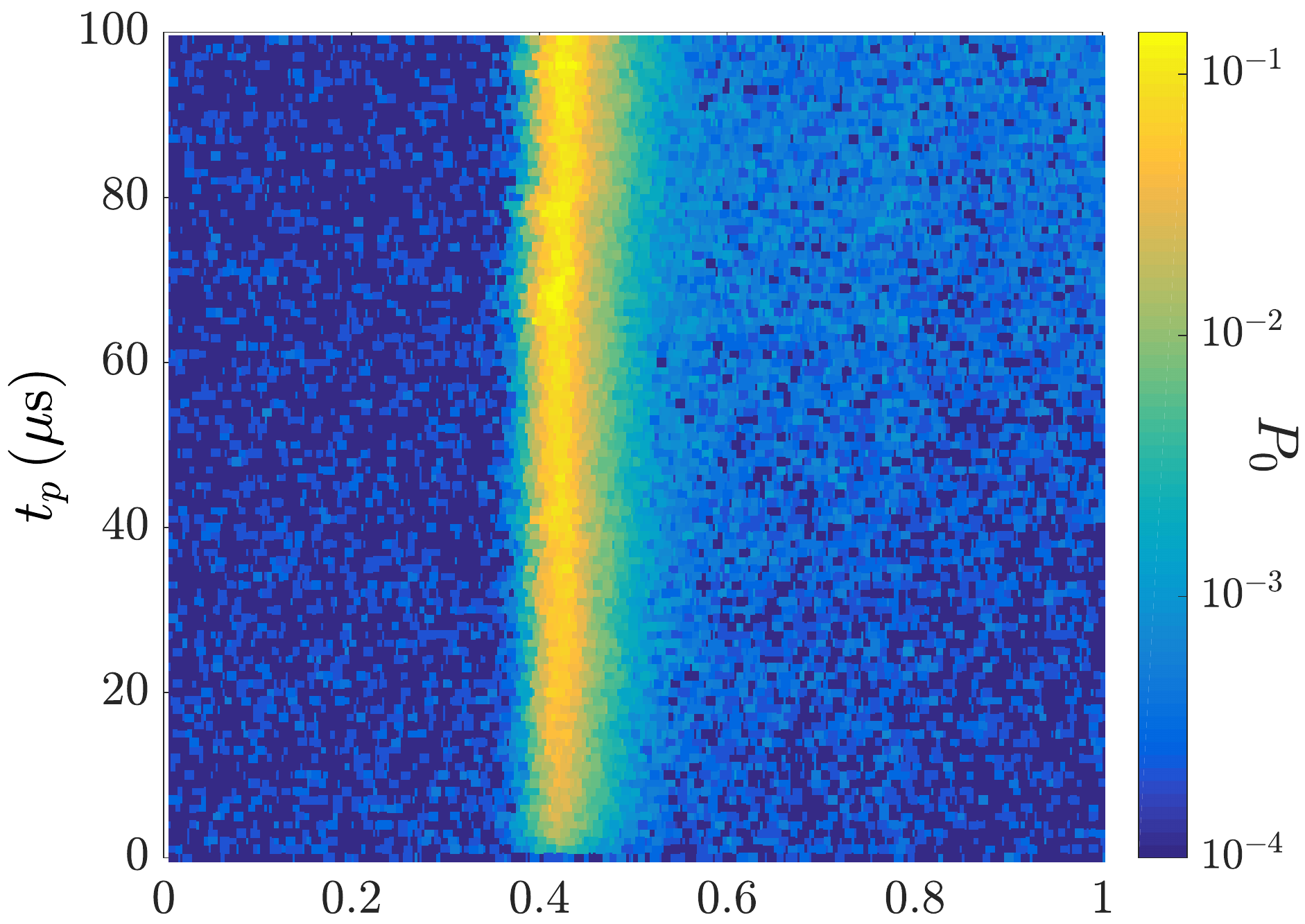}
\includegraphics[width=0.9\columnwidth]{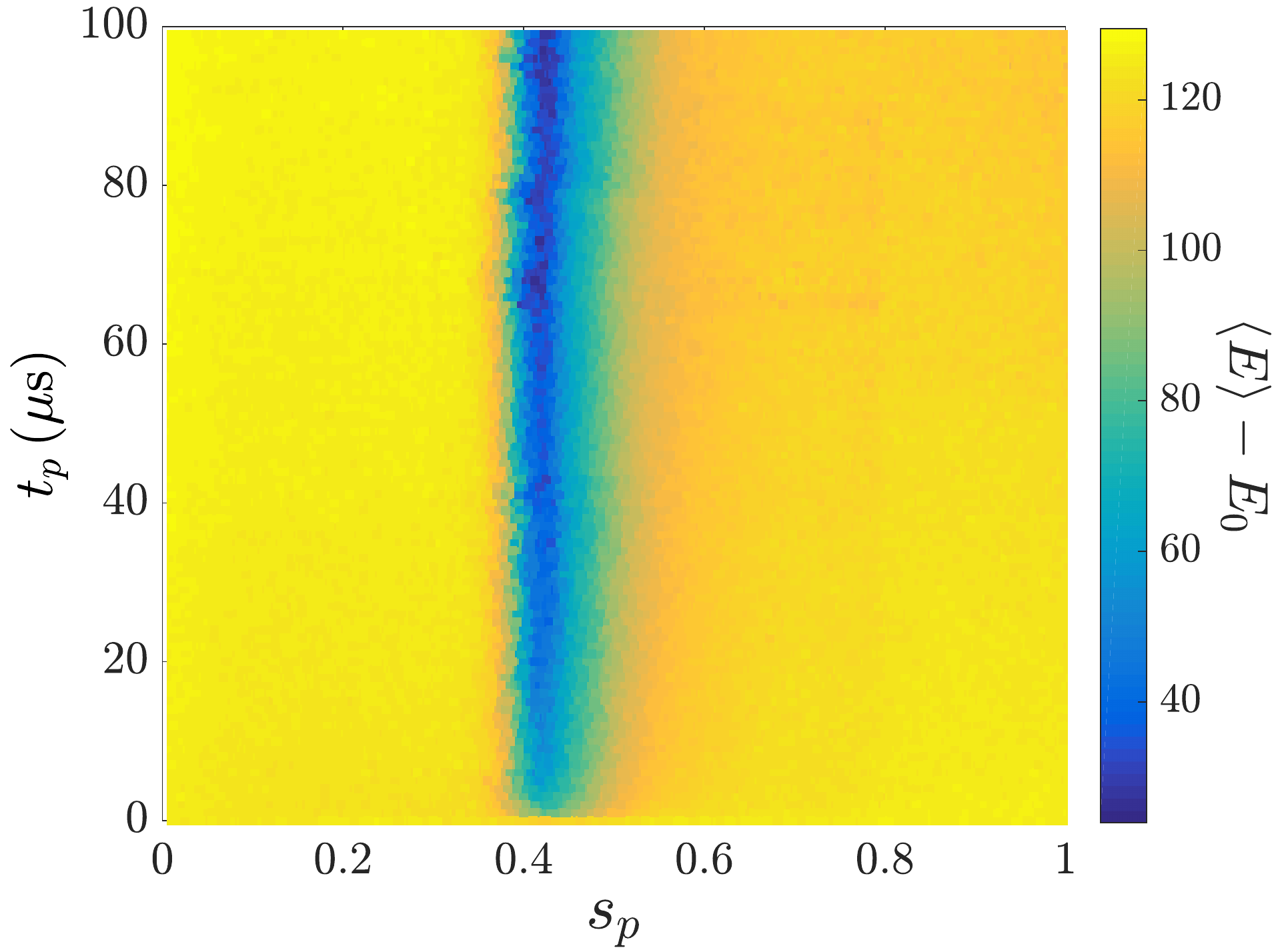}
\caption{Forward annealing with a pause for a single 783 qubit (planted-solution) problem instance ($\mathcal{I}_{783}$). The top figure shows the success probability with respect to pause length $t_p$, and the location of the pause $s_p$. The total evolution time (aside from the pause) is 1$\mu$s. The corresponding bottom figure shows the average energy (in arbitrary units) returned by the annealer. Each data point is averaged from 10000 anneals with 5 different choice of gauge.
In the absence of a pause, $P_0  \approx 10^{-4}$.}
\label{fig:zoom-pause}
\end{center}
\end{figure}

Fig.~\ref{fig:zoom-pause} shows that this pause dramatically effects the samples returned from the annealer. While Fig.~\ref{fig:zoom-pause} is for one instance, almost all  instances we tested exhibit this behavior, including a strong peak in the success probability when a pause is inserted into the regular annealing schedule within a narrow band of values of $s_p$.
Fig.~\ref{fig:zoom-pause} shows that the corresponding average energy returned is also significantly reduced.
We define the `optimal pause point', $s_p^{\mathrm{opt}}$, as the point in the anneal for which a pause returns the lowest average energy returned from many samples \cite{note-sopt} (just after $s_p=0.4$ in this example).

\begin{figure}
\begin{center}
\includegraphics[width=0.9\columnwidth]{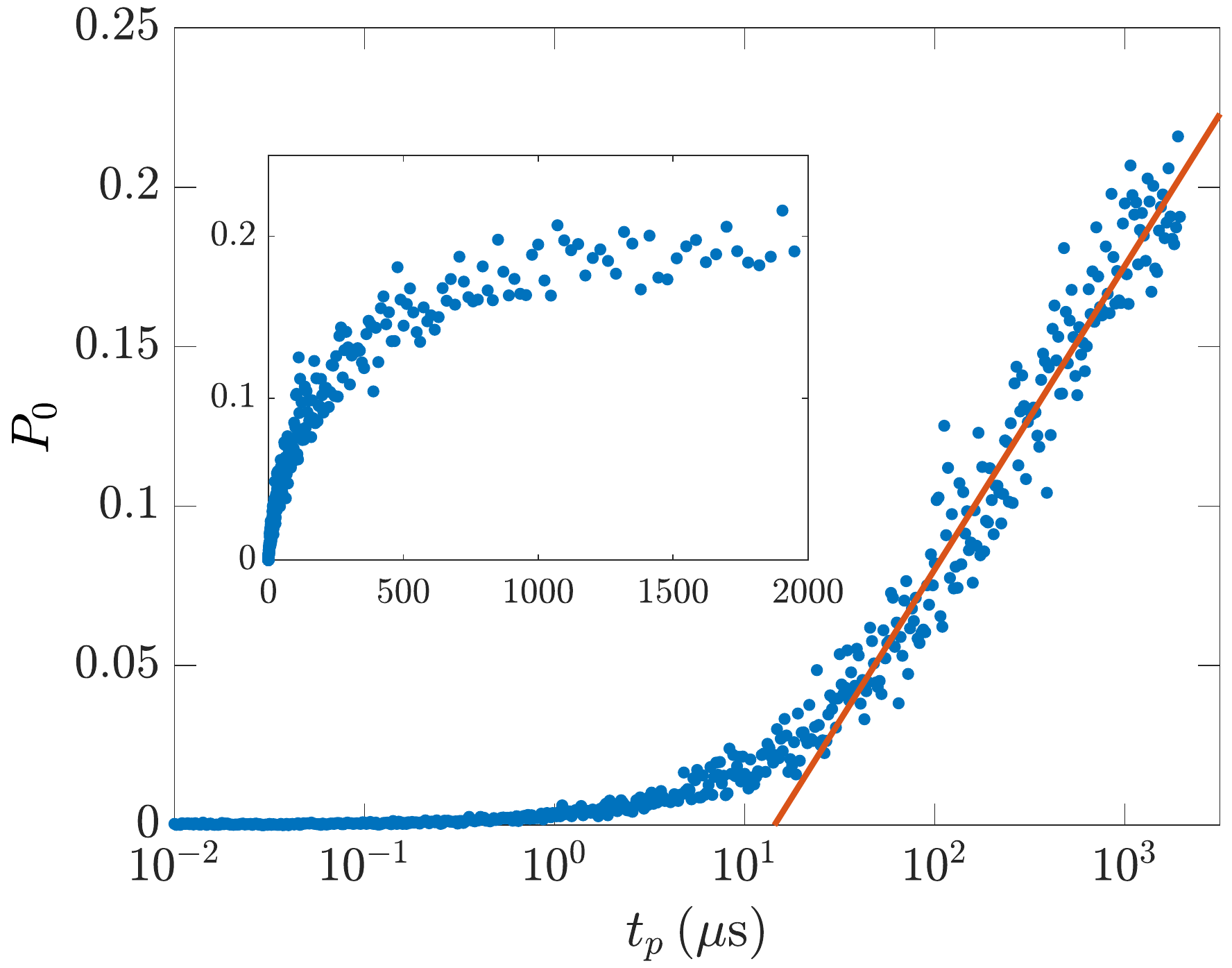}
\caption{Dependence of success probability $P_0$ on pause length $t_p$, for the same problem considered from Fig.~\ref{fig:zoom-pause} ($\mathcal{I}_{783}$), where we fix $s_p=0.44$ (corresponding to the peak in Fig.~\ref{fig:zoom-pause}). We see increasing the pause length corresponds to a larger success probability (although it mostly saturates around 500$\mu$s). In the absence of a pause the success probability is $P_0\approx 10^{-4}$, which increases by several orders of magnitude to around 20\%. Red solid line is linear fit to tail end ($t_p>10^{1.5}\mu$s) Inset: Same as main figure, but with linear scale on $t_p$-axis. Each data point is averaged from 10000 anneals with 5 different choice of gauge.
}
\label{fig:pauselength}
\end{center}
\end{figure}

Fig.~\ref{fig:pauselength} shows that the longer the pause, the greater the increase in the success probability. Here, we see that the success probability, for a pause at re-scaled time $s_p=0.44$, increases from the baseline ($\approx 0.01\%$) to over 10\% for pause lengths longer than around 500$\mu$s, and approaches 20\% as the pause length approaches 2ms (the longest allowable pause length on the D-Wave machine), although saturating around 1ms (in the logarithmic regime). That is, an increase of around three orders of magnitude. 
This behavior gives us new insight into the time-scales involved in these annealers. It shows that even a 10$\mu$s pause (inserted within a default schedule with $t_a=1\mu$s) can dramatically effect the nature of the samples returned from the annealer.

These observations are consistent with the thermalization picture mentioned in the previous section, and the cartoon in Fig.~\ref{fig:cartoon};
we attribute the purple region in Fig.~\ref{fig:cartoon} to the region where the huge spike in success probability is observed, since the system can effectively re-populate  the low lying energy levels on a time-scale comparable with the pause length.
After this (e.g. the blue region in Fig.~\ref{fig:cartoon}), the effect is much weaker (dropping off exponentially) as the relaxation time scale increases: notice in Fig.~\ref{fig:zoom-pause} that late in the anneal, the relative increase in success probability is much less, or non-existent, as compared to during the region around $s_p=0.4$.

\begin{figure}
\begin{center}
\includegraphics[width=0.9\columnwidth]{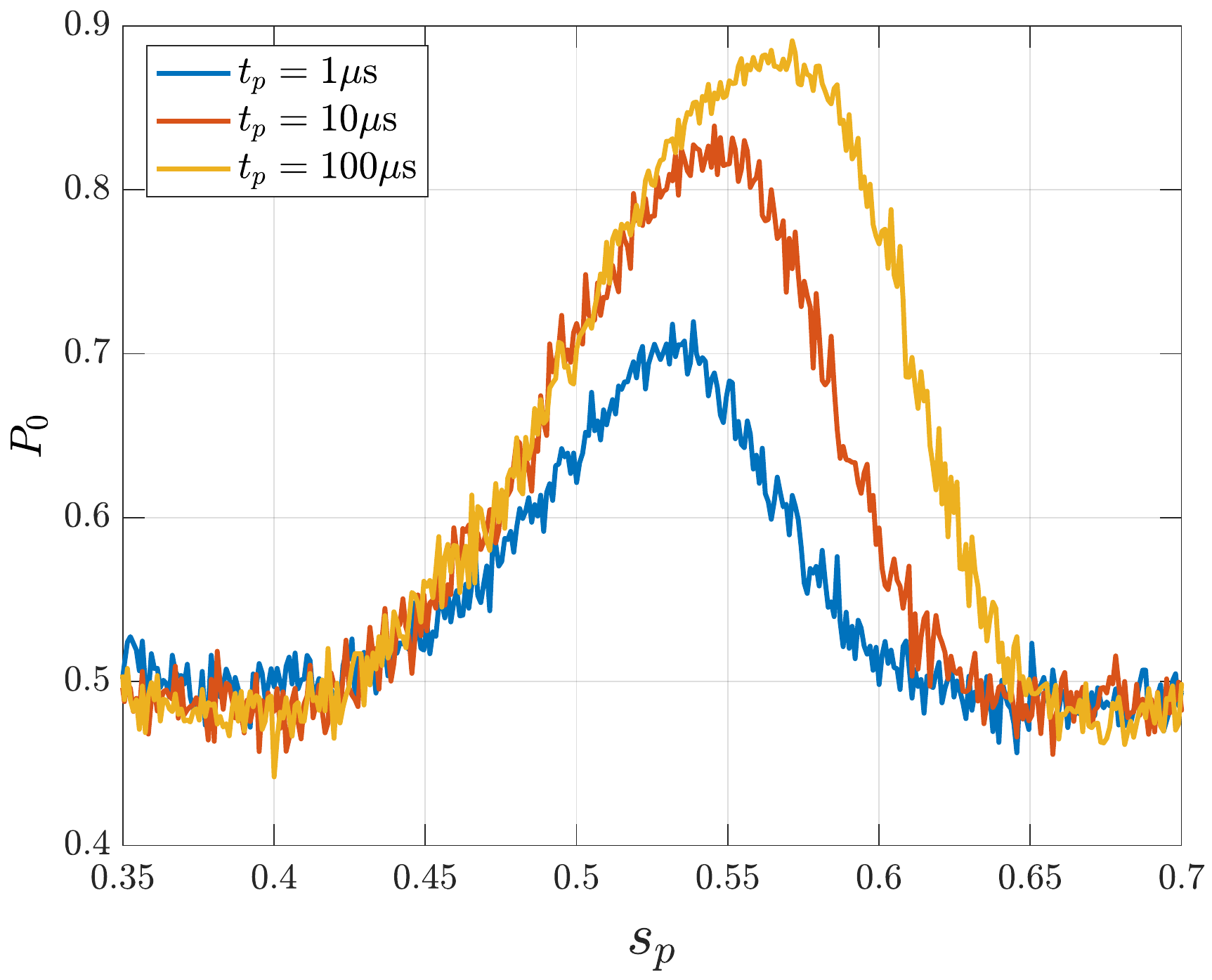}
\caption{Effect of changing the pause length for a 12-qubit problem instance ($\mathcal{I}_{12}^0$). Each data point is from 10000 annealing runs using 5 different gauges. The anneal time $t_a=1\mu$s for all data sets shown.}
\label{fig:vary_tp}
\end{center}
\end{figure}

\begin{figure}
\begin{center}
\includegraphics[width=0.9\columnwidth]{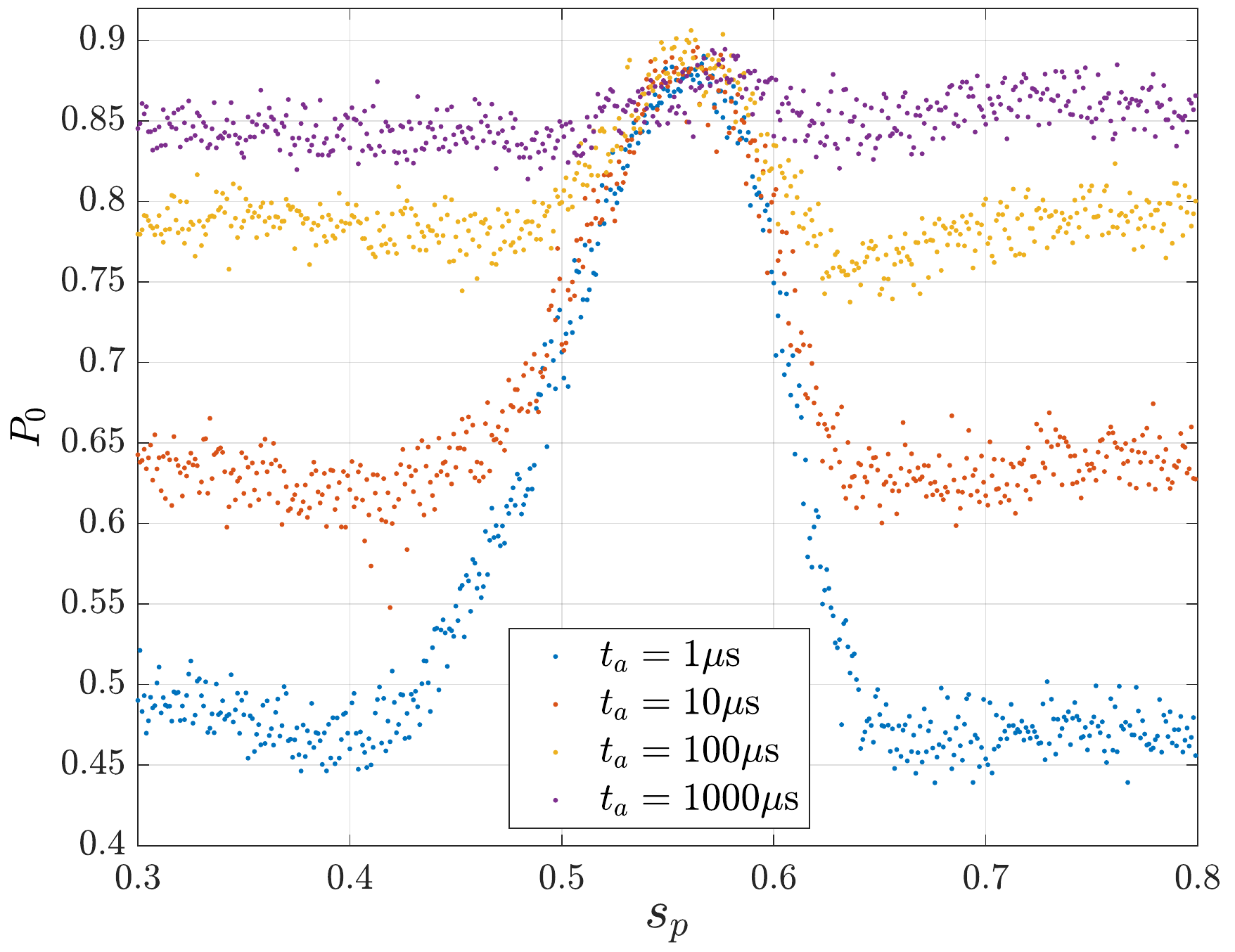}
\caption{Effect of changing the anneal time for $\mathcal{I}_{12}^0$.
Each data point is from 10000 annealing runs using 5 different gauges. The pause length $t_p=100\mu$s for all data sets shown.
}
\label{fig:vary_ta}
\end{center}
\end{figure}

We observe similar phenomena for the second problem class we study (with $J_{ij}\in[-1,1]$), as in Figs.~\ref{fig:vary_tp}, \ref{fig:vary_ta}. In these figures we show the effect of changing the pause length, and the total anneal time respectively.

An effect we observe upon increasing the pause length is that the width of the peak increases, as shown in Fig.~\ref{fig:vary_tp}. 
Notice that in this figure all curves start to show an increase in success probability at the same pause point $s_p$ (just after 0.4), but come back to the baseline probability at later points for longer pause lengths. That is, the region of interest is slightly extended to the right.
This also fits in with the model discussed in Sect.~\ref{sect:theory} and the cartoon picture Fig.~\ref{fig:cartoon}, where increasing $t_p$ increases the size of the purple region by extending it to the right (i.e. the region where $t_H < t_r \lesssim t_p$).
The location of the peak $s_p^{\mathrm{opt}}$ we posit to be around the point when $t_r \approx t_p$, the last point in the anneal for which thermalization can effectively occur during a pause of length $t_p$.
Indeed we experimentally observe (in Fig.~\ref{fig:vary_tp}) that increasing the pause length shifts the peak to later in the anneal (and also increases in size $P_0$ in accordance with this picture \footnote{For a given problem, the later one can thermalize, the greater the ground state success probability, assuming the gap $\omega_{01}$ opens up so that e.g. the ratio $P_1/P_0 \propto \exp(-\beta \omega_{01})$ is reduced.}).

If one instead increases the anneal time (hence $t_H$), the peak narrows (and flattens), and eventually disappears, as observed in Fig.~\ref{fig:vary_ta}. 
Note, in accordance with Fig.~\ref{fig:cartoon}, the location of $s_p^{\mathrm{opt}}$ does not change upon increasing $t_a$ (since this should reduce the size of the purple region from the left).
We also show a corresponding heat map of this effect in Appendix \ref{sect:figs} (Fig.~\ref{fig:heat_ta}).

\begin{figure}
\begin{center}
\includegraphics[width=0.9\columnwidth]{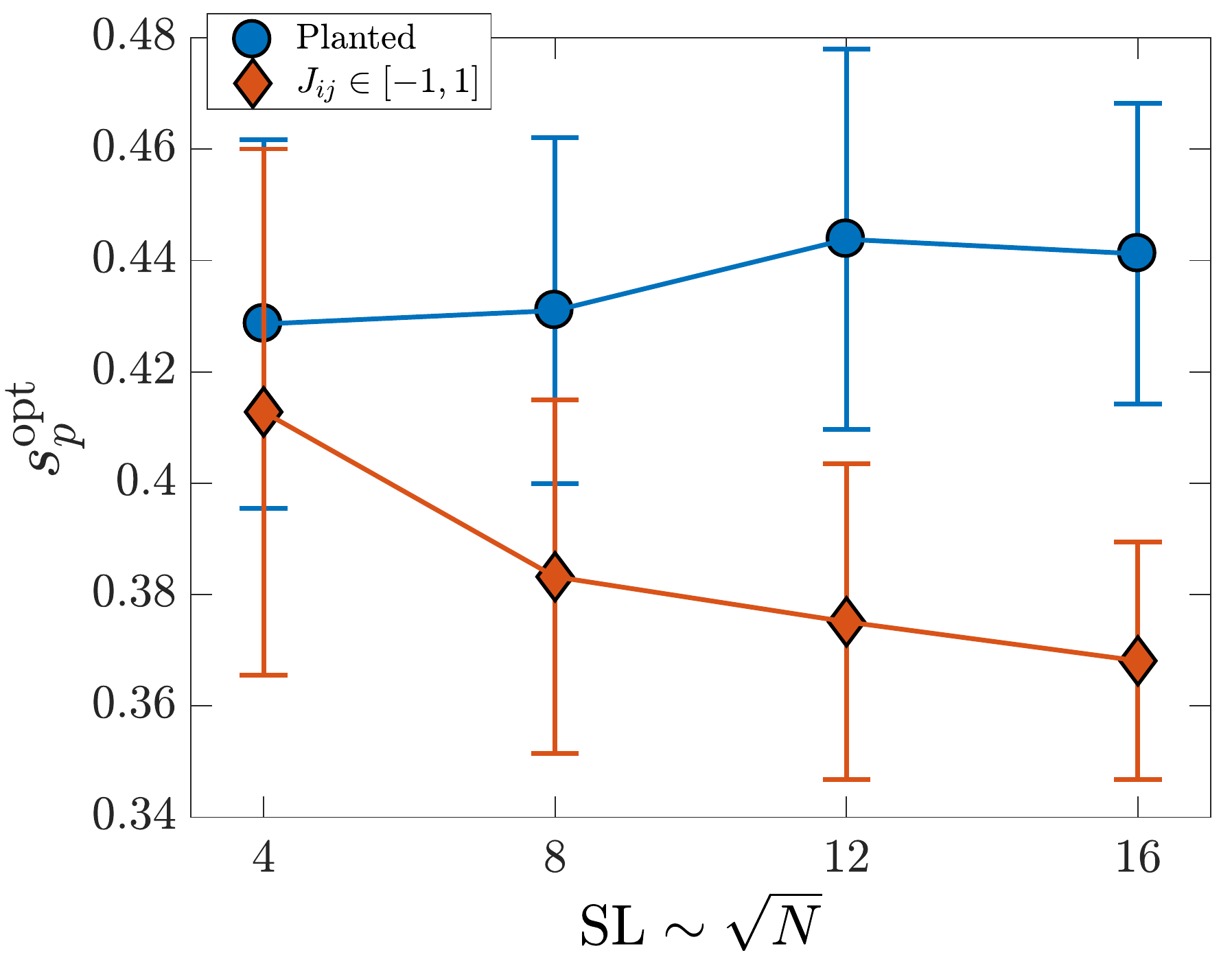}
\caption{The optimal pause point $s_p^{\mathrm{opt}}$ as a function of problem size, for two different problem classes (see legend).
The problems were generated on a square subgraph of the chimera with `side-length' SL, consisting of SL$\times$SL unit cells each containing 8 qubits (see Fig.~\ref{fig:chimera} of Appendix \ref{sect:figs}).
Each SL shown $[4,8,12,16]$ corresponds to (taking account for dead qubits) $N=[127,507,1141,2031]$ respectively.
Each data point shown is an average over (at least) 50 instances. Error bars represent the standard deviation.
Each instance (for each $s_p$ tested) is averaged from 10000 anneals with 5 different choice of gauge, with $t_a=1\mu$s (not including the pause time), and $t_p=100\mu$s.}
\label{fig:scaling_SL}
\end{center}
\end{figure}

It is remarkable that the peak is extremely well defined, occurs in such a concentrated region, and exists for almost all problems we studied (the only exception being some small problem instances, which we discuss below).
For problems of the planted-solution type, there seems to be little dependence on problem size for the position of this peak.
Fig.~\ref{fig:scaling_SL} plots the optimal pause point, $s_p^{\mathrm{opt}}$, which does not vary much with problem size and, in addition, the deviation (i.e., the error bars in the figure) in the samples appears more or less constant.
For problems generated with $J_{ij}\in[-1,1]$ (uniformly random), we see a mild effect with increasing problem size, in which the optimal pause point seems to decrease, and concentrate in location (i.e. the error bars in the figure are decreasing with problem size). This effect would presumably saturate with large enough $N$ (note, in the figure, SL=16 is the largest possible problem size available).

\begin{figure}
\begin{center}
\includegraphics[width=0.9\columnwidth]{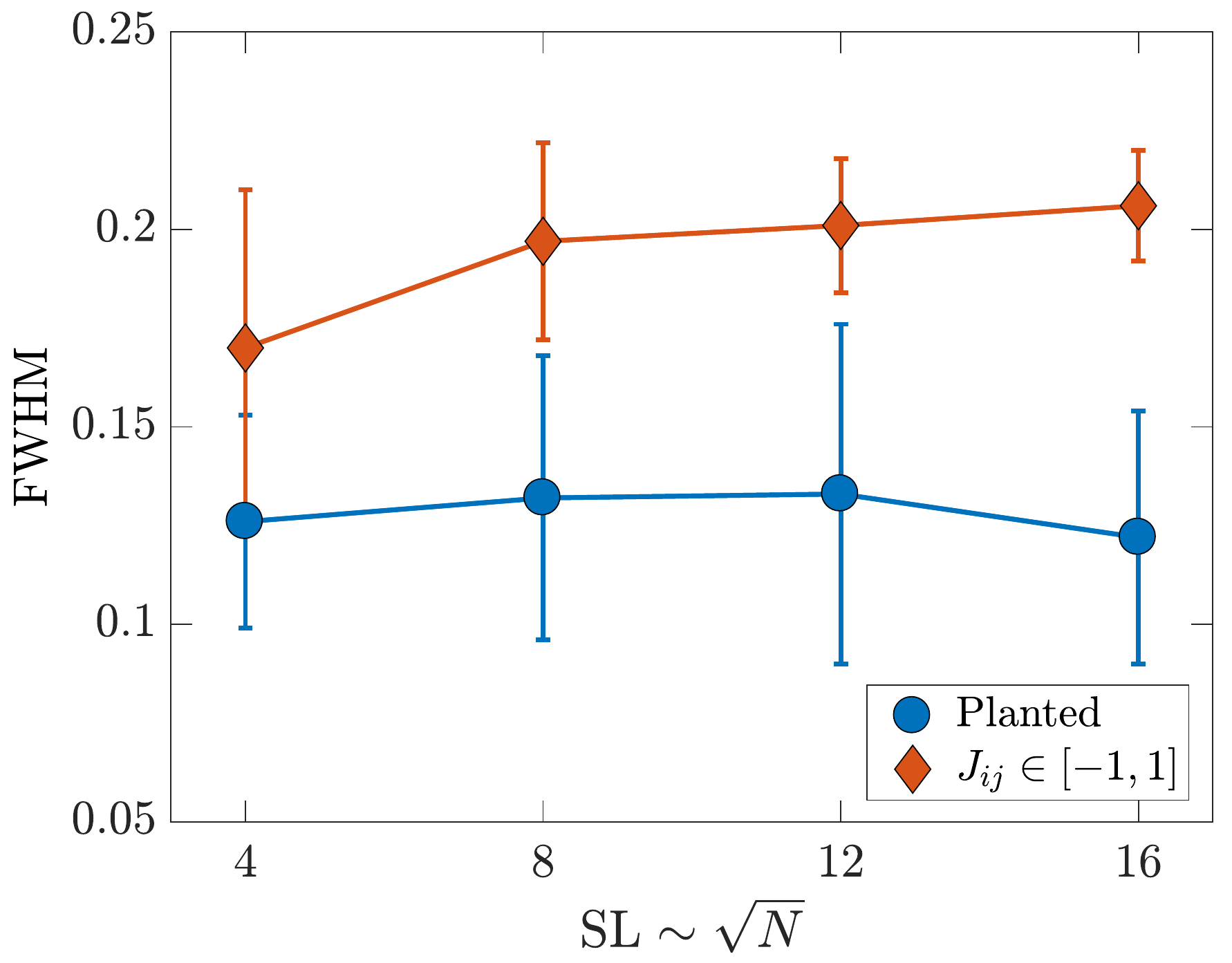}
\caption{The width of the peak on a graph of $|\langle E \rangle |$ Vs. $s_p$, as a function of problem size. The `full-width-at-half-maximum' (FWHM) is the width of the peak in the energy curve $|\langle E \rangle|$ as a function of $s_p$, at $(|E(s_p^{\mathrm{opt}})| + |E_{\mathrm{BG}}|)/2$ where $E_{\mathrm{BG}}$ is the background average energy (i.e., the average energy returned from the annealer in the absence of a pause), and $E(s_p^{\mathrm{opt}})$ the (mean) energy returned by the annealer at the optimal pause point.
Note, modulus is used since all energies observed negative.
The problems (same as in Fig.~\ref{fig:scaling_SL}) were generated on a square subgraph of the chimera with `side-length' SL, consisting of SL$\times$SL unit cells each containing 8 qubits (see Fig.~\ref{fig:chimera} of Appendix \ref{sect:figs}).
Each SL shown $[4,8,12,16]$ corresponds to (taking account for dead qubits) $N=[127,507,1141,2031]$ respectively.
Each data point shown is an average over (at least) 50 instances. Error bars represent the standard deviation.
Each instance (for each $s_p$ tested) is averaged from 10000 anneals with 5 different choice of gauge, with $t_a=1\mu$s (not including the pause time), and $t_p=100\mu$s.}
\label{fig:fwhm}
\end{center}
\end{figure}

\begin{figure}
\begin{center}
\includegraphics[width=0.9\columnwidth]{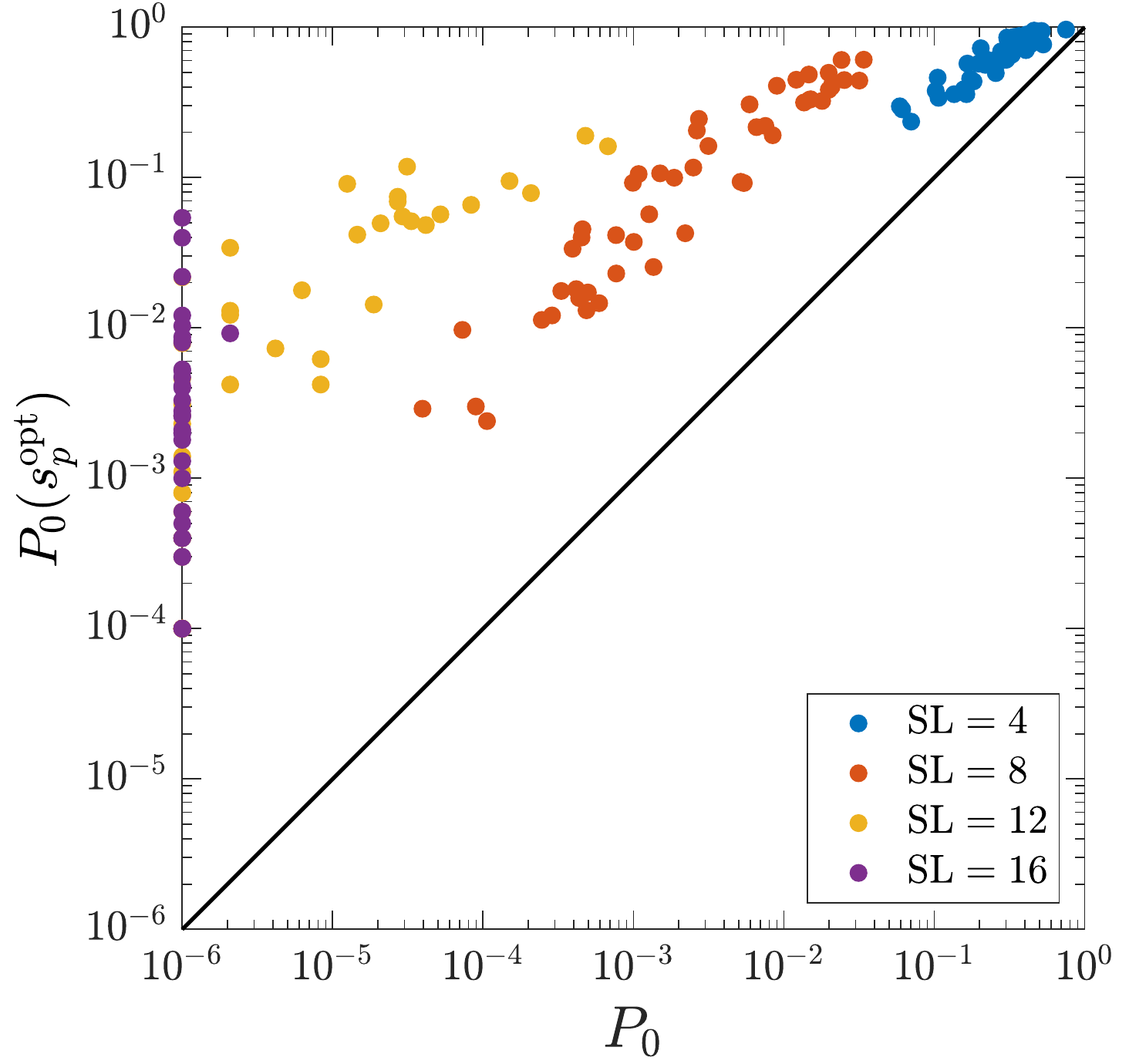}
\caption{For the planted problems of Fig.~\ref{fig:scaling_SL} (50 instances per side-length), we plot the success probability under the default annealing schedule (with $t_a=1\mu$s) $P_0$ as compared to the success probability for an anneal with pause ($t_p=100\mu$s) at the optimal pause point, $P_0(s_p^{\mathrm{opt}})$. We see typically orders of magnitude improvement for the hard problems. Note, the problems which are plotted vertically on the $y$-axis have $P_0=0$ (i.e. were not solved once over all anneals). Note, some problems which were not solved once, even under the pause, are not included.}
\label{fig:p0}
\end{center}
\end{figure}

The simple observations demonstrated here show that one may be able to design more efficient annealing schedules by inserting a short pause intermediately.
In Fig.~\ref{fig:fwhm}, we see the width of the peak only depends very weakly (or not at all) on the problem size; this suggests that for most problems, regardless of size, there is a fairly large window in which one can pause and observe an increase in success probability, potentially by orders of magnitude, see Fig.~\ref{fig:p0}. In this figure, we also see many instances which are never solved once without the pause inserted into the schedule, but are solved up to $\sim 10\%$ of the time with an optimized pause schedule. A natural question is of course whether or not this type of schedule can be used to increase performance by, for example, reducing the time-to-solution. To do this, one would of course need to take into account the time required to estimate optimal parameters $s_p, t_a,$ and $t_p$ for a problem class, in order to assess whether such a protocol advantageous.  This poses an interesting direction for future study into optimal designs for quantum annealers. In Appendix \ref{sect:figs} (Fig.~\ref{fig:default_vs_pause}) we also provide a comparison of a paused schedule versus an unpaused schedule, with same total `run-time', demonstrating further the possibility of using these alternative schedules for application purposes.

\subsection{Reverse annealing with a pause \label{sect:RA}}
Before proceeding with a statistical analysis we briefly present some relevant results using the reverse annealing protocol, with a pause, the general protocol of which is demonstrated graphically in Fig.~\ref{fig:pause-example}. This allows us to identify some of the key regions during an anneal, which, as explained above, depend on the ratios of the various energy scales involved and associated time-scales.

\begin{figure}
\begin{center}
\includegraphics[width=0.9\columnwidth]{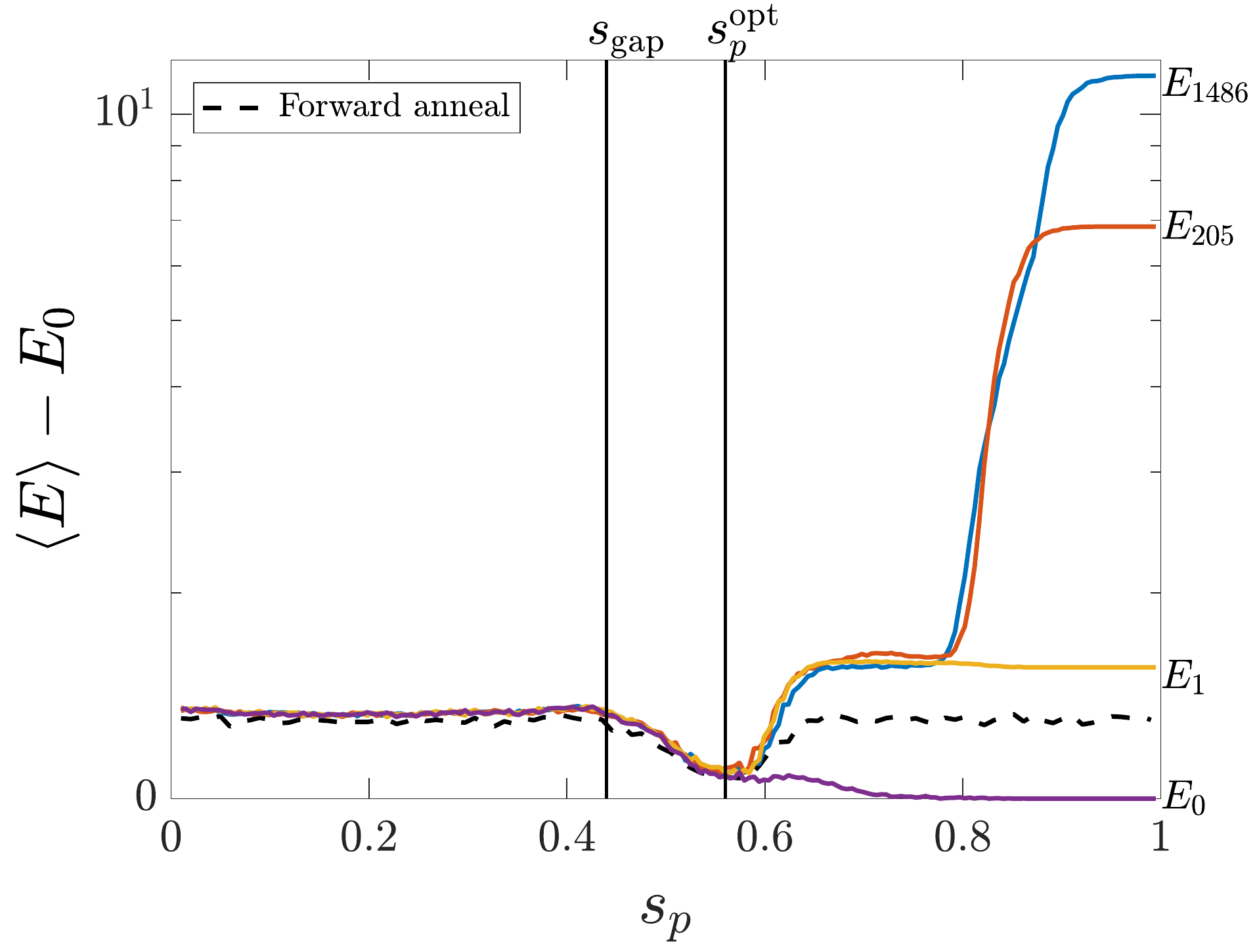}
\caption{Reverse annealing with pause at $s_p$ (solid lines), for a four different initial configurations for a 12-qubit problem $\mathcal{I}_{12}^{0}$.
We plot the average energy (arbitrary units) returned from 5000 anneals, evolved at rate $\mathrm{d}s/\mathrm{d}t=1\mu$s$^{-1}$, and $t_p=100\mu$s.
We consider ground and first excited state configurations, as well as two highly excited energy levels.
The energy level  $E_i$ is indicated on the right hand side.  We also show the corresponding forward anneal curve (black-dash line). This problem has a minimum gap at $s=0.44$ indicated in the figure. Note a sample of the spectrum for this problem is shown in Fig.~\ref{fig:spectrum} in Appendix \ref{sect:schrodinger}.}
\label{fig:RA}
\end{center}
\end{figure}

We show some of our findings in Fig.~\ref{fig:RA} where we identify 4 regions of interest.
1) $s_p<s_{\mathrm{gap}}$. The system has been evolved (from $s=1$) past the minimum gap, and been paused at a point where $Q > O(1)$, allowing for mixing between energy levels in the computational basis. There is no memory of the initial configuration.
2) In the region just after the minimum gap, up to around $s_p^{\mathrm{opt}}$, where the lowest energy solutions are found, and corresponding to the purple region in Fig.~\ref{fig:cartoon}. Here there is no memory of the initial state, and no clear difference between forward and reverse annealing. 
We expect thermalization is able to occur effectively on times scales comparable with $t_p$, i.e., the transition rates between energy levels $\Gamma_{ij} \ge 1/t_p$.
3) After $s_p^{\mathrm{opt}}$, where there is a clear difference between forward annealing and reverse annealing, there is `memory' of the initial configuration. Thus the state returned by the annealer at $s=1$ depends heavily on the system state at the pause point $s_p$ suggesting different time-scales and transition rates are important here. 
In this region, as $Q\rightarrow 0$ and $\langle E_i |\sigma^z |E_j\rangle \rightarrow 0$, the $g_{x,y}$ couplings may play more of a role, leading to  qualitatively different thermalization mechanisms and time-scales. Here some transition rates $\Gamma_{ij}$ may be comparable to $1/t_p$, whereas others much less.
4) Very late in the anneal, with $Q \ll C \ll 1$, almost no dynamics occur (the state returned from the annealer is almost always the same as the one initialized), i.e. $\Gamma_{ij} \ll 1/t_p$.

We mention that these general observations seem to be fairly generic, and not specific to this particular example. Indeed, recent work experimenting reverse annealing for embedded instances on the same DW2000Q obtained results compatible with our findings and models~\cite{venturelli2018reverse}.
With this in mind, we proceed with a statistical analysis, demonstrating to what extent the picture outlined in Fig.~\ref{fig:cartoon} holds.

\subsection{Correlation with the minimum gap \label{sect:gap}}
Typical folklore of (open system) quantum annealing dictates that around the location of the minimum gap, thermal excitations from the ground state to excited energy levels may occur, and that after the gap, thermal relaxation will allow some of the excited population to fall back to the ground state \cite{DWave-16q} (of course, this is heavily dependent on the nature of the minimum gap, and hence on $H_p$ itself, as well as the temperature and annealing schedule). This general idea is also demonstrated in Fig.~\ref{fig:cartoon}.
This framework would suggest that a pause in the annealing schedule some (problem dependent) time after the minimum gap may lead to an increase in the success probability (that is, the population in the ground state at time $s=1$).

\begin{figure}
\begin{center}
\includegraphics[width=0.9\columnwidth]{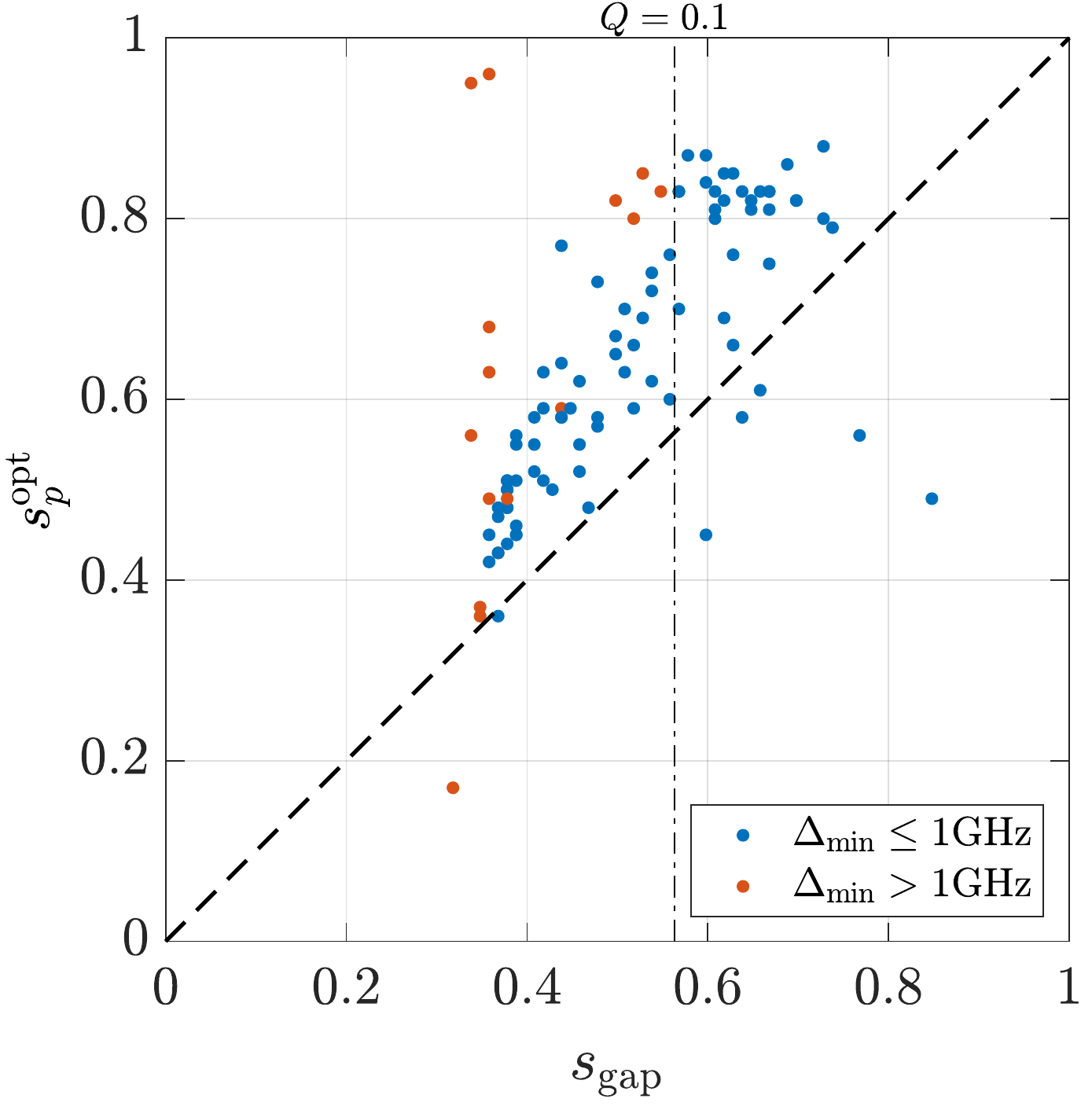}
\caption{Correlation between the location of the minimum gap, $s_{\mathrm{gap}}$, and the optimal pause point $s_p^{\mathrm{opt}}$ for 100 problems of size 12 qubits. These problems have $J_{ij}\in[-1,1]$ (uniformly random) and $h_i=0$. 
We divided the data into two groups based on their minimum gap $\Delta_{\mathrm{min}}$ (see legend), and we also indicate by the dash-dot (vertical) line the location corresponding to $Q(s_{\mathrm{gap}})=0.1$.
We fixed the pause length to $t_p=1000\mu$s, and total anneal time (excluding the pause) to $t_a=1\mu$s.
Data from the annealer is averaged over 10000 anneals, with 10 different choices of gauge. 
}
\label{fig:sgap_spause}
\end{center}
\end{figure}

Working with 12-qubit problems with $J_{ij}\in[-1,1]$ (uniformly random), we indeed find such a correlation between the location of the minimum gap, and the optimal pause point, $s^{\mathrm{opt}}_p$, where for over 90 of the 100 problems tested the best place to pause is after the minimum gap.
This is demonstrated in Fig.~\ref{fig:sgap_spause}, where on average $s^{\mathrm{opt}}_p \approx s_{\mathrm{gap}} + 0.14$. 

We comment briefly on some of the outliers (e.g. with $s_{\mathrm{gap}} < s_p^{\mathrm{opt}}$, or not in the main cluster of points) in the data set. For some of these 12-qubit problems, they are solved almost 100\% of the time by the D-Wave annealer (i.e. they are extremely easy optimization problems). These are problems that have large minimum gaps $\Delta_{\mathrm{min}}>1$GHz, and we indicate them in red in the plot (also see Fig.~\ref{fig:gap_P0} in Appendix \ref{sect:figs}).
For these instances, we typically do not observe a well defined optimal pause point; since the gap is so large for all $s$, we expect very little thermal excitation to occur at all, hence pausing has little effect. We see these red points have a fairly random spread in the $s_p^{\mathrm{opt}}$-axis.

A second set of outliers are (some of the) instances which have minimum gaps relatively late in the anneal. 
We observe these instances do not have well defined optimal pause points, and expect this is due to quirks of their individual spectra (one such example is shown in Fig.~\ref{fig:outlier} of Appendix \ref{sect:figs}).
In particular when the minimum gap occurs late in the anneal (e.g. after the point when $Q<0.1$), either the transition time scale $t_r$ may already be too large for effective thermalization to occur during a pause (since the $\sigma^z_i$ matrix elements driving the transitions in Eq.~(\ref{eq:rate}), governed by $Q$, are negligible), or the gap doesn't open up enough before the end of the anneal to transfer enough population out of the excited states.

It is interesting to note that the observation of instances without well defined optimal pause points appears to be a small size effect, since all of the (both planted, and uniformly random) problems tested in Fig.~\ref{fig:scaling_SL} (which all had over 100 qubits) had well defined optimal pause points, i.e. all exhibited a well defined minima in the energy as a function of pause point (and peak in the success probability as a function of pause point, for the planted problems).

Nevertheless, the overall trend is clear, with the majority of problem instances exhibiting an optimal pause point in a narrow region shortly after the location of the minimum gap.

We similarly study the same problems where we re-scale the problem Hamiltonian by a constant factor (2,4,8). This has two effects; 1) it shifts the location of the minimum gap to later in the anneal, and it also reduces the size of the minimum gap (as an explicit example, see Fig.~\ref{fig:gap_scale} in Appendix \ref{sect:figs}).
We indeed see that correspondingly, the location of the optimal pause point shifts to later in the anneal (see inset of Fig.~\ref{fig:sgap_spause_scale}). We show this explicitly for a single problem in Fig.~\ref{fig:scale} in Appendix \ref{sect:figs}.

\begin{figure}
\begin{center}
\includegraphics[width=0.9\columnwidth]{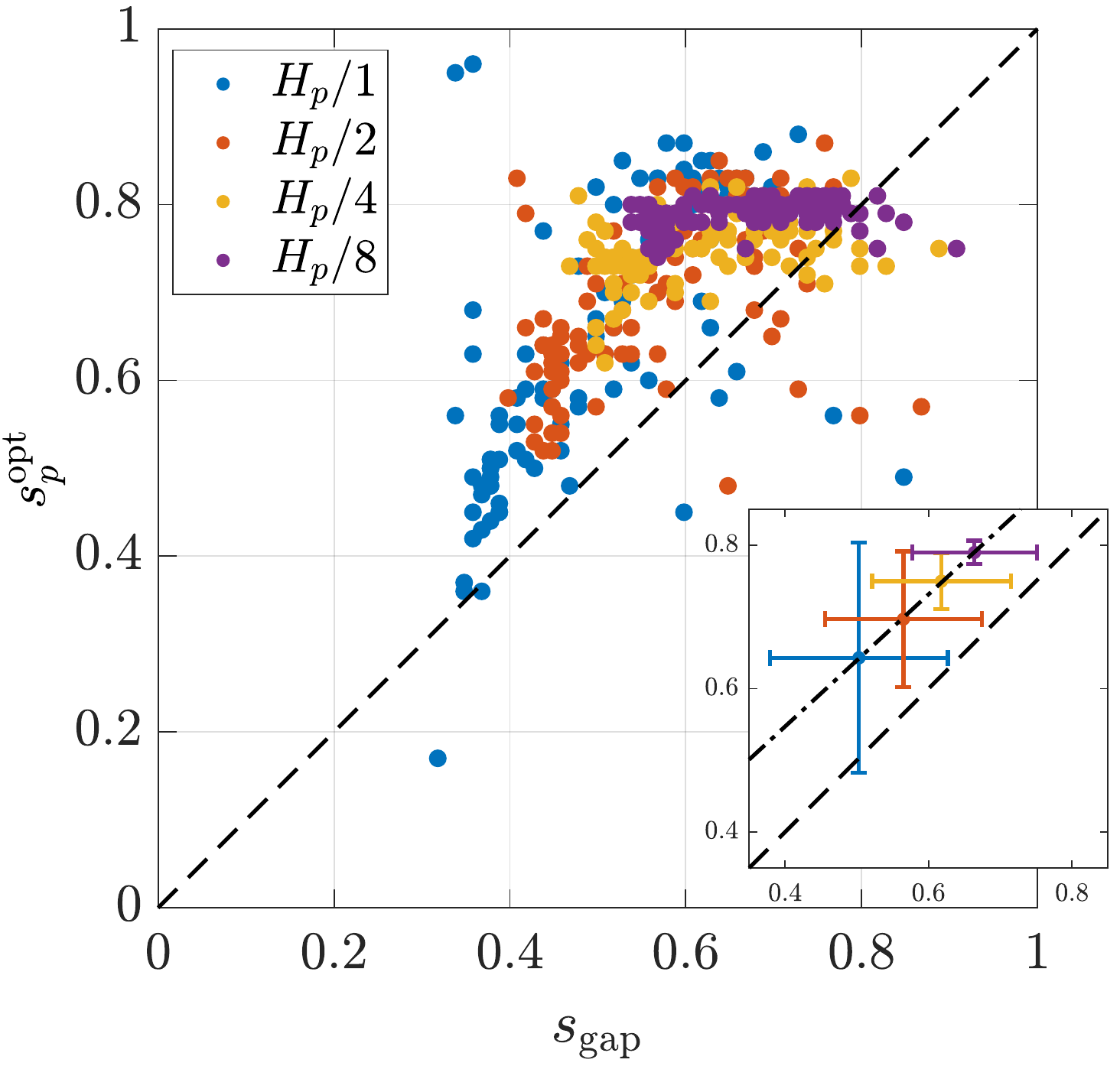}
\caption{Effect of reducing problem energy scale (for the same 100 problems studied in Fig.~\ref{fig:sgap_spause}). We divide the problem Hamiltonian $H_p$ by 1,2,4,8 (see legend).
Inset: Mean data point for each group of 100 instances upon dividing $H_p$ by 1,2,4,8 (see legend). Dash-dot line is least squares fitting to the median data point. Error bars are the standard deviation.
}
\label{fig:sgap_spause_scale}
\end{center}
\end{figure}

Interestingly, we also observe that the location of $s^{\mathrm{opt}}_p$ concentrates (thus becoming less correlated with $s_{\mathrm{gap}}$) upon reducing the energy range of the problem; notice how in Fig.~\ref{fig:sgap_spause_scale}, the purple points ($H_p/8$) are almost perfectly aligned close to $s_p^{\mathrm{opt}}=0.8$. We also see this by noting that the error bars  (standard deviation) in the $s_p^{\mathrm{opt}}$-axis decrease (inset of figure).

We explain this behavior by pointing out that by dividing $H_p$ by a large enough factor, $\beta\omega_{01}(s) < 1$ for $s > s_{\mathrm{gap}}$ where $\omega_{01} (s):= E_1(s)-E_0(s)$ is the gap between the ground and first excited state, and $\beta$ the inverse temperature of the annealer. This implies the system should be able to effectively thermalize until very late in the anneal, until the matrix elements in Eq.~(\ref{eq:rate}) become small enough (determined by $Q$ being small enough).
This picture means the start of the region when the dynamics become frozen (purple region in Fig.~\ref{fig:cartoon}) is not determined so much by the problem itself  (i.e. the exact spectrum as a function of $s$), but the annealing schedule (i.e. $Q(s)$). Thus different problems may exhibit very similar optimal pause points.

It would be worth while to explore this in more detail to confirm this hypothesis.

\subsection{Quantum Boltzmann distribution \label{sect:qbm-main}}
For a set of 10 problem instances of size 12-qubits $\mathcal{I}_{12}^{0-9}$ with $J_{ij}\in[-1,1]$ (uniformly random), each with well defined optimal pause points, with $\Delta_{\mathrm{min}}<1$GHz, $Q(s_{\mathrm{gap}})>0.1$, we compare the returned statistics to the instantaneous quantum Boltzmann distribution $\rho \sim \exp(-\beta(T) H(s))$ (projected to the $z$-eigenbasis), for various choices of $T$; we vary the temperature $T$ from $\frac{1}{4}T_{\mathrm{DW}}$ to $4T_{\mathrm{DW}}$, in increments of $\frac{1}{4}T_{\mathrm{DW}}$. Here $T_{\mathrm{DW}}=12.1$mK is the operating temperature of the annealer. We outline these calculations in Appendix~\ref{sect:qbm}.

We study these problems, with a very long pause length of $t_p=1000\mu$s to allow enough time to thermalize, and run with short anneal time (excluding pause time) $t_a=1\mu$s, i.e. the quickest possible rate $\frac{\mathrm{d}s}{\mathrm{d}t}$, so that we can try to isolate the effect of the pause by coming as close as possible within the current D-Wave constraints to `quenching' the system.

\begin{figure}
\begin{center}
\includegraphics[width=0.9\columnwidth]{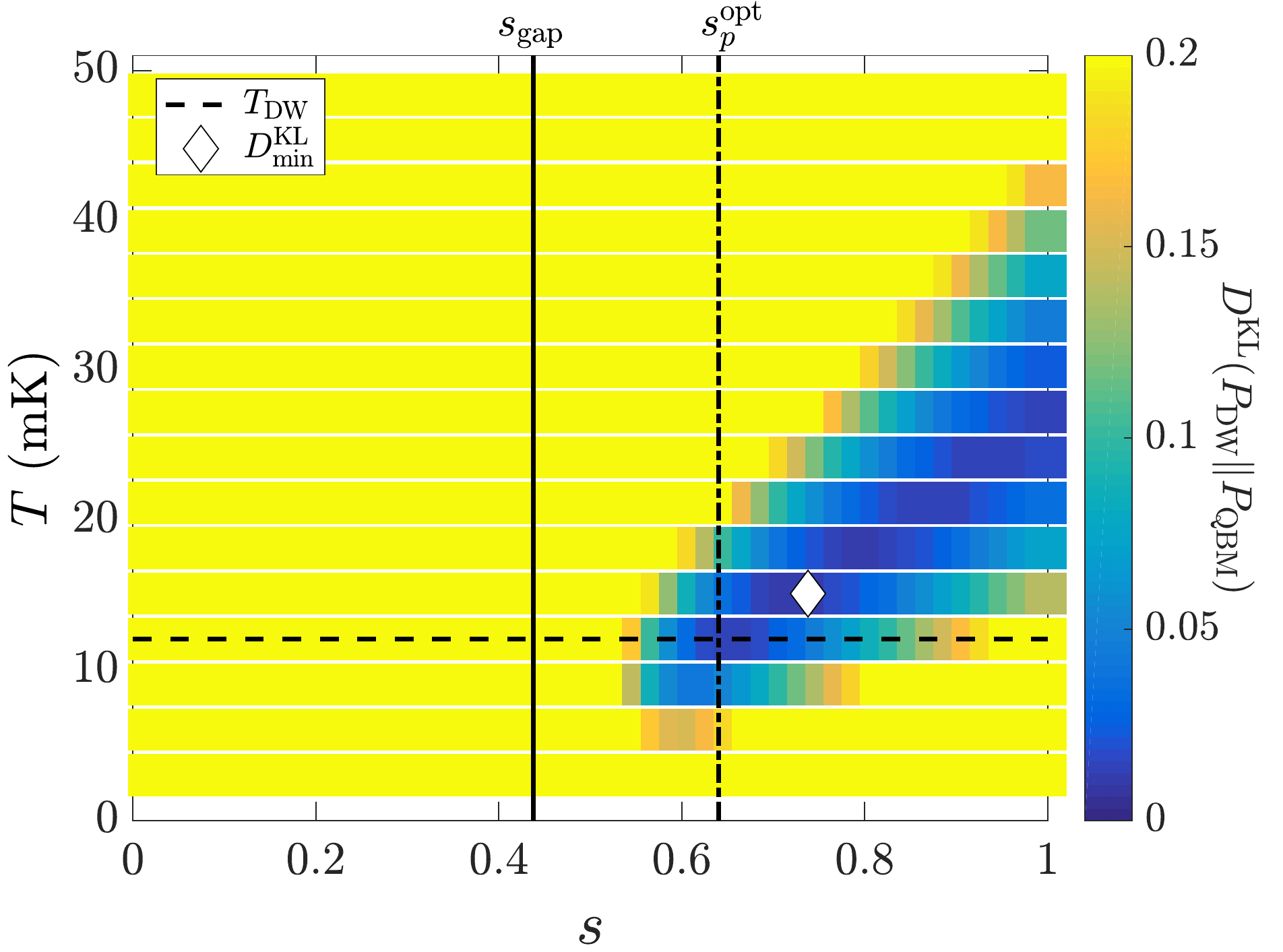}
\caption{KL divergence $D^{\mathrm{KL}}$ between data from the annealer $P_{\mathrm{DW}}$ and the Boltzmann distribution $P_{\mathrm{QBM}}$ (projected onto the computational basis) for various choices of $(s,T)$, for a single 12-qubit instance ($\mathcal{I}_{12}^1$). 
The D-Wave data is sampled from from an anneal with a pause of length $t_p=1000\mu$s at $s_p^{\mathrm{opt}}$, with $t_a=1\mu$s (from 10000 anneals and 10 choices of gauge).
We indicate in the plot three key parameters; the physical temperature $T_{\mathrm{DW}}=12.1$mK, the location of the minimum gap $s_{\mathrm{gap}}$, and the optimal pause point $s_p^{\mathrm{opt}}$.
The white diamond corresponds to the minimum value of $D^{\mathrm{KL}}$ over all $(s,T)$, and is equal to $D_{\mathrm{min}}^{\mathrm{KL}} = 0.01$ bits of information.
Note, to be able to distinguish the features in the plot, we set the upper limit on the plot to be $D^{\mathrm{KL}}=0.2$ (any value above this is mapped to this color).}
\label{fig:KL-heat}
\end{center}
\end{figure}

An example of this analysis for a single instance is shown in Fig.~\ref{fig:KL-heat}, though the pattern looks much the same for all of the instances studied.
Here, we focus on the distribution returned from the annealer with a pause at the optimal pause point $s_p^{\mathrm{opt}}$, and compare this to a distribution of the form $\exp(-\beta(T) H(s))$, leaving $(s,T)$ as parameters to optimize over.
We do indeed observe a correspondence between the projected quantum Boltzmann distribution at $(s_p^{\mathrm{opt}},T_{\mathrm{DW}})$, and the data sampled from the experimental device; over all of the instances $D^{\mathrm{KL}}(s_p^{\mathrm{opt}},T_{\mathrm{DW}}) = 0.076 \pm 0.065$, where $s_p^{\mathrm{opt}}= 0.57\pm 0.05$.

We observe, however, that the experimental distribution is best described by a hotter Boltzmann distribution at a later point in the anneal than the optimal pause point.
For these problems, the optimal parameters $(s^*,T^*)$ such that $D^{\mathrm{KL}}$ is minimized, correspond to $s^*=0.78 \pm 0.10$ and $T^* = 26.1 \pm 8.8$mK (and up to 4 times the physical temperature). 
The corresponding optimal KL-divergence over all problems is $D^{\mathrm{KL}}_{\mathrm{min}} =0.016 \pm 0.015$, markedly better than at $(s_p^{\mathrm{opt}},T_{\mathrm{DW}})$.

\begin{figure}
\begin{center}
\includegraphics[width=0.9\columnwidth]{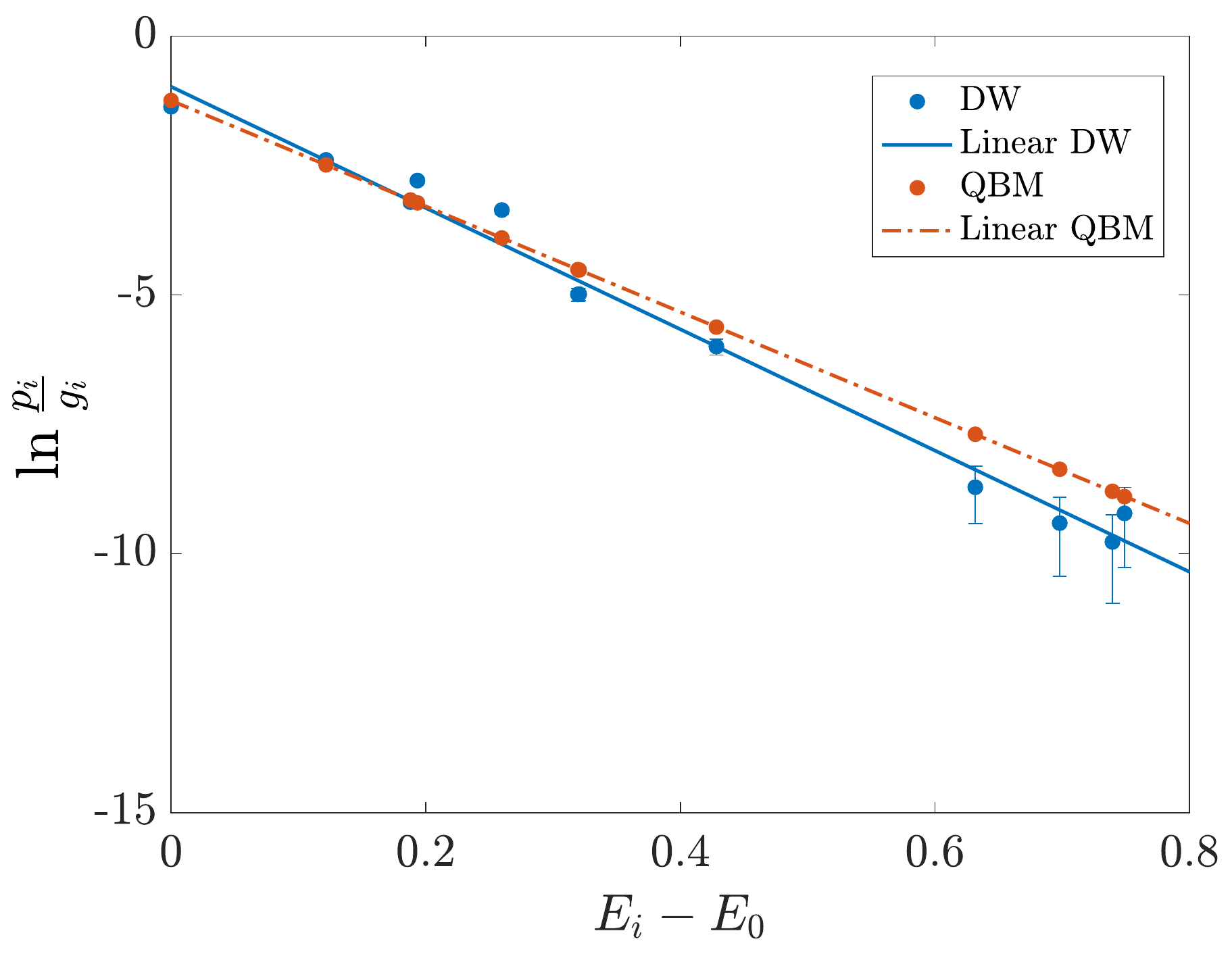}
\caption{Comparison of the D-Wave probability distribution and the closest fit (as measured by KL-divergence) to a quantum Boltzmann (QBM) distribution of the form $\exp(-\beta(T) H(s))$, where the fit is over the parameters $(s,T)$ for $\mathcal{I}_{12}^2$. Optimal values for this problem are $(s^*,T^*) = (0.76,18.5\mathrm{mK})$ (corresponding to $D^{\mathrm{KL}}_{\mathrm{min}}=0.03$).
Here $p_i$ is the probability of observing a configuration with energy $E_i$, and $g_i$ is the degeneracy of that energy level.
The solid and dash-dot lines are least-squares fit for the annealer data and the QBM data respectively. The annealer data is from a schedule with a pause at the optimal pause point ($s_p^{\mathrm{opt}}=0.59$), over 50 samples, each of 10000 anneals (and 10 gauges), with $t_a=1\mu$s, $t_p=1000\mu$s. Error bars are standard deviation over the 50 samples.}
\label{fig:pdf}
\end{center}
\end{figure}

At values of $s^*$ this late in the anneal, with  $Q(s^*) \approx 10^{-3}$, the distribution would be expected to be effectively a classical Boltzmann distribution (of $H_p$); that is, of the form $ \exp(-\tilde{\beta}H_p)$, where $\tilde{\beta}$ is an effective inverse temperature.
In Fig.~\ref{fig:pdf},
we compare the D-Wave distribution of a single instance sampled from an anneal with pause at $s_p^{\mathrm{opt}}$, to the optimal found over all $(s,T)$ in $\exp(-\beta(T) H(s))$. 
We plot on a logarithmic scale to show the similarity to a classical Boltzmann distribution of $H_p$, for which $\ln \frac{p_i}{g_i} = -\tilde{\beta}E_i - \ln Z$ (i.e. a straight line on this graph). 
The experimental data and the computed Boltzmann distribution both fit reasonably well to a straight line. This good fit indicates classical thermalization is occurring to some extent. Whether the two distributions correspond to one and the same is not clear however; there are clearly some large divergences (the $y$ axis is a logarithmic scale).

These general observations explain the diagonal `streak' in Fig.~\ref{fig:KL-heat}: If the returned experimental data does indeed fit a Boltzmann distribution at a late point in the anneal when $Q(s^*)\ll 1$, then the distribution at $s^*$ should be a classical Boltzmann distribution, specifically $\exp (-\beta(T^*)H(s^*))\approx \exp(-\beta(T^*)B(s^*)H_p)$.
Therefore there are a range of values $(s^*,T^*)$ which give a similar distribution, i.e., so long as $B(s^*)/k_B T^*$ is constant. This means that if $s^*$ (hence $B(s^*)$) is larger, $T^*$ should also be larger to compensate, as we see in this figure.

This explains, to some extent, the variability in the values of $(s^*,T^*)$ mentioned above. Most likely there are many values of $(s^*,T^*)$ which provide a very good fit to the experimental data. It is perhaps therefore not meaningful to speak of a single optimal value $(s^*,T^*)$.
We do however universally find the data does fit better to a hotter Boltzmann distribution, at a later point in the anneal than $(s_p^{\mathrm{opt}},T_{\mathrm{DW}})$. A hotter temperature at the qubits could explain this to some extent, but these results  also suggest that non-trivial dynamics may occur after after the pause at $s_p^{\mathrm{opt}}$, and that therefore the approximate quench is not sufficiently quenching the dynamics.
That is, 
dynamics after the pause may allow the system to experience some further classical thermalization.

In the absence of a pause we do not see any clear correspondence between the D-Wave distribution and a Boltzmann distribution. 
The same analysis as above for an anneal without a pause, shows the KL-divergence varying wildly, with  $D^{\mathrm{KL}}_{\mathrm{min}} = 0.19 \pm 0.15$. Over the range of $(s,T)$ for which we computed $\exp(-\beta(T) H(s))$, there is typically no good choice of $(s,T)$, and it is not clear what the distribution is.
These results make sense, since in the absence of a pause, we expect that the system will not be able to thermalize appropriately during the anneal (e.g. because the anneal time $t_a$ is too quick), so we have no reason to expect the output distribution to be close to a Boltzmann distribution.
We conjecture that after the optimal pause point non-trivial dynamics still occur. The intermediate pause helps the D-Wave distribution equilibriate to the instantaneous thermal distribution, and after this, although the thermalization time-scale is `large' (relative to $t_{p,H}$), some dynamics, will inevitably occur.

Though we have presented direct evidence that the pause indeed plays a key role in allowing the system to thermalize, more work is required to understand precisely the distribution at the optimal pause point. The main constraint in our experiments prohibiting us from probing this behavior further is the maximum annealing rate $\frac{ds}{dt}$ which is limited to 1$\mu \mathrm{s}^{-1}$ on the present annealer. Even though we hope to be approximately quenching the system from after the the pause, in reality non-trivial dynamics (e.g. driven by $Q$) likely still occur during the approximate quench.
Were we able to perform a true quench, we should be able to precisely verify the picture presented here: 
We would expect to find a region in which the system is instantaneously thermalizing, corresponding to the green region in Fig.~\ref{fig:cartoon}, and unique, unambiguous optimal fitting values $(s^*,T^*)$ in the purple region, in which, without a pause, the system no longer thermalizes. 

We provide some more intriguing evidence in the next section, where we analyze results on a set of large-scale problem instances. These results strongly suggest that classical thermalization is occurring, specifically thermalization to a Boltzmann distribution of $H_p$ .

\subsection{Classical Boltzmann distribution \label{sect:cbm-main}}
The extent to which annealers sample from a classical Boltzmann distribution at some point $s^*$ late in the anneal is a hotly contested issue \cite{Amin:2015,Amin:boltzmann,perdomo,fairInUnfair,marshall-rieffel-hen-2017}.
Sampling from a Boltzmann distribution is NP-hard, so annealers would not be expected to efficiently solve this problem across the board, but they could have advantage over the best classical methods. Quantum computing is known to have advantages over classical computing for sampling from certain distributions \cite{Aaronson11,Bremner11}, including Gibbs distributions \cite{brandao2017quantum}, but whether quantum annealing has an advantage remains an open question.
If problems for which machine learning is applicable (e.g. in the restricted Boltzmann machine paradigm) freeze-out at a point late in the anneal, when $Q\ll 1$, then annealers may have an advantage over classical samplers 
\cite{perdomo,adachi,Amin:boltzmann}.

With the advent of a new entropic sampling technique based on population annealing \cite{Barash-2018-entropic_PA}, we were able to accurately estimate the degeneracies for 225 planted-solution  instances containing 501 qubits (that is the estimated ground and first excited state degeneracies are within 5\% of the known values found by planting). For more information on these techniques see Refs.~\cite{marshall-rieffel-hen-2017,Barash-2018-entropic_PA}, and Appendix \ref{sect:cbm}. Due to the large size of these problems, we are of course not able to compute the Boltzmann distribution of the full Hamiltonian $H(s)$ as we did in the previous section.

However, having accurate values for the degeneracies allows us to calculate the classical (problem Hamiltonian) Boltzmann distribution $\rho \sim \exp(-\tilde{\beta} H_p)$, where $\tilde{\beta}$ is an instance-dependent effective inverse temperature, i.e., a fitting parameter, which depends on the physical temperature, and the strength problem Hamiltonian $B(s)$ (and in principle anything else effecting the distribution returned from the annealer such as various noise sources).

If the distribution returned from the D-Wave machine is indeed a classical Boltzmann distribution at freeze-out point $s^*$ late in the anneal (when $Q(s^*)\ll 1$), then one would expect (ignoring any other noise sources) to find $\tilde{\beta} = \beta B(s^*)/J_{\mathrm{max}}$, where $\beta=1/k_B T$ is the physical inverse temperature, $B(s^*)$ is the problem Hamiltonian strength at the freeze-out point, and $J_{\mathrm{max}} = \max |J_{ij}|$ is a normalization parameter (since the $J_{ij}$ programmed into the quantum annealer are restricted to the range $[-1,1]$).

We make two key observations. 1) almost all of these problems tested exhibit a well defined optimal pause point in a fairly narrow region $s_p^{\mathrm{opt}} \in [0.35,0.46]$ (i.e., much less varied than the 12-qubit instances studied above). 2) The data returned from the annealer, for all of these problems, resembles functionally a classical Boltzmann distribution for $H_p$, but at a higher temperature than the operating temperature of the annealer (at least 1.5 times higher). This is in accordance with the results of Ref.~\cite{marshall-rieffel-hen-2017} where calculated freeze-out points for most large problems were very early in the annealing schedule (equivalent to a higher than expected temperature), although in this work, a pause during the anneal to more directly study thermalization was not available.
We demonstrate these points below.

\begin{figure}
\begin{center}
\includegraphics[width=0.9\columnwidth]{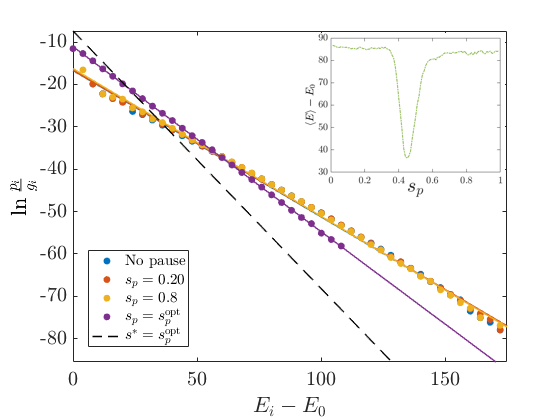}
\caption{Fitting experimental data to linear fit for a single 501 qubit instance ($\mathcal{I}_{501}^0$). We show data for three different pause points (and from the standard annealing schedule). We see the standard schedule is almost indistinguishable from the case where the anneal is paused at $s=0.2,0.8$ (slope $\tilde{\beta}\approx 0.35$), however, when pausing at $s=s_p^{\mathrm{opt}}=0.44$ for this instance, we see a change in the distribution returned (purple, with slope $\tilde{\beta}=0.44$). We also plot the corresponding classical Boltzmann distribution expected (black-dash line) if the system were thermalizing to $H_p$ at $s^*=s_p^{\mathrm{opt}}$, at temperature 12.1mK (with slope $\tilde{\beta}=0.61$). Note, freeze-out at $s^*=1$ would give $\tilde{\beta}=2.52$.
Inset: Average energy returned by the annealer as a function of pause point for the same instance. The curve has a minimum value at $s_p^{\mathrm{opt}}=0.44$. Experimental data obtained from 10000 anneals with 5 choices of gauge, with $t_a=1\mu$s, and $t_p=100\mu$s.}
\label{fig:linear-boltzmann}
\end{center}
\end{figure}

First consider Fig.~\ref{fig:linear-boltzmann} where for a single instance we plot $\ln \frac{p_i}{g_i}$ against $E_i$, where $p_i$ is the probability of observing a configuration with energy $E_i$, and $g_i$ is the degeneracy of that energy level. One can see the data returned from the annealer corresponds closely to a linear fit (for all problems, and all pause points $s_p$, we find the $R^2$ (coefficient of determination) value is greater than 0.97, and up to 0.9999).
That is, the data from annealer seems to fit to $p_i = \frac{g_i}{Z}e^{-\tilde{\beta}E_i}$, for some constants $Z$, and $\tilde{\beta}$ (which can be determined by least-squares fitting).

Though the results are clear, the correct interpretation of them is not. For example, if we obtain the effective inverse temperature of the distribution $\tilde{\beta}$ from the least squares fitting, and set it equal to $\tilde{\beta} = \beta B(s^*)/J_{\mathrm{max}}$, with knowledge of $\beta = 1/k_B T$, and $J_{\mathrm{max}}$, we can calculate $B(s^*)$. If one does this however, the value returned corresponds to an extremely early point during the anneal, even earlier than $s_p^{\mathrm{opt}}$ (e.g. with $Q \approx 1$, or equivalently $s\approx 0.35 $).
If one however assumes the thermalization picture presented in Sect.~\ref{sect:theory}, which suggests the dynamics should freeze when the relaxation time scale is longer than the system time scale, i.e. approximately around the optimal pause point, $s_p^{\mathrm{opt}} = 0.42 \pm 0.01$ for these problems, the temperature required for the fit is $>1.5$ times higher than the physical temperature $T = 19.8 \pm 1.1$mK (compared to $T_{\mathrm{DW}} = 12.1 \pm 1.4 $mK).

It is however not clear whether $\exp(-\beta H(s^*))$ with $s^* = 0.42$ and $Q(s^*) \approx  0.5$ would indeed correspond to a classical Boltzmann distribution (of $H_p$) since the off diagonal driver is still relatively strong in magnitude. 
Indeed, this is a somewhat similar result as from the previous section where we observed the optimal parameter value for $s$ was in fact slightly after $s_p^{\mathrm{opt}}$ (and $T$ larger than the physical temperature). If there are in-fact still dynamics after $s_p^{\mathrm{opt}}$, the dynamics will be frozen somewhat later in the anneal (when $Q(s)$ is smaller), and the associated temperature of the fit will be larger.
We discuss some implications of this in the next section. For now we compare the samples from the optimal pause point to those from outside of it.

\begin{figure}
\begin{center}
\includegraphics[width=0.9\columnwidth]{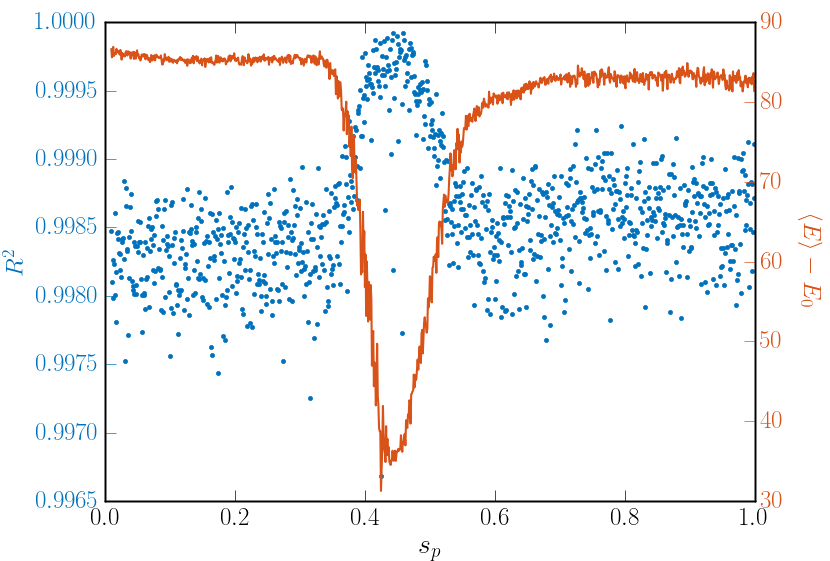}
\caption{Quantifying accuracy of linear regression using $R^2$ for $\ln \frac{p_i}{g_i}$ as a function of $E_i$ (as demonstrated in Fig.~\ref{fig:linear-boltzmann}) for a single 501 qubit instance ($\mathcal{I}_{501}^0$). For each data point shown, we obtain a least-squares fitting from the distribution returned by the annealer from which we calculate the $R^2$ value.
The solid line (red, right-axis) is the average energy returned from the annealer.
We see the peak in $R^2$ corresponds to the region around $s_p^{\mathrm{opt}}$ (just after $s_p=0.4$).
 Each data point is obtained from 10000 anneals with 5 choices of gauge, with $t_a=1\mu$s, and $t_p=100\mu$s.}
\label{fig:R2_sp}
\end{center}
\end{figure}

In Fig.~\ref{fig:R2_sp} we plot the $R^2$ value found by the least squares fitting for a typical instance as a function of pause point $s_p$. We see that the peak corresponds closely to the optimal pause point under the pause. Moreover, we see a very similar trend for almost all of the problems we tested, with the pause point for which the largest $R^2$ value is observed differing by just a few percent of the annealing schedule from the optimal pause point; $s_{R^2}^{\mathrm{max}} = s_p^{\mathrm{opt}} \pm 0.03$.
This indicates that the data returned from a pause at this critical point, fits better to a classical Boltzmann distribution, as compared to the rest of the data (or indeed, from the distribution returned in the absence of a pause in the schedule). 
If indeed the problems are thermalizing to a classical Boltzmann distribution, this work shows that by pausing the anneal at a particular (instance-dependent) point allows a more complete thermalization to occur. In particular, it thermalizes at a later point in the anneal, when $H(s)$ is more diagonal (`classical').
This result is also similar to that found in the previous section, and is largely explainable, qualitatively, through the model introduced in this work.

\begin{figure}
\begin{center}
\includegraphics[width=0.9\columnwidth]{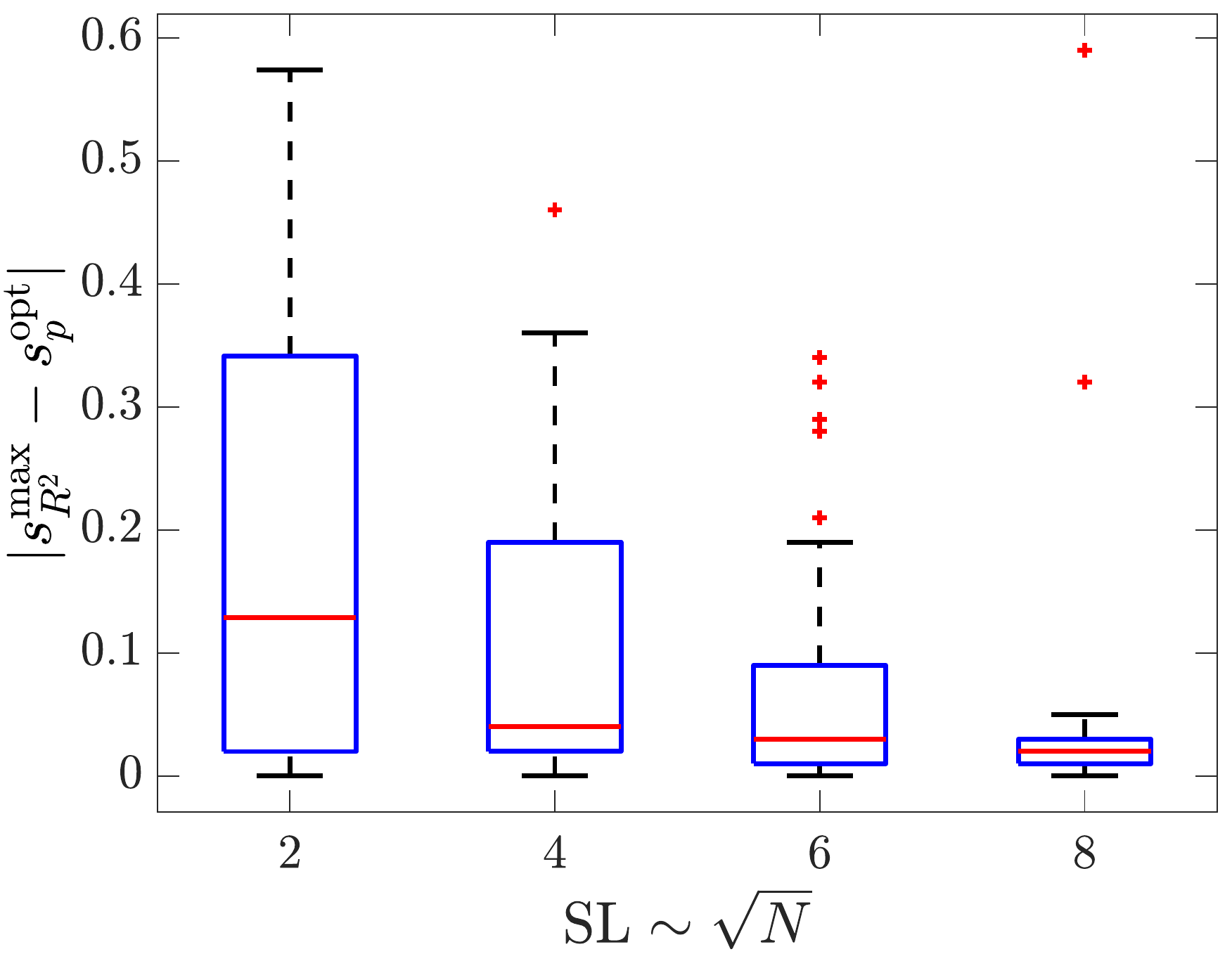}
\caption{Box plot of the difference between the pause point for which the maximal $R^2$ value is found $s_{R^2}^{\mathrm{max}}$ (i.e., the closest to a Boltzmann distribution), and the optimal pause point $s_p^{\mathrm{opt}}$, with problem size.
The problems are defined on square sub-graphs of the full D-Wave chimera with side-length SL.
Each SL contains at least 55 instances.
Plotted in the box is median, lower ($q_1$) and upper ($q_3$) quartiles. Whiskers are minimum and maximum values across entire data set, excluding outliers. Outliers are red crosses, determined by being larger [smaller] than $q_3 + 1.5(q_3-q_1)$ [$q_1 - 1.5(q_3-q_1)$].
 To collect our data we used 10000 anneals with 5 choices of gauge, with $t_a=1\mu$s, and $t_p=100\mu$s, for each problem.}
\label{fig:s_ratio}
\end{center}
\end{figure}

We performed a similar analysis for three other problem sizes ($N=31,125,282$ qubits), and find that increasing the problem size in general increases the mean $R^2$ value for a fit to a classical Boltzmann distribution. For $N=31,125,282,501$, the corresponding values are $\langle R^2 \rangle = 0.911,0.994,0.995,0.997$, where the average (median) is over all instances and all pause points $s_p$ tested.
Moreover, the correlation between $s_p^{\mathrm{opt}}$ and $s_{R^2}^{\mathrm{max}}$ seems to also increase with problem size, as demonstrated in Fig.~\ref{fig:s_ratio}, which shows the variation between different instances decreases with problem size, and seems to suggest that $s_p^{\mathrm{opt}} \approx s_{R^2}^{\mathrm{max}}$, for large $N$ (i.e., the optimal pause point and the pause location for which the best fit to a Boltzmann distribution is observed, coincide for large problems).

We summarize and interpret some of these findings in the next section.

\section{Summary and Outlook \label{sect:summary}}

Our work demonstrates that even enabling simple adaptions of the default annealing schedule, such as pausing, can result in a rather stark change in performance. For the two classes of problems we studied, we found a critical region for which if one pauses, even for a relatively short time (say $10\mu$s), a drastically different distribution of solutions is returned. In particular, annealing with a pause in this region more effectively samples low lying energy states, resulting in a larger probability of success.

The best place to pause is around 10-15\% of the total anneal time after the minimum gap (for the problem class we studied).
This effect does not make sense in a closed-system scenario, as discussed in Appendix~\ref{sect:schrodinger}, but can be explained, at least qualitatively, in terms of open system dynamics, particularly  thermalization.
The qualitative picture we discussed suggests that after the minimum gap -- when thermal excitations may allow a significant fraction of the population to leave the ground state -- population can begin to thermally relax back into the ground state. 
This picture also explains the sharp-peaked nature of our observations; if one pauses just a little too late, since the transition rates depend exponentially on the size of the instantaneous energy gaps 
they quickly drop off, and the pause length becomes too small (relative to the instantaneous relaxation time) for effective thermalization to occur.
The exact behavior depends heavily on the spectrum of each individual instance, since this determines the transition rates.

Our results provide positive evidence for the Boltzmann nature of the distributions returned from the annealer, even for the majority of problems for which, as was shown in Ref.~\cite{marshall-rieffel-hen-2017}, a sensible freeze-out point does not exist. 
To obtain further insight into the nature of the distributions at the end of the anneal, we performed two studies. On small ($12$-qubit) problems for which we are able to compute the quantum Boltzmann distribution for every point in the anneal, we compare the final distribution to the projected quantum Boltzmann distributions. For larger problems, it is not feasible to compute the quantum Boltzmann distributions throughout the anneal, but with the aid of recent entropic sampling techniques 
that enable accurate estimation of the eigenspectrum degeneracies for a class of planted solution problems, we were able to fit the final distribution, for multiple pause locations, to a classical Boltzmann distribution for $H_p$ with the effective temperature as a fitting parameter.

The first study showed that the best performance occurs when the pause takes place after the minimum gap, confirming our qualitative picture. Further, the fit between the projected quantum Boltzmann distribution and the final distribution is poor except when the pause is in the region shortly after the gap, strongly suggesting that the pause contributes significantly to thermalization.   The best fit between the final distribution and the projected quantum Boltzmann distribution occurs somewhat after this pause point, and was much higher than the physical temperature.
It is unclear from our picture as to where one would expect the best fit to be because, as we demonstrated, there may be significant dynamics after the optimal pause point, since the annealing rate is limited on the device. 
Further, there may be discrepancies between the computed quantum Boltzmann distribution and the distribution that would be predicted if we had better knowledge of the device. The computed effective temperature depends on the device temperature, which is sampled only coarsely in time, and may not be the temperature at the qubits, and may also fluctuate within a single anneal. Control errors on the $J_{ij}$ can also have significant effect on the distribution \cite{analog-errors}. Future work with more flexible and instrumented annealers, as well as theoretical advances, will further deepen our understanding of open system dynamics and quantum annealing.

The second study showed that for larger problems the best fit between the final distribution and a classical Boltzmann distribution for $H_p$ occurs when the pause is at the optimal pause point, indicating these samples are thermalizing more completely, and again confirming our qualitative picture. In this case, an excellent fit between a classical Boltzmann distribution for $H_p$ and the final distribution occurs at all points, 
and even in cases in which no pause has been inserted. Our results further suggest that the larger a problem is, the more likely it is to have a good fit to a classical Boltzmann distribution. By varying the pause point, one can effectively vary the temperature at which one samples from the classical Boltzmann distribution, with potential implications for machine learning, such as the use of quantum annealing in restricted Boltzmann machines.  
The fits are obtained with effective temperature as a free parameter. We find that generally the effective temperature is substantially higher than predicted. As with the small problem study, discrepancies between the device temperature and the qubit temperature, fluctuations in this temperature during an anneal, and control errors masquerading as higher temperatures may all contribute to a higher effective temperature. While the device temperature is sampled only infrequently, the fluctuations in fitted temperature did fairly consistently track those of the actual device.

A further  difficulty in interpreting this data is that the optimal pause point seems slightly too early in the anneal to observe a classical Boltzmann distribution. 
The extent to which these devices can be used as effective thermal samplers remains a somewhat open question, and most likely depends on the application; on the one had as it is known that  solution sampling can be biased on these types of annealing devices \cite{mandra-biased-sampling,fairInUnfair}, but on the other hand, several works \cite{adachi,perdomo,DW2x-gibbs-ML} suggests that for practical applications, such as in the context of machine learning from a thermal distribution, these devices can indeed be used effectively, even if the temperature is unknown.
Again, hardware and theory advances are needed to fully clarify this picture.

An obvious next step would be to investigate the effect of pausing on other problem classes, particularly embedded problems related to applications. Key questions include the extent to which pausing improves performance in other problem classes, and the robustness of the region in which pausing is effective across instances within each class and as the size of the problems grow. A natural question to ask is the extent to which yet more flexibility in the annealing schedule can provide yet greater improvements in performance. A significant challenge to the field is to provide more theory guidance, beyond the qualitative picture presented here, in order to (i) better explain the results we presented, (ii) suggest which new features would be most effective to add in the next generations of quantum annealers, and (iii) guide the design of effective annealing schedules on current and future generations of these devices. Early challenges for such a theory would be to predict the location and width of the effective pause regime, and particularly how they scale with problem size, and to characterize the output distributions, particularly the effective temperature of the best fit Boltzmann distribution for $H_p$, deviations for Boltzmann, and scaling with problem size.

One possible avenue of study along these lines is to attempt to optimize the annealing schedule as in Ref.~\cite{opt-schedule-qa}. In this work, the rate at which the anneal is performed is related to the specific heat, governing the amount of influence the quantum dynamics have on the sampling. Whether or not this picture and other thermodynamic quantities can also explain some of the observations in this work is another interesting question.

Similarly, more detailed simulations of the pause effect -- for example, in the master equation setting \cite{adiabaticME,albash-2015-entanglement} -- would also lead to a deeper understanding.

Based on the study we present here, useful features of future quantum annealers would include both a faster quench and more accurate temperature data. On the D-Wave 2000Q, no part of the schedule can be traversed faster than $\frac{ds}{dt}=1\mu\mathrm{s}^{-1}$. We would like to more precisely understand the distribution at the optimal pause point. To do so, one would need to be able to more effectively quench the system. Enabling faster traversal of the schedule, particularly in the 3rd and 4th regions of the anneal (see Fig.~\ref{fig:cartoon}), would enable experiments giving greater insights into the distributions during the anneal than are possible currently. 
An alternate means for performing a fast quench or the addition of probes mid anneal would also provide insight. 
With regard to the temperature, it would be helpful to have further insight into the extent to which (i) the temperature at the qubits differs from that of the measured device temperature, (ii) the temperature changes during the course of the anneal, and (iii) control errors, such as deviations in the values of the $J_{ij}$ on the device from the intended values. While deep probing of the temperature is experimentally challenging, lighter probing should be possible, and even access to finer-grained temperature data than the current rate of once every few hours would help fill out the picture.

The qualitative picture we give to explain our results suggests that similar behaviour should be found in problem classes more closely tied to applications, both in machine learning and optimization, assuming the spectral gaps open up enough towards the end of the anneal (in the third region of Fig.~\ref{fig:cartoon}). 
Since we find empirically that the optimal pause region is fairly consistent across a problem class, it suggests that good heuristics may be derivable for adapted annealing schedules including a pause, through trialing just a small representative set of problems. Of course, this is assuming that problem classes of interest exhibit similar properties as those demonstrated in this work, which can only be verified through further testing.
Whilst the region in which a pause is beneficial is only a fraction of the anneal schedule, that region is substantially wider than that of the minimum gap, suggesting both greater robustness and greater ease in finding effective locations in which to pause or slow down than in closed system adiabatic quantum computing. 
More generally, a deeper open systems understanding would give greater insight into the design of both annealing schedules and quantum hardware, from annealers to universal quantum computers.

\section{Take Away Points}
We provide the following summary of our results in a concise manner:
\begin{itemize}
    \item We considered a simple adaptation to the annealing schedule in an experimental quantum annealer, in which the anneal is paused at an intermediate point for a certain amount of time (up to 2ms).
    \item Typically each problem instance has is an `optimal pause point' $s_p^{\mathrm{opt}}$: a location during the anneal which a pause has a dramatic effect on the output distribution of the device. In particular, we observe a more efficient sampling of the ground state and low lying energy levels.
    \item We experimentally observed orders of magnitude improvement in performance with respect to unpaused annealing for certain problems of the planted-solution type by the use of a pause at the optimal location.
    \item These observations, as well as some other qualitative features are largely explainable through a phenomenological model of non-equilibrium thermalization, with three relevant time-scales: the relaxation time $t_r$, the Hamiltonian evolution time $t_H$ and the pause time $t_p$.
    \item $s_p^{\mathrm{opt}}$ is found to occur after the location of minimum gap, as expected; we conjecture pausing after the minimum gap allows for the ground state to re-populate after dissipative transitions which occur during the region of the minimum gap.
    \item By studying the output probability distribution of the annealer, we provide evidence suggesting thermalization to a classical Boltzmann distribution is occurring in problems containing up to 500 qubits. The temperature of this classical distribution however is different from the physical temperature of the device. More accurate temperature estimates and a faster annealing rate would allow for a deeper investigation.
    \item We observed $s_p^{\mathrm{opt}}$ is relatively robust across a problem class; that is, the optimal pause point occurs in a similar, fairly narrow region for different problems of the same problem class, even for different sizes.
\end{itemize}

\section{Data acquisition}
For all instances labeled as $\mathcal{I}_N^s$ (where $N$ is the qubit number, and $s$ a serial number when applicable), we provide the problem instance itself as part of the ancillary files.
Any other problem instances (and data) available upon request.
\\

\acknowledgments
We thank Lev Barash for computing the degeneracies of the 500 qubit problems in Sect.~\ref{sect:cbm-main}. E.G.R. and D.V. would like to acknowledge support from the NASA Transformative Aeronautic Concepts program and the NASA Ames Research Center. D.V. was supported by NASA Academic Mission Services (NAMS), contract number NNA16BD14C, and J.M. by the USRA Feynman Quantum Academy under NAMS. E.G.R. and D.V. were also supported in part by the AFRL Information Directorate under grant F4HBKC4162G001 and the Office of the Director of National Intelligence (ODNI) and the Intelligence Advanced Research Projects Activity (IARPA), via IAA 145483. I.H. was partially supported by the Office of the Director of National Intelligence (ODNI), Intelligence Advanced Research Projects Activity (IARPA), via the U.S. Army Research Office contract W911NF-17-C-0050.
The views and conclusions contained herein are those of the authors and should not be interpreted as necessarily representing the official policies or endorsements, either expressed or implied, of ODNI, IARPA, AFRL, or the U.S. Government. The U.S. Government is authorized to reproduce and distribute reprints for Governmental purpose notwithstanding any copyright annotation thereon.

\bibliography{biblio.bib}

\begin{thebibliography}{39}
\bibitem [{not({\natexlab{a}})}]{note-cartoon}%
  \BibitemOpen
  \href@noop \bibinfo {{The time-scales,
  $t_{r,H,p}$, shown in Fig.~\ref{fig:cartoon}, are drawn arbitrarily for
  illustrative purposes.}}\BibitemShut {Stop}%
\bibitem [{not({\natexlab{b}})}]{note-sopt}%
  \BibitemOpen
  \href@noop \bibinfo {{Throughout this work
  the optimal pause point $s_p^{\mathrm{opt}}$ is evaluated by finding the
  lowest average energy (from many thousands of anneals) returned from the
  D-Wave device. In principle one could use the ground state success
  probability as a metric, however this is often not practical since 1) some
  hard problems are not solved even once by the D-Wave, and 2) for randomly
  generated instances (i.e., not of the planted-solution type), the ground
  state energy is typically unknown.}}\BibitemShut {Stop}%
\end{thebibliography}

\clearpage

\appendix

\section{Expected closed-system dynamics\label{sect:schrodinger}}
We consider the effect of an intermediate pause under the closed system (Schr{\"o}dinger) evolution alone, studying 12-qubit problems, with $J_{ij}\in [-1,1]$ (uniformly random) and $h_i=0$.

In the simulation, we estimate the exact (closed system) unitary evolution for an anneal, with anneal time $t_a$, and pause of length $t_p$ at $s=s_p$, which can be written,
\begin{equation}
    U(s_p) = \mathcal{T}e^{-\frac{it_a}{\hbar}\int_{s_p}^{1}H(s)ds} e^{-\frac{it_p}{\hbar}H(s_p)}\mathcal{T}e^{-\frac{it_a}{\hbar}\int_0^{s_p}H(s)ds}
\end{equation}
where $\mathcal{T}$ represents time-ordering.
In particular, we evolve the initial ground state $|\Psi(0)\rangle$ at $s=0$ to $s=s_p$ (by `Trotterization'), evolve unitarily for a time $t_p$ under Hamiltonian $H(s_p)$, and then evolve from $s=s_p$ to $s=1$ (again, by Trotterization). This computes the full unitary evolution, with an intermediate pause, $|\Psi(1)\rangle = U(s_p)|\Psi(0)\rangle$.
After this we compute (for example) the ground state success probability 
\begin{equation}
P_0(s_p) = \sum_{z_0 : H_p|z_0\rangle = E_0 |z_0\rangle}|\langle z_0 |U(s_p)| \Psi(0)\rangle |^2, 
\end{equation}
where $E_0$ the ground-state eigenvalue of $H_p$.

\begin{figure}
\begin{center}
\includegraphics[width=0.9\columnwidth]{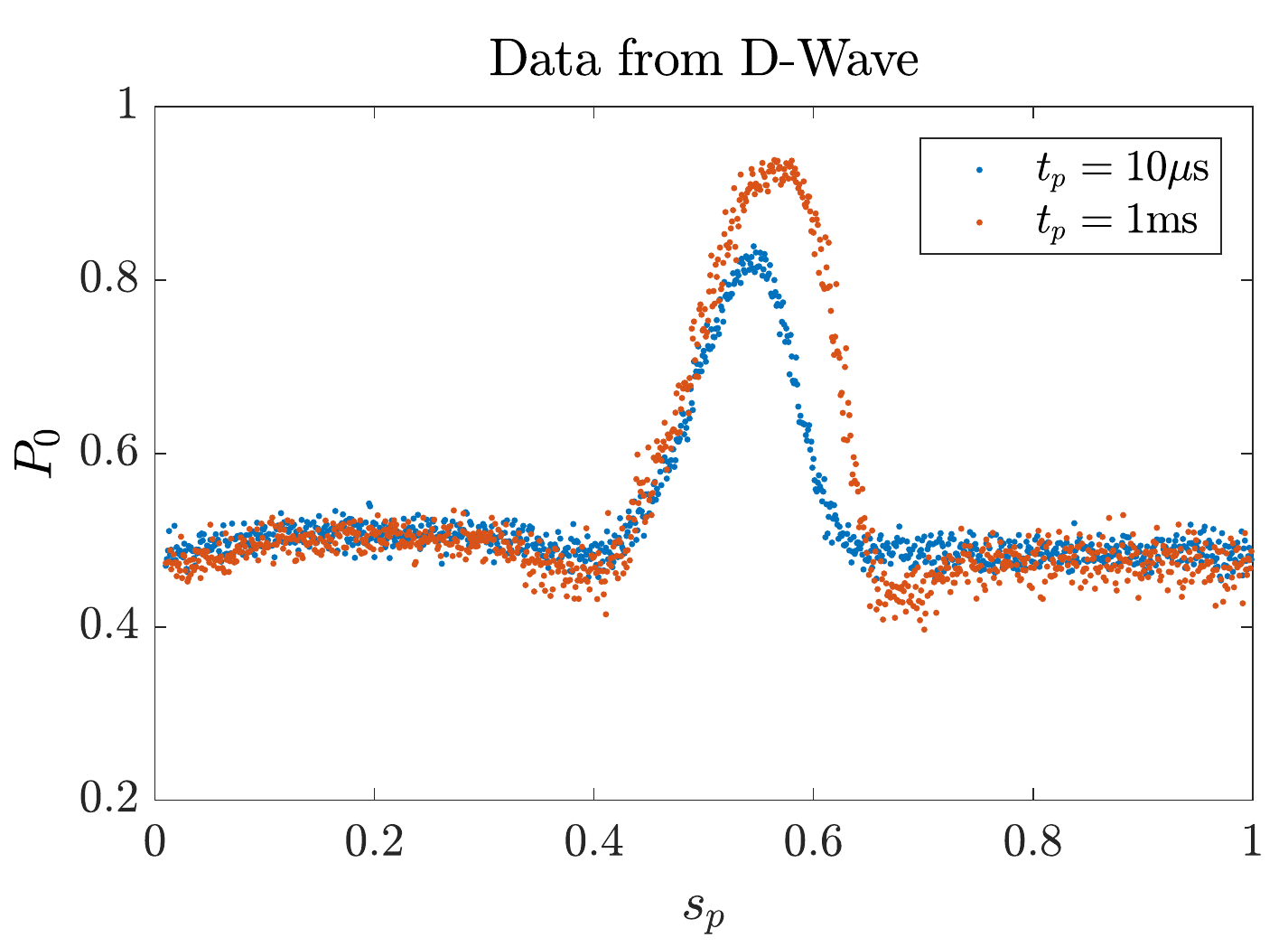}
\includegraphics[width=0.9\columnwidth]{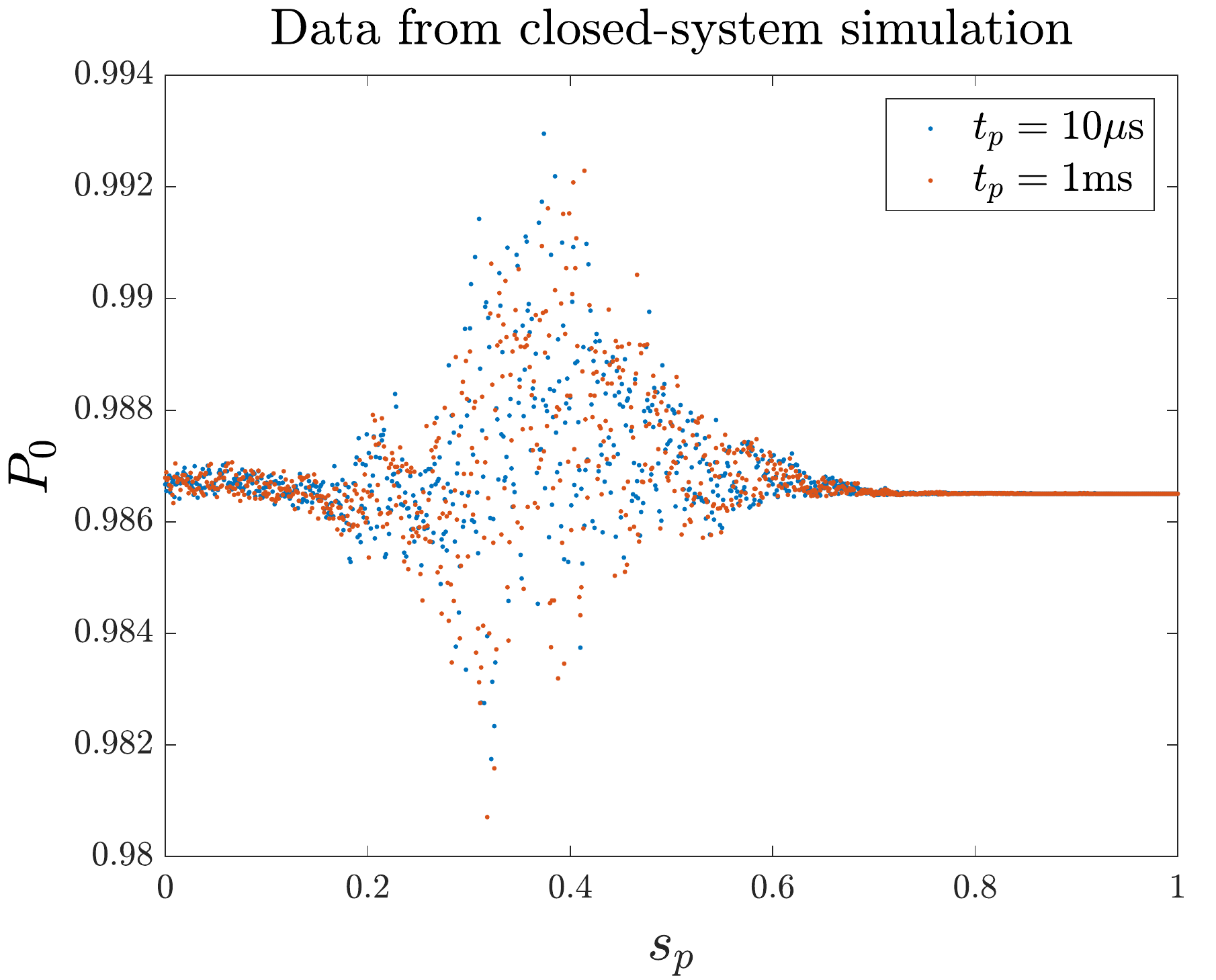}
\caption{Comparison between results from the experimental D-Wave annealer (top), and a closed-system Schr{\"o}dinger evolution (bottom) for a single 12-qubit problem instance ($\mathcal{I}_{12}^0$). Both plots show the success probability $P_0$ against the pause point, for two different pause lengths ($t_p=10,1000\mu$s) as shown by the legend. The annealer data is from 10000 annealing runs, using 5 different gauges. Each plot contains 1000 data points evenly distributed in $s_p\in[0,1]$. Both have an annealing time of $t_a=1\mu$s (in addition to the pause time).
The simulation uses 1000 time steps in the Trotterization of the evolution operator.}
\label{fig:DW_vs_schrodinger}
\end{center}
\end{figure}

\begin{figure}
\begin{center}
\includegraphics[width=0.9\columnwidth]{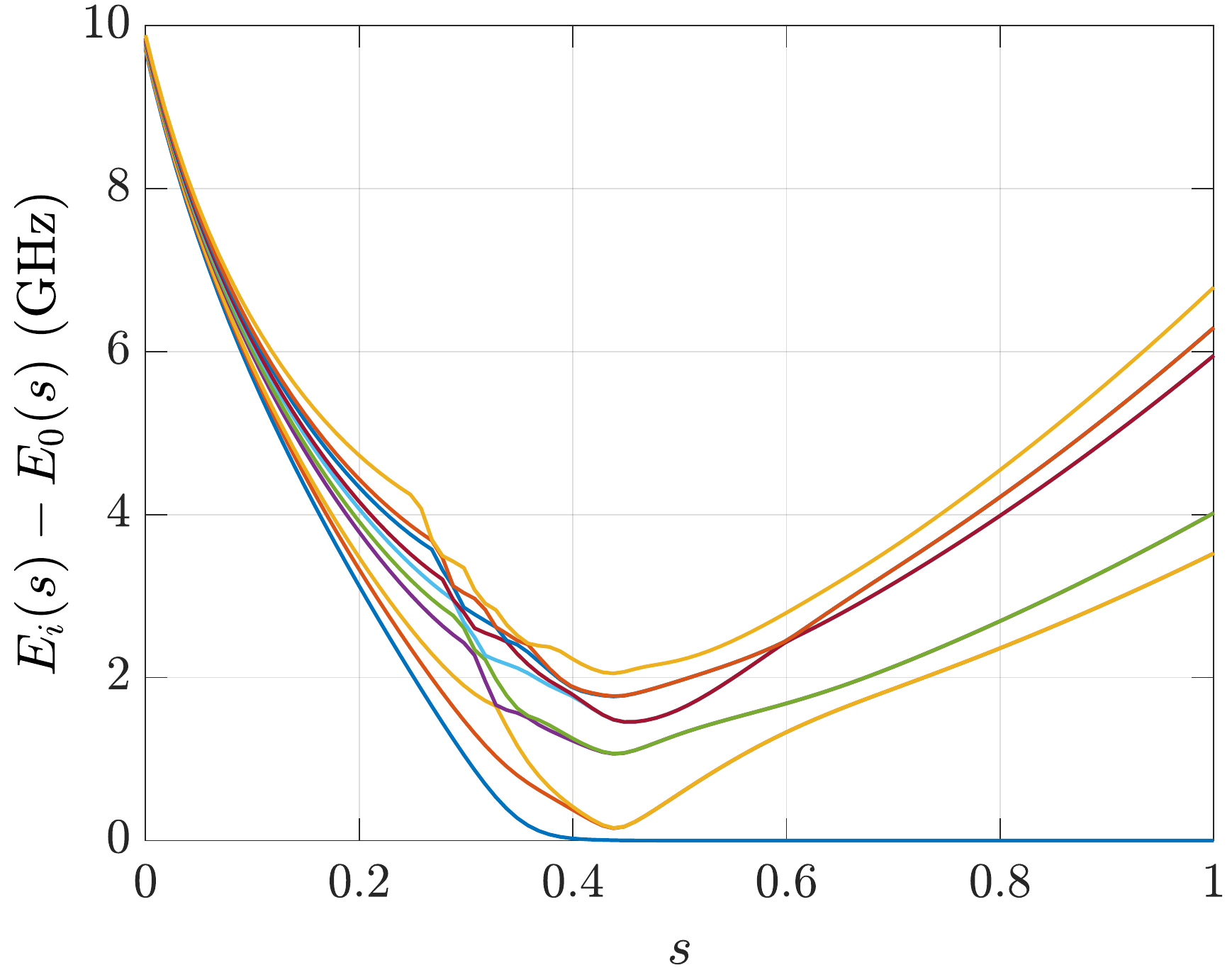}
\caption{Spectrum (lowest 10 energy levels) of the problem considered in Figs.~\ref{fig:vary_tp}, \ref{fig:vary_ta}, \ref{fig:RA}, \ref{fig:DW_vs_schrodinger} ($\mathcal{I}_{12}^0$). This problem has a small, well defined minimum gap of 0.15 GHz located at $s=0.44$. The units are defined with $h=1$.}
\label{fig:spectrum}
\end{center}
\end{figure}

It is interesting to note that the effect of pausing the anneal does have a noticeable effect, as demonstrated in Fig.~\ref{fig:DW_vs_schrodinger}, even in the closed system case. We believe this is essentially caused by Rabi oscillations during the pause, and we note it does not match the observed output from the D-Wave quantum annealer. We explain below.

We consider the three regions in Fig.~\ref{fig:DW_vs_schrodinger} (bottom). 1) During the evolution, when $s<0.2$ the state is almost entirely in the instantaneous ground state, $|\Psi(s)\rangle \approx |E_0(s)\rangle$. Thus, when the system is paused, and evolved under $H(s_p)$, very little happens since just an overall global phase is acquired. 2) A little later on, when the energy gap starts to close between $s\in[0.2,0.4]$ (see Fig.~\ref{fig:spectrum}), diabatic transitions to excited energy levels may occur. Once a non-negligible amount of the population has been transferred to excited states, a pause will give rise to Rabi oscillations between the eigen-states of $H_p$, hence directly affecting the success probability at the end of the anneal. 3) Late in the anneal, after around $s=0.7$, the driver Hamiltonian is essentially negligible, so can not drive any transitions between energy levels, hence a pause, will only change the relative phases of eigen-states of $H_p$, but not affect the probabilities upon measurement in the computational basis.

We describe three fundamental differences between the simulation, and the results from the experimental annealer (in addition to the large difference in success probability). Firstly it is evident there is much less structure in the closed system case; although the success probability does seem to increase on average, there is much more variability. This is due to the sensitivity of the period of the Rabi oscillations to the energy gaps (and hence to the location of the pause $s_p$). Second, there is seemingly no qualitative difference between a short and long pause in the closed system case, as compared to the observed phenomena which has the peak increasing with pause length. We believe this is due again to the nature of the Rabi oscillations. If the gap between energy levels is of the order of 1 GHz (see Fig.~\ref{fig:spectrum}), the time-scale of the Rabi oscillations is much shorter than the pause lengths considered here (e.g. 1ns compared with 100$\mu$s). Lastly, the location at which this Rabi dynamics occurs does not correspond precisely to that observed in the experiment (it is seemingly shifted slightly earlier in the schedule).

\begin{figure}
\begin{center}
\includegraphics[width=0.9\columnwidth]{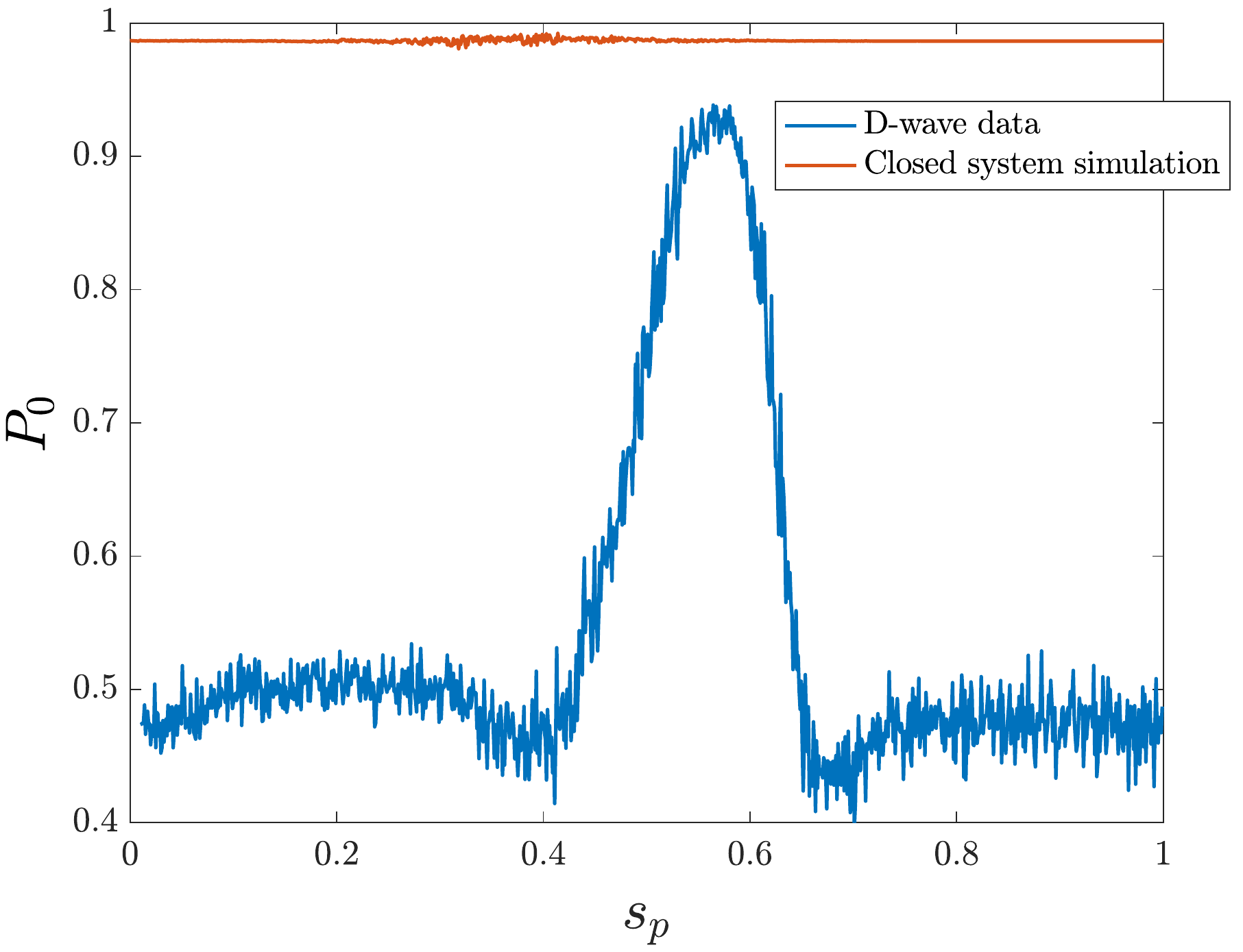}
\caption{Same as Fig.~\ref{fig:DW_vs_schrodinger}, but with both data sets on the same plot. Both data sets correspond to $t_p=1000\mu$s, and $t_a=1\mu$s.}
\label{fig:schrodinger-same_plot}
\end{center}
\end{figure}

We also plot both data sets on the same axis in Fig.~\ref{fig:schrodinger-same_plot} for reference. This shows the Rabi induced oscillations are effectively negligible compared to the effect observed on the physical annealer.

\section{Computing the quantum Boltzmann distribution \label{sect:qbm}}
We wish to compare the distribution returned from the D-Wave annealer, i.e., the probability that a configuration  with energy $E$  is returned $P_{\mathrm{DW}}(E)$, to what would be expected if the annealers were instantaneously thermalizing to the quantum Boltzmann distribution $\rho(s,T) := \frac{1}{Z}e^{-\beta H(s)}$, where $Z = \mathrm{Tr} e^{-\beta H(s)}$, and $\beta = 1/k_B T$ with $T$ a temperature parameter.

We note that the D-Wave annealer can only measure in the computational ($z$) basis, and assuming the system can be quenched appropriately, to compare the probability distributions we compute
\begin{equation}
P^{(s,T)}_{\mathrm{QBM}}(E) = \sum_{z\, : \, H_p|z\rangle = E|z\rangle} \langle z | \rho(s,T) |z\rangle
\end{equation}
for $s\in[0,1]$ (steps of 0.01), and $T\in[\frac{1}{4},4]T_{\mathrm{DW}}$ (steps of $\frac{1}{4}T_{\mathrm{DW}}$).

\section{Computing the classical Boltzmann distribution for large problems \label{sect:cbm}}
In Sect.~\ref{sect:cbm-main} we analysed planted-solution type problems \cite{hen:15}, of four different problem sizes, $N\in \{31,125,282,501\}$. These were (a subset of) the same instances as studied in Ref.~\cite{marshall-rieffel-hen-2017}, and are generated on sub-graphs of the full chimera of side-length $\mathrm{SL} \in\{2,4,6,8\}$ respectively.
Each $N$ group tested consisted of (at least) 55 problem instances, with a random number of sub-Hamiltonian loops chosen (as described in more detail in Ref.~\cite{marshall-rieffel-hen-2017}).

The benefit to using this problem type is that one knows in advance the general form of the spectrum of $H_p$, and one can calculate exactly the degeneracy of the ground and first excited states. The description of this algorithm is outlined in Ref.~\cite{fairInUnfair}.

Knowledge of the exact ground and first excited state degeneracies is extremely powerful as it allows one to verify any estimated degeneracy values from entropic sampling techniques (such as the well known Wang-Landau method \cite{wang:01,wang:01a}).

We used a newly devised algorithm for estimating the density of states based on population-annealing \cite{Barash-2018-entropic_PA} to obtain accurate estimates of the degeneracies for the largest instances tested (501 qubits), for which traditional (e.g. Wang-Landau) approaches failed. Here `accurate' implies neither the ground nor first excited state degeneracy estimate differed by more than 5\% of the exact values.
For the 125 and 282 qubit problems we were able to use the Wang-Landau algorithm to obtain accurate estimates of the degeneracies.
For the 31 qubit instances we used exact enumeration to compute the degeneracies.

\section{Supplemental figures \label{sect:figs}}

\begin{figure}
\begin{center}
\includegraphics[width=0.95\columnwidth]{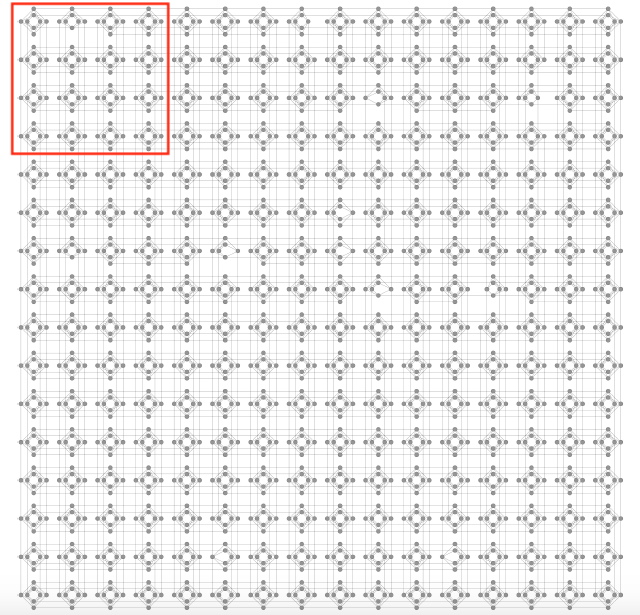}
\caption{The D-Wave 2000Q `chimera' graph which we conducted our experiments on (the machine is housed at NASA Ames Research Center). There are $16\times 16$ unit cells each containing 8 qubits. Dead (malfunctioning) qubits are not shown on the graph. Top left bordered in red is an example of a square subgraph of side-length SL$=4$.}
\label{fig:chimera}
\end{center}
\end{figure}

In Fig.~\ref{fig:chimera} we show the full (working) D-Wave 2000Q hardware graph. All of our experiments were conducted on this graph.
 
\begin{figure}
\begin{center}
\includegraphics[width=0.8\columnwidth]{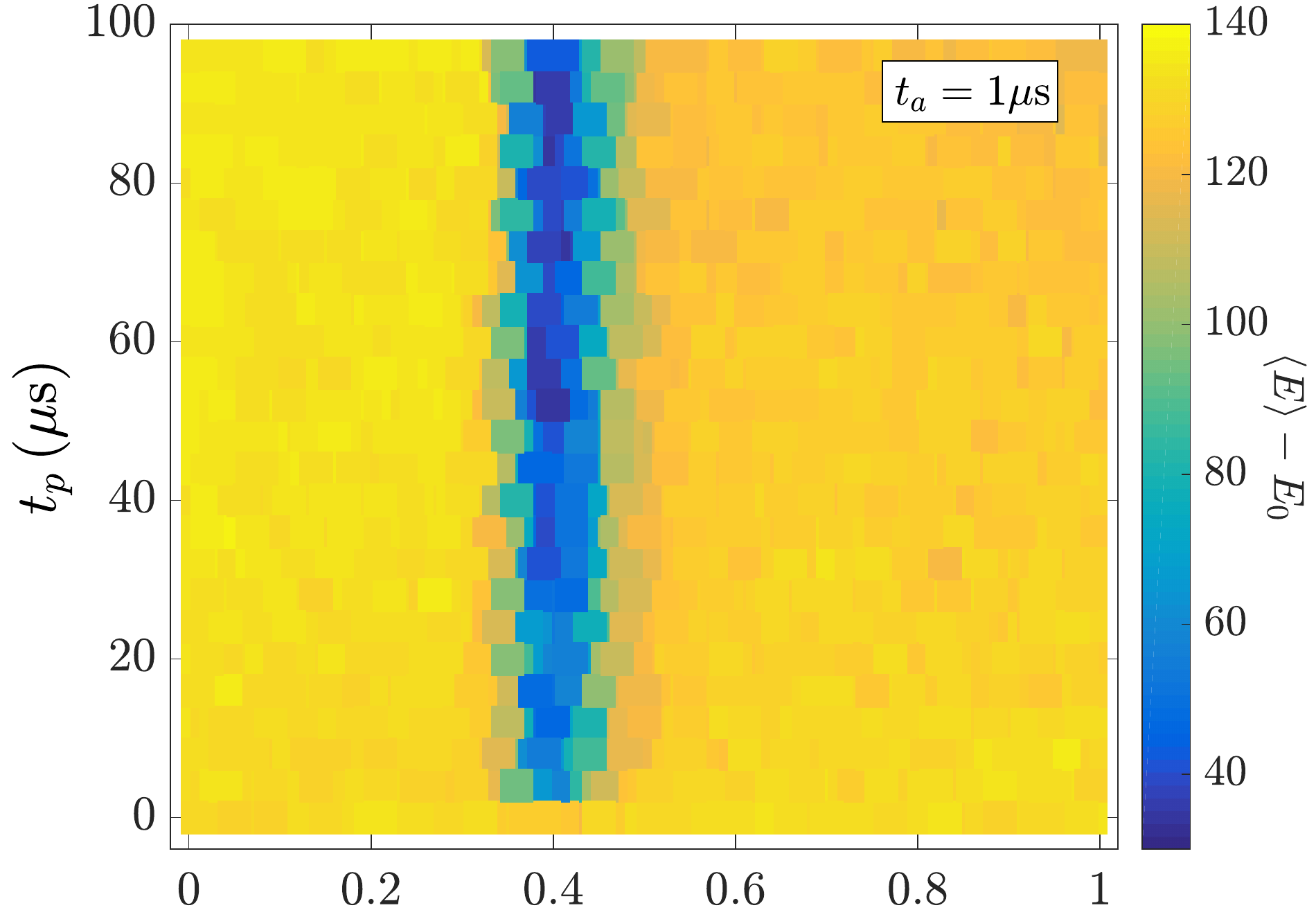}
\includegraphics[width=0.8\columnwidth]{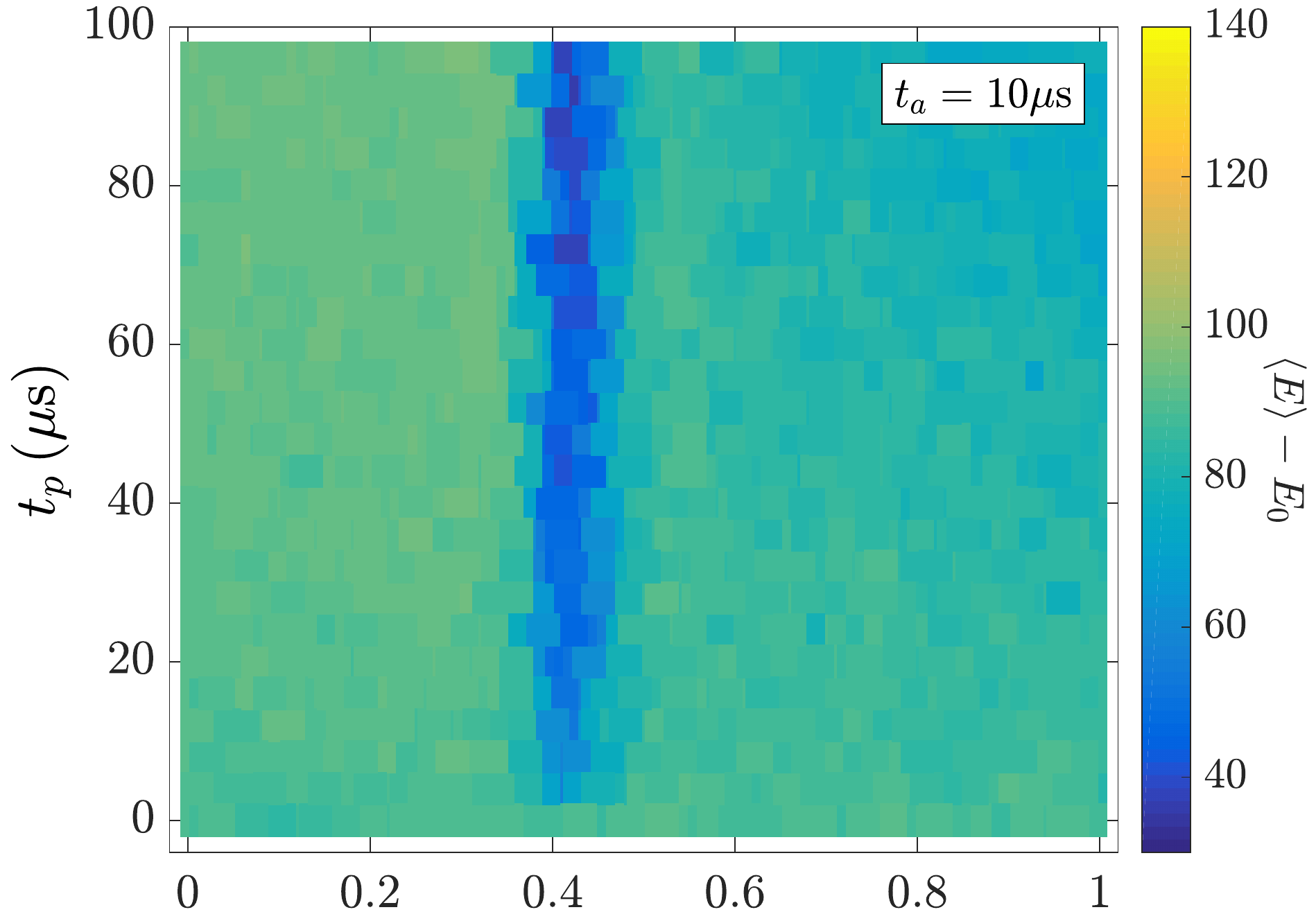}
\includegraphics[width=0.8\columnwidth]{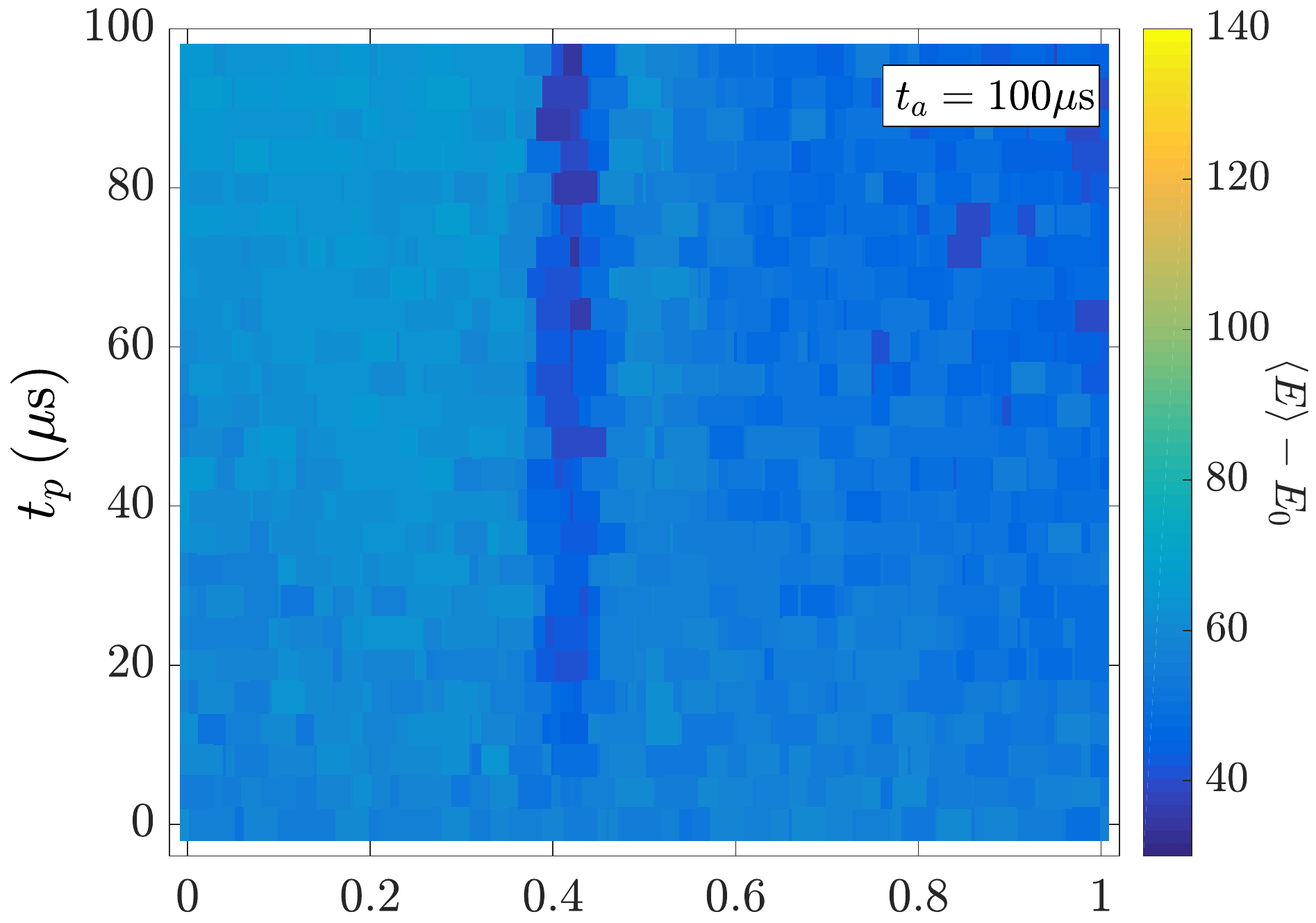}
\includegraphics[width=0.8\columnwidth]{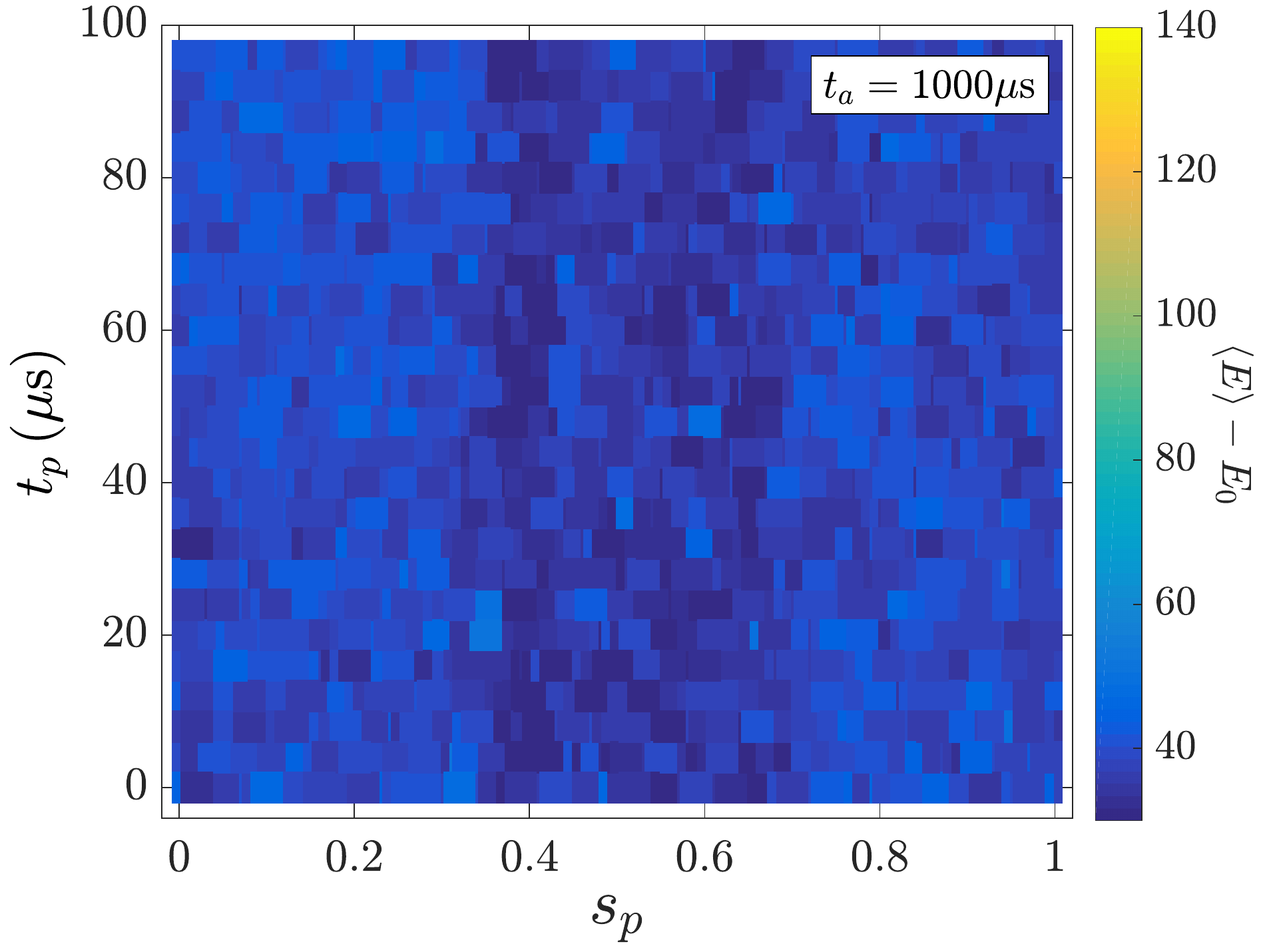}
\caption{Effect of changing the total annealing time (not including the pause time), $t_a$, for a 501 qubit planted problem instance ($\mathcal{I}_{501}^1$). The heat map color corresponds to the average energy, $\langle E \rangle -  E_0$ (arbitrary units) returned from the annealer. From top to bottom the total anneal time $t_a = 1,10,100,1000\mu$s (see legend).
 In the bottom figure, with longest anneal time, the pause has little to no effect.
Each data point is averaged from 5000 anneals with 5 different choice of gauge.}
\label{fig:heat_ta}
\end{center}
\end{figure}

In the main text (Sect.~\ref{sect:FA}), one effect we studied was varying the total anneal time $t_a$, but keeping the pause length $t_p$ constant (e.g.~Fig.~\ref{fig:vary_ta}). Here, in Fig.~\ref{fig:heat_ta} we show the corresponding heat map (i.e. where $t_p$ is also varied for each choice of $t_a$).
One sees that the pause is seemingly an efficient way of increasing the success probability (lowering average energy) without increasing the anneal time; notice for a short anneal time, $t_a=1\mu$s, with a pause of around $20\mu$s at $s\approx 0.4$ gives approximately the same average energy as an anneal for time of $1$ms.

Following on from this observation, for the planted problems of Sect.~\ref{sect:FA} we provide a basic comparison of two anneal schedules with the same total run-time ($t_a+t_p$), for one schedule with a pause, and one without. We see for all problems, the pause schedule outperforms the non-paused schedule, indicating further the possibility that this type of schedule could be used to (for example) reduce the time-to-solution. As mentioned in the main text, to fully determine any increased efficiency would require a more detailed analysis, for example, taking into account the time to estimate the optimal pause point for a particular problem class.

\begin{figure}
\begin{center}
\includegraphics[width=0.98\columnwidth]{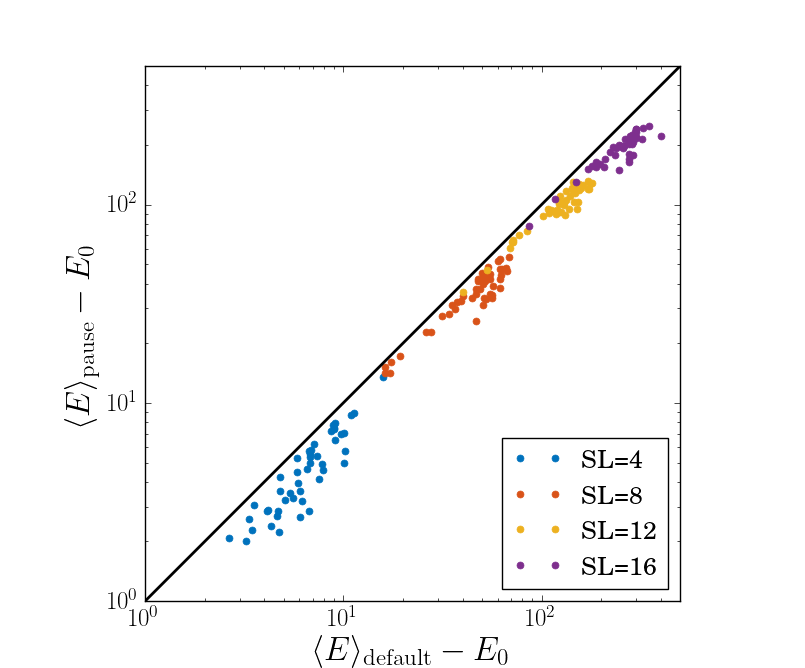}
\includegraphics[width=0.98\columnwidth]{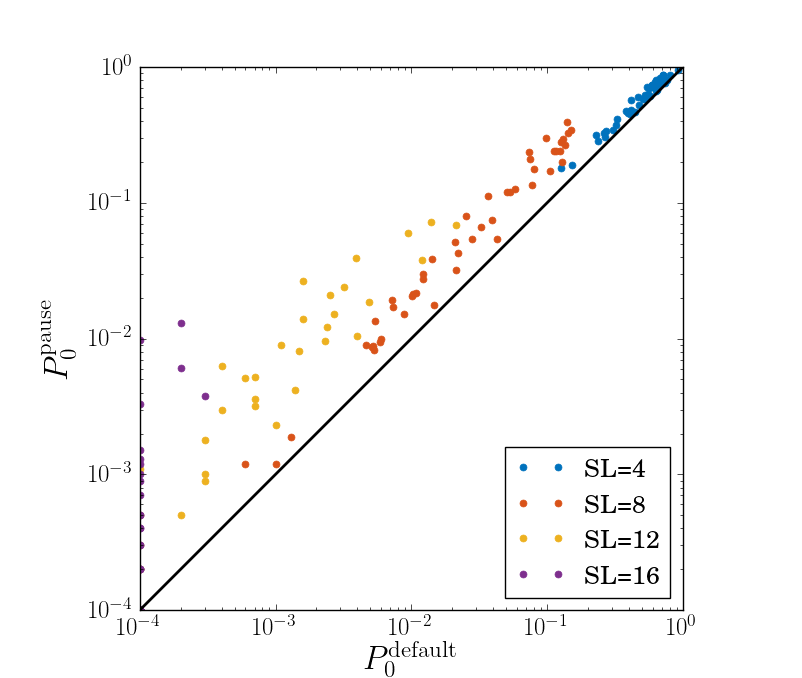}
\caption{Comparison of default annealing schedule against schedule with pause at optimal pause point, for the planted problems of Fig.~\ref{fig:scaling_SL}. We use a pause of $t_p=5.5\mu$s inserted into a schedule otherwise with a total anneal time $t_a=5.5\mu$s (total execution time $11\mu$s). The default schedule is run with $t_a=11\mu$s, and no pause. Data for each point is an average over 50000 anneals, with 50 gauges. Top: comparison of average energy (arbitrary units) returned by the device. We see the schedule with a  pause consistently returns samples with lower average energies. Bottom: comparison of success probability. Note, those instances on the vertical axis were not solved once by the default schedule. Instances not solved by either machine are not included in the plot. We see the schedule with a pause returns samples with greater success probability.}
\label{fig:default_vs_pause}
\end{center}
\end{figure}

\begin{figure}
\begin{center}
\includegraphics[width=0.9\columnwidth]{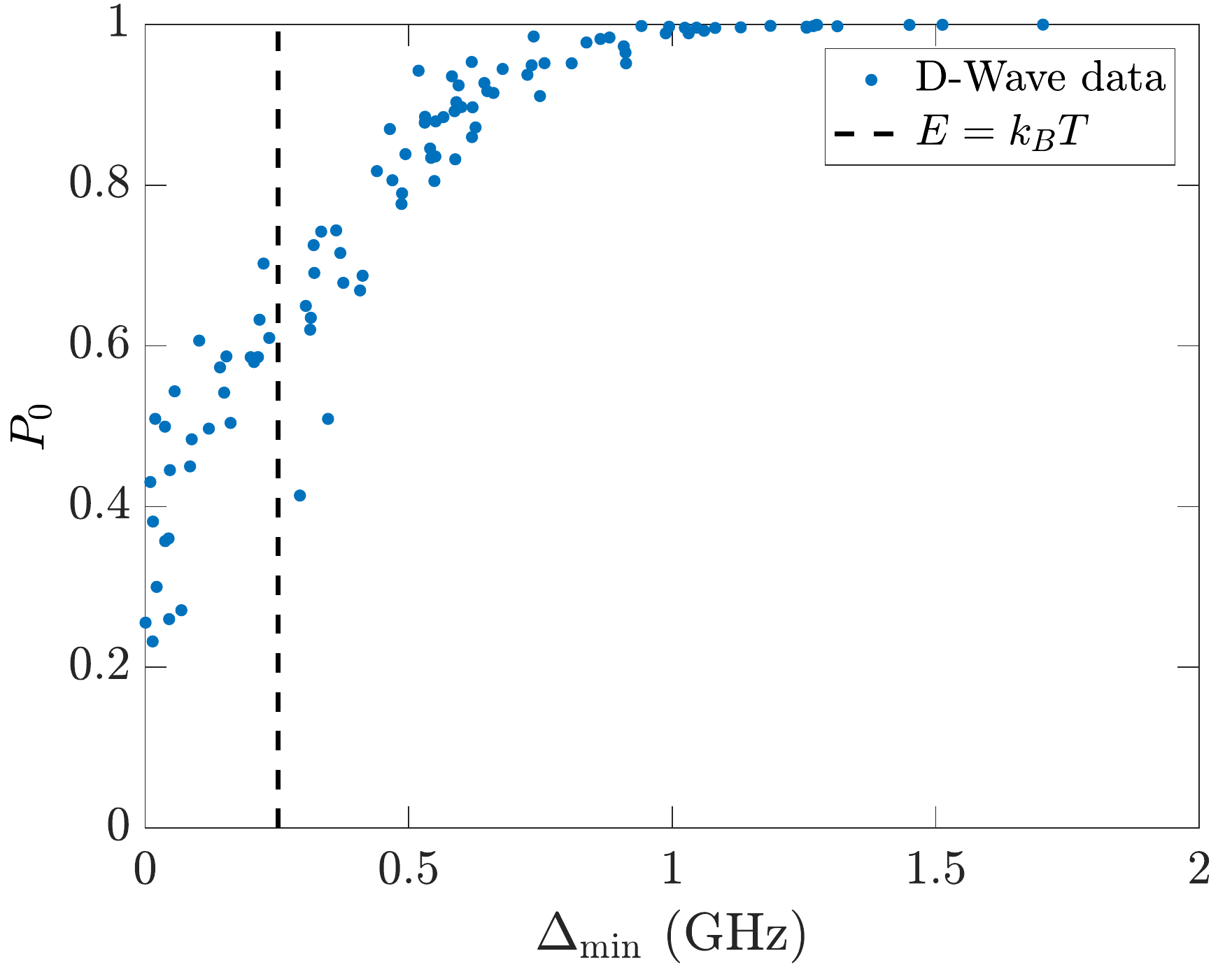}
\caption{Success probability $ P_0 $ (under the default annealing schedule with $t_a=1\mu$s) as a function of minimum gap $\Delta_{\mathrm{min}}$ for the 100 problem instances of size 12 qubits reported on in the main text. We also plot the operating temperature of the annealer (black-dash line). The units are defined with $h=1$. The data is from 10000 anneals with 10 choices of gauge.}
\label{fig:gap_P0}
\end{center}
\end{figure}

In Sect.~\ref{sect:gap} we analysed 100 problems of size 12 qubits for which $J_{ij}\in[-1,1]$ (uniformly random), and $h_i=0$. In Fig.~\ref{fig:gap_P0} we show how the success probability of these problems depends on the minimum gap $\Delta_{\mathrm{min}}$. For problems for which $\Delta_{\mathrm{min}}$ is larger than around 1GHz, the problems are solved with nearly 100\% success probability.

\begin{figure}
\begin{center}
\includegraphics[width=0.8\columnwidth]{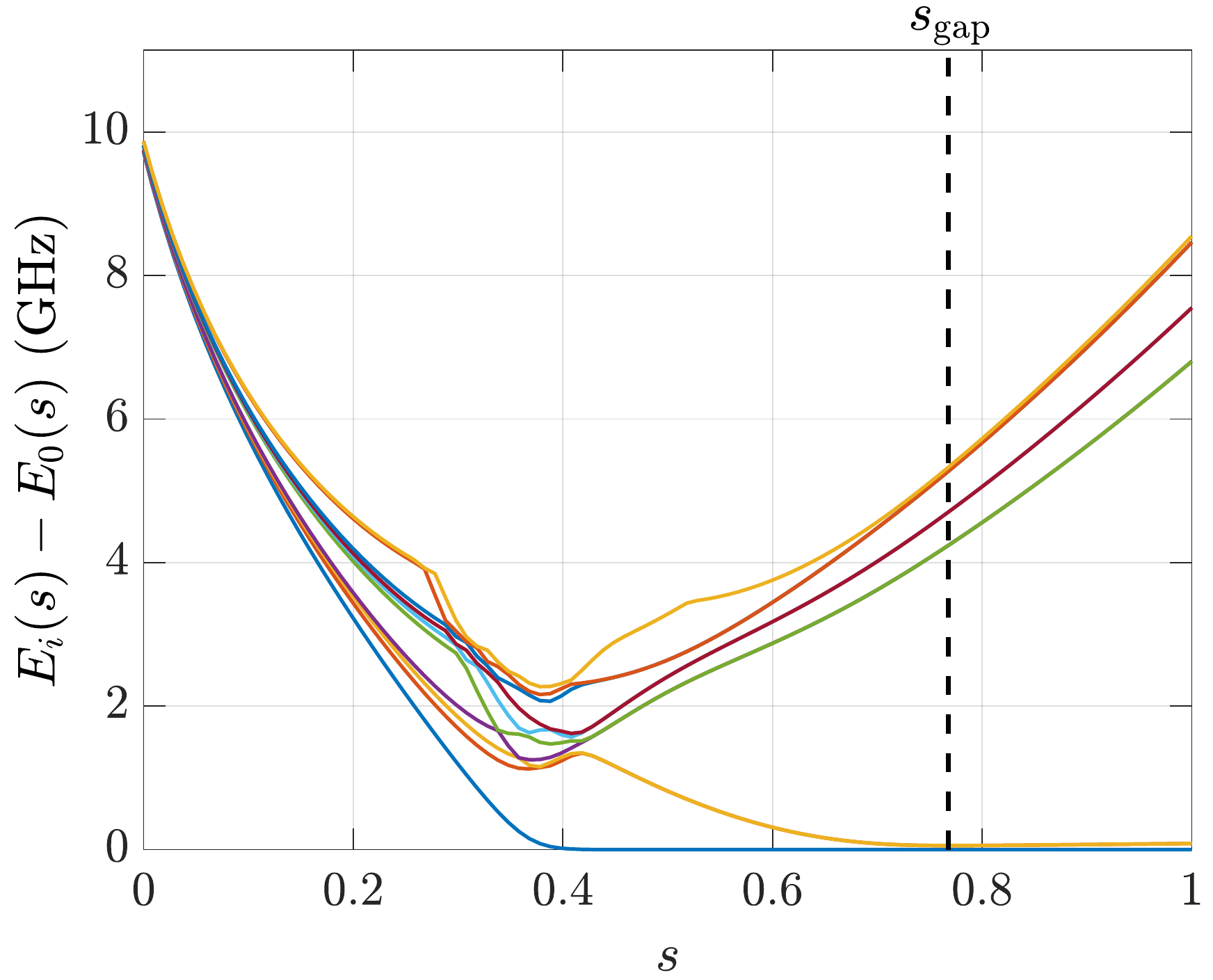}
\includegraphics[width=0.8\columnwidth]{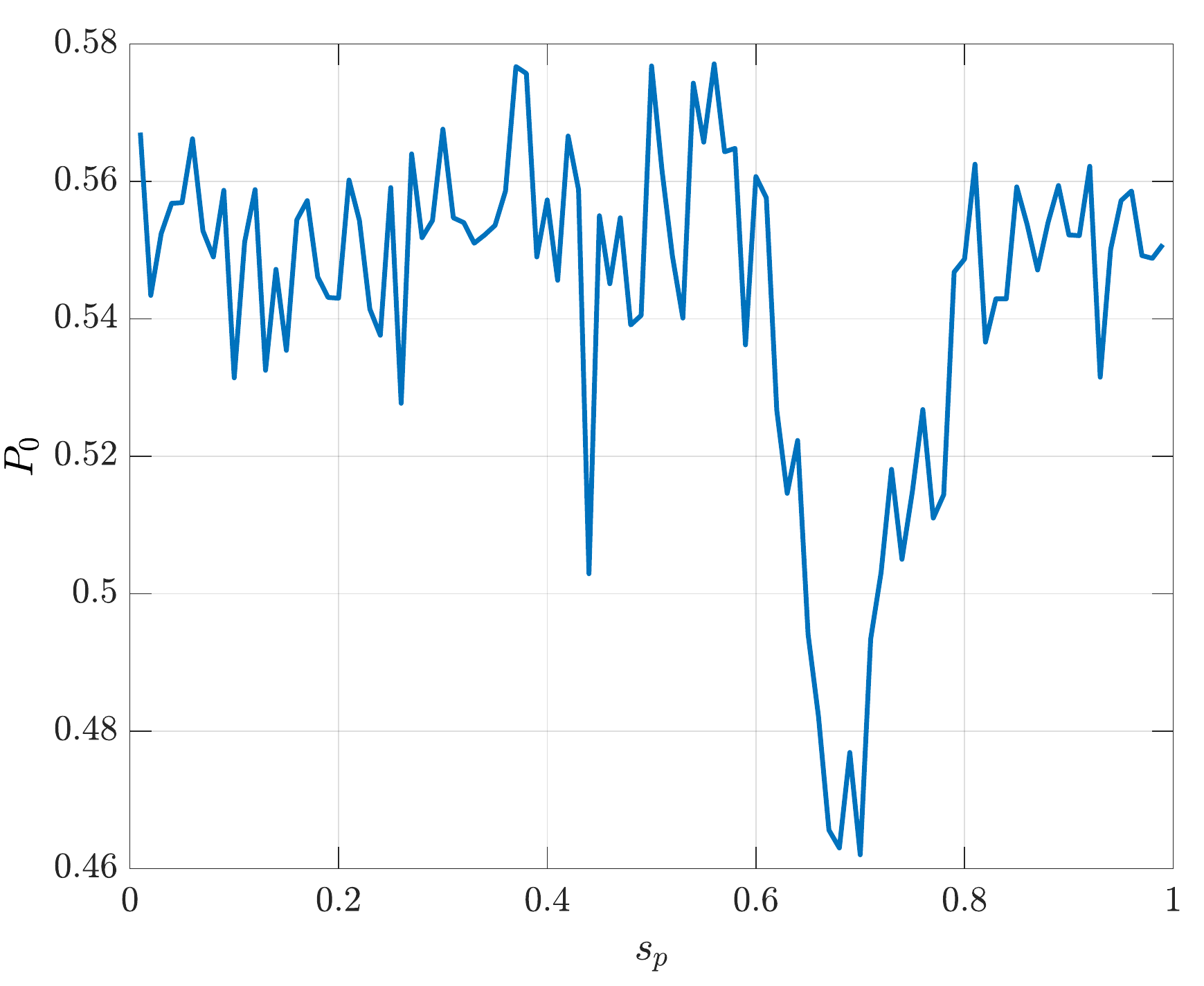}
\caption{(Top) Spectrum of one of the `outliers' $\mathcal{I}_{12}^{10}$ discussed in Sect.~\ref{sect:qbm-main}, which has a minimum gap late in the anneal ($s_{\mathrm{gap}}=0.77$), as shown in the figure. Also notice that the first excited energy level remains very close the the ground state all the way until $s=1$. Note, the blue line which approaches $E_0$ around $s=0.4$ is a ground state of $H_p$ (doubly degenerate ground state).
(Bottom) Corresponding success probability plot for the same instance where a pause of length $t_p=1000\mu$s is inserted into the anneal at point $s_p$. We see there is no clear peak (i.e. no optimal pause point), and in fact if one pauses around the location of the minimum gap there is a reduction in success probability, most likely due to excitations which do not have time to relax back down due to the gap occurring so late in the anneal. 10000 anneals, 10 gauges.
}
\label{fig:outlier}
\end{center}
\end{figure}

In Fig.~\ref{fig:outlier} we show the spectrum of one of the 12-qubit problems from Fig.~\ref{fig:sgap_spause} which we classed as an `outlier', due to it not having a well defined optimal pause point. We see the spectrum is quite different from one which has a well defined optimal pause point, e.g.~Fig.~\ref{fig:spectrum}, since the minimum gap occurs very late in the anneal, and does not open up much by the end of the anneal.
It is interesting to note that all of the larger ($>100$ qubits) problems tested had well defined optimal pause points, and therefore this is likely a small size effect.

\begin{figure}[H]
\begin{center}
\vspace{0.2cm}
\includegraphics[width=0.9\columnwidth]{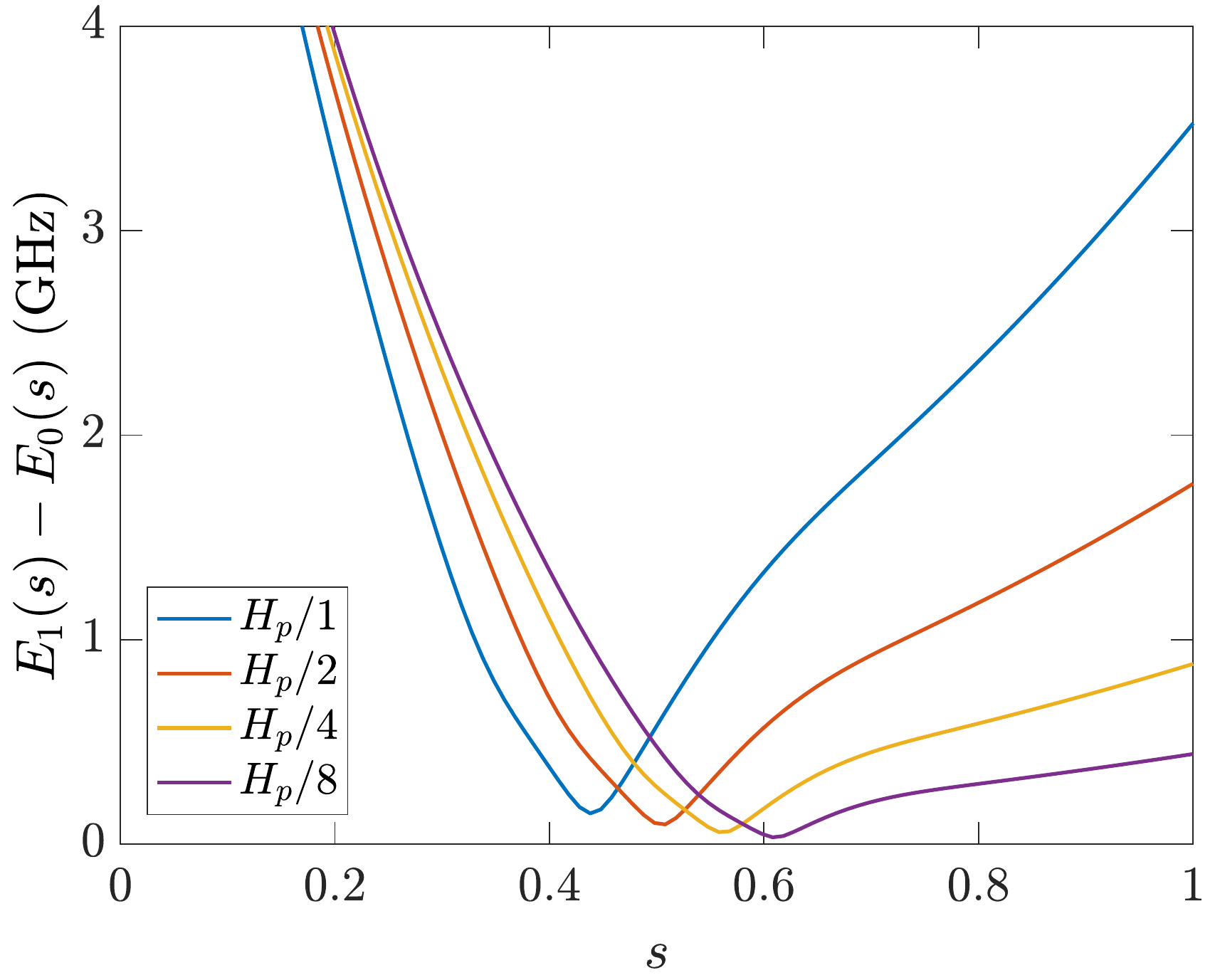}
\caption{Changing of the spectral properties upon re-scaling the problem. This is the same 12-qubit problem studied in Fig.~\ref{fig:scale} ($\mathcal{I}_{12}^0$). The location of the minimum gap changes as $s_{\mathrm{gap}} = [0.438,0.508,0.558,0.608]$, and the minimum gap itself changes accordingly as $\Delta_{\mathrm{min}}=[0.15,0.10,0.06,0.03]$ GHz, when the problem Hamiltonian is re-scaled by $[1,2,4,8]$ respectively (see legend). Energy units defined via $h=1$.}
\label{fig:gap_scale}
\end{center}
\end{figure}

\begin{figure}
\begin{center}
\includegraphics[width=0.8\columnwidth]{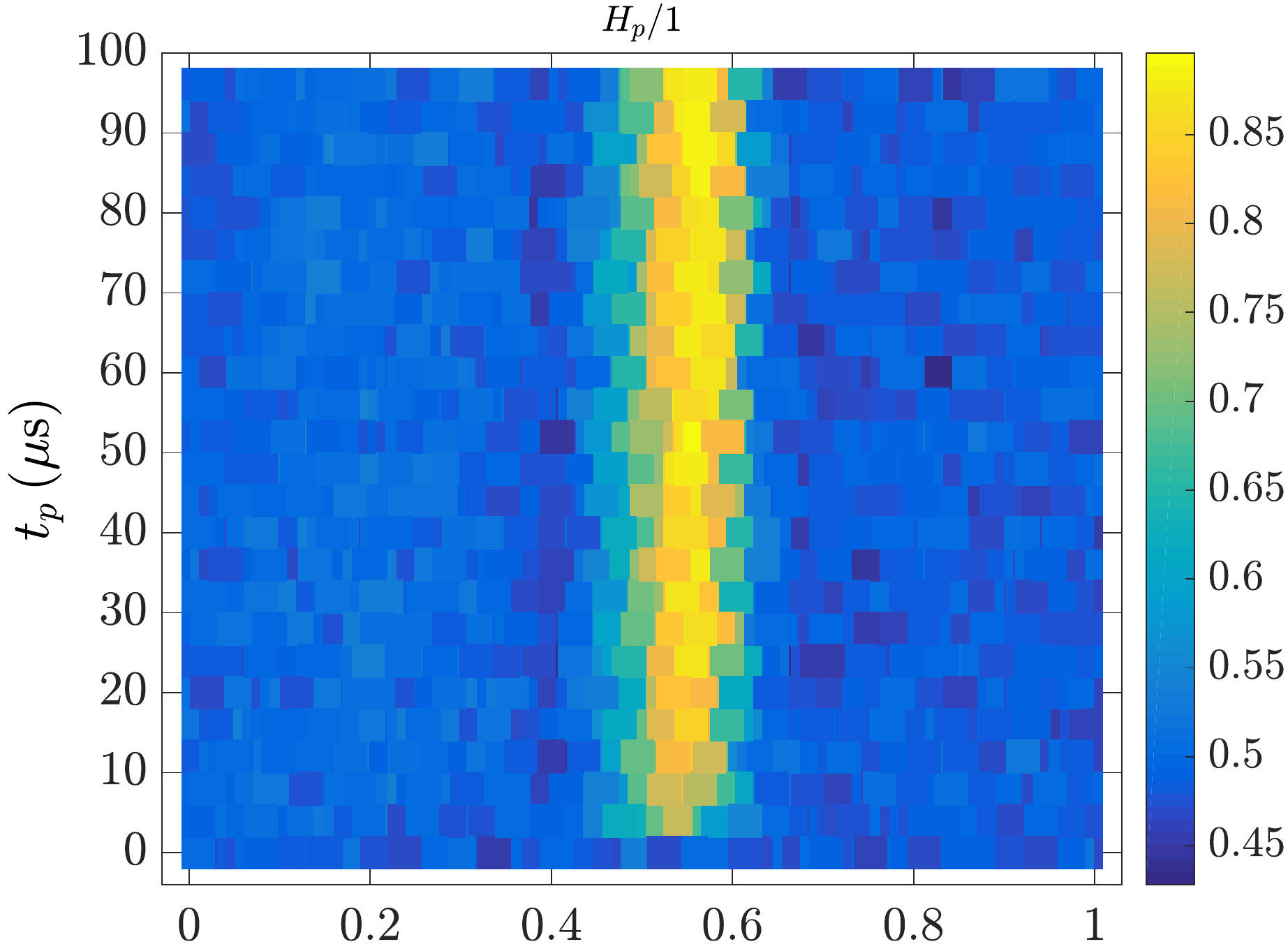}
\includegraphics[width=0.8\columnwidth]{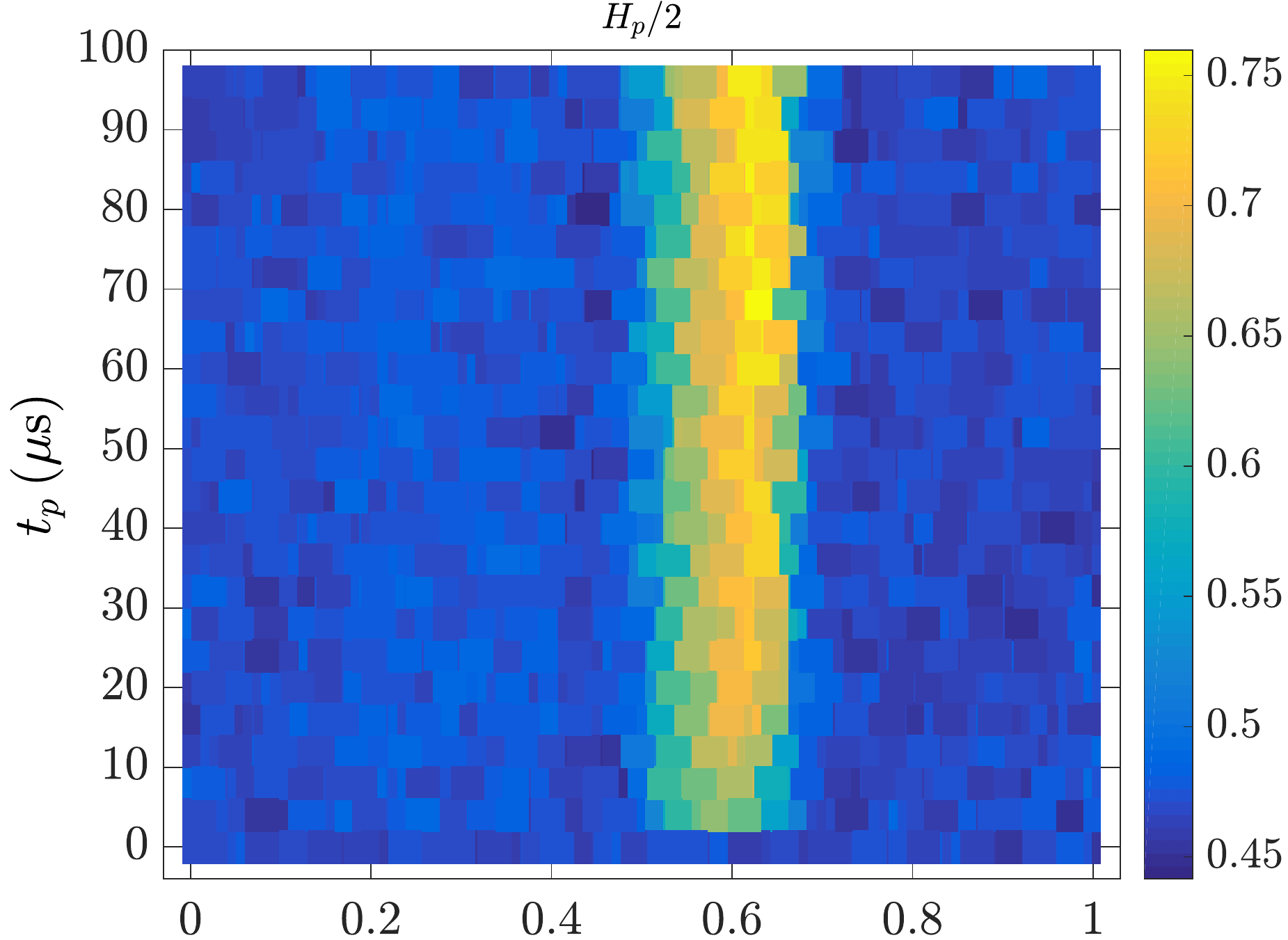}
\includegraphics[width=0.8\columnwidth]{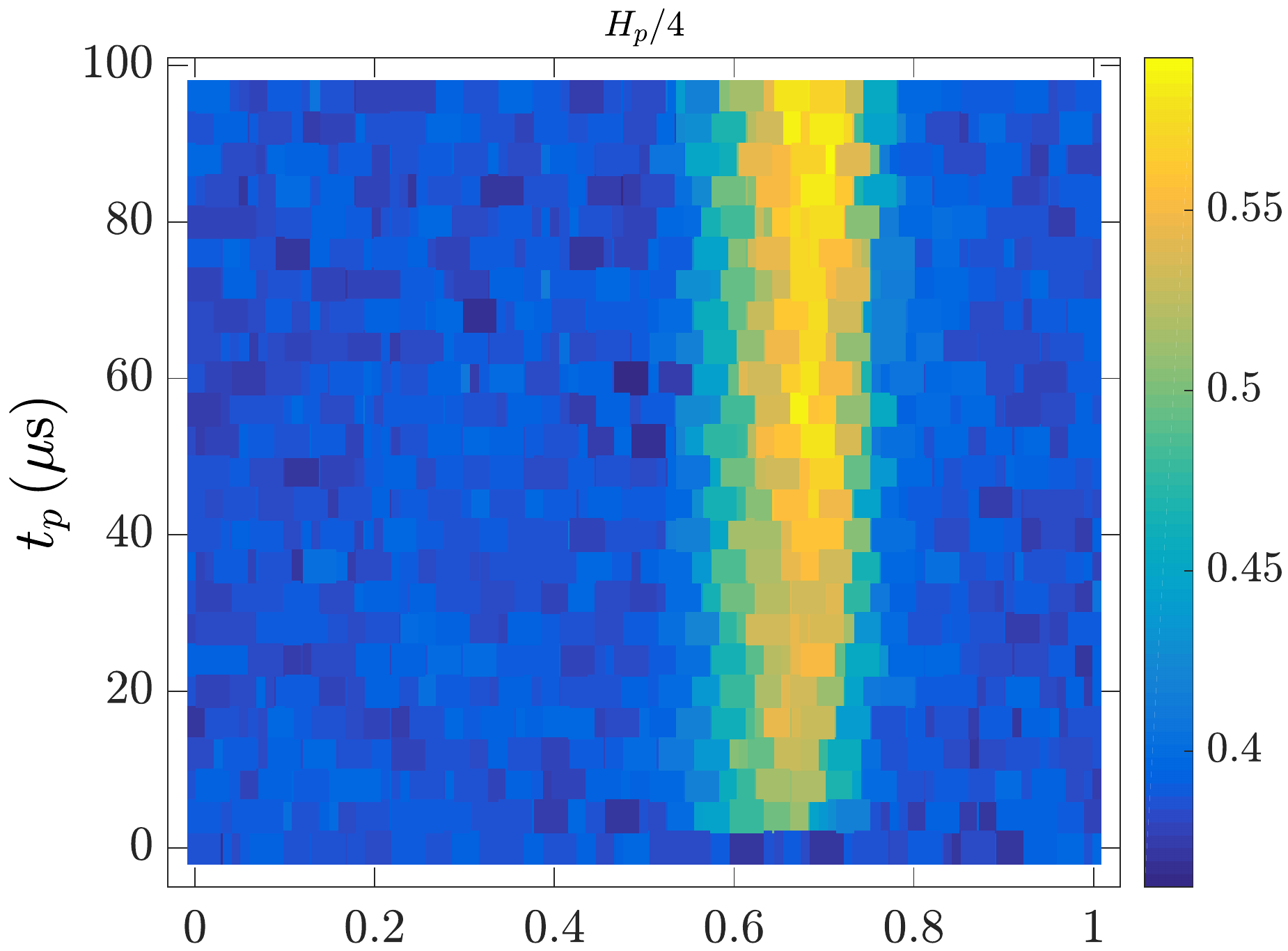}
\includegraphics[width=0.8\columnwidth]{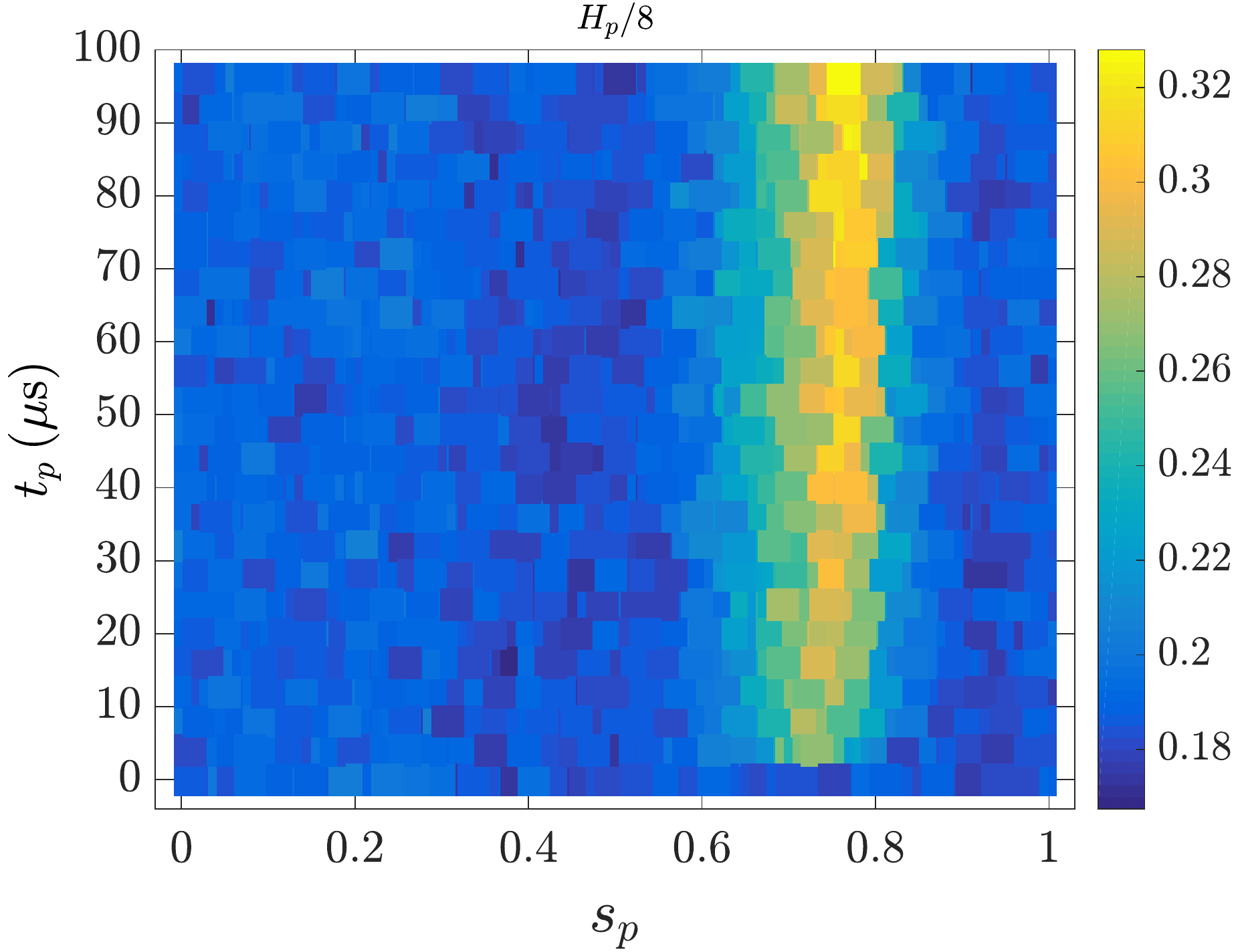}
\caption{Success probability $P_0$ heat map for a single 12-qubit instance ($\mathcal{I}_{12}^0$) where the annealing schedule has a pause of length $t_p$ inserted at $s_p$, where from top to bottom the problem Hamiltonian $H_p$ has been re-scaled by a factor of 1,2,4,8 (that is, $H_p \rightarrow H_p/C$, where $C=1,2,4,8$). See Fig.~\ref{fig:gap_scale} for the corresponding minimum gap plot. Each data point is an average from 10000 anneals with 5 gauges. Notice the change in the color bar scale between the different images.}
\label{fig:scale}
\end{center}
\end{figure}

Another interesting observation in Sect.~\ref{sect:gap} was that by dividing the energy scale of the problem Hamiltonian, $H_p\rightarrow H_p/C$ for (e.g.) $C=1,2,4,8$, was that the peak in the success probability shifts to later in the anneal. This was partly due to the minimum gap which shifts to later in the anneal (see Fig.~\ref{fig:gap_scale}), but we also related this in the main text to the diminishing quantum fluctuations $Q$. In Fig.~\ref{fig:scale} we show the corresponding heat map for a single problem instance upon dividing the problem energy scale.

\end{document}